\documentclass[11pt]{article}
\usepackage{jheppub}
\usepackage{amsmath,amssymb,amsfonts,graphicx,slashed,amsthm,mathtools,upgreek, enumerate, tensor}
\usepackage[dvipsnames]{xcolor}
\usepackage{arydshln}

\usepackage{comment}
\usepackage{hyperref}
\usepackage[utf8]{inputenc}
\usepackage[titletoc]{appendix}

\usepackage{cleveref}

\usepackage{subcaption}

\usepackage{braket}
\usepackage{bm}
\usepackage{cancel}

\numberwithin{equation}{section} 
\allowdisplaybreaks

\allowdisplaybreaks

\newcommand{\be}{\begin{equation}}
\newcommand{\ee}{\end{equation}}

\newcommand{\bea}{\begin{eqnarray}}
\newcommand{\eea}{\end{eqnarray}}
\newcommand{\ba}{\begin{align}}
\newcommand{\ea}{\end{align}}

\newcommand{\ra}{\rangle}
\newcommand{\beq}{\begin{equation}}
\newcommand{\eeq}{\end{equation}}

\newcommand{\al}{\alpha}

\DeclareMathOperator{\tr}{tr}

\title{
Black Hole Interior and Quantum Error Correction with Dynamical Gravity
}

\author[a]{Akihiro Miyata}
\author[b]{\! Tomonori Ugajin}

\affiliation[\,a]{Center for Gravitational Physics and Quantum Information (CGPQI), \\
Yukawa Institute for Theoretical Physics (YITP), Kyoto University,\\
	Kitashirakawa Oiwakecho, 
    Sakyo-ku, Kyoto 606-8502, Japan}
\affiliation[\,b]{Department of Physics, Rikkyo University, Toshima, Tokyo 171-8501, Japan}

\emailAdd{akihiro.miyata@yukawa.kyoto-u.ac.jp}
\emailAdd{ugajin@rikkyo.ac.jp}

\abstract{
According to the island formula, information in the code subspace defined in the black hole interior is embedded in the Hawking radiation after the Page time. At first sight, this embedding suggests that operations acting on the Hawking radiation could modify the information in the code subspace, potentially leading to an apparent violation of causality. Indeed, in previous studies based on the PSSY model, which incorporates only the topological degrees of freedom of gravity, it was shown that when the error is sufficiently large, a violation of causality can arise, as indicated by a nonvanishing mutual information.

In this paper, we investigate the situation in which dynamical gravity also acts on the Hawking radiation. In this case, operations on the Hawking radiation induce nontrivial backreaction on the bulk spacetime appearing in the gravitational path integral for the mutual information — an effect that is absent when the Hawking radiation is non-gravitating. We find that this backreaction renders the relevant mutual information vanishing. This result implies that, in theories with dynamical gravity, the apparent violation of causality is resolved.
}


\preprint{}

\begin{document}

\maketitle

\parskip=10pt

\section{Introduction}

Quantum error correction and holography are deeply intertwined. Within the framework of the AdS/CFT correspondence \cite{Maldacena:1997re}, the idea of entanglement wedge reconstruction \cite{Almheiri:2014lwa,Harlow:2016vwg} asserts that the bulk region reconstructible from a boundary subregion A is precisely the region bounded by A and its Ryu-Takayanagi surface \cite{Ryu:2006bv,Ryu:2006ef}. For this to hold, the Hilbert space of the bulk effective QFT must be understood as a quantum error-correcting code embedded within the Hilbert space of the boundary CFT \cite{Almheiri:2014lwa,Harlow:2016vwg}. 

Moreover, it has been argued that a similar error-correcting code structure also emerges in the case of evaporating black holes \cite{Hayden:2007cs,Verlinde:2012cy,Yoshida:2021xyb,Nakayama:2023kgr}. As a concrete example, one may consider the Hayden–Preskill protocol \cite{Hayden:2007cs}. Let us take a black hole that evaporates by emitting Hawking radiation. After the Page time, such a black hole is approximately maximally entangled with the Hawking radiation \cite{Page:1993wv,Page:2013dx}. In this setup, Hayden and Preskill demonstrated, using tools of quantum information theory and modeling the black hole dynamics by a Haar-random unitary, that the information of an object (a diary) thrown into the black hole becomes encoded in the Hawking radiation almost immediately \cite{Hayden:2007cs}. By regarding the Hilbert space of the diary as a code subspace embedded in the Hilbert space of the Hawking radiation, the system can be viewed as a quantum error-correcting code with respect to the erasure of the black hole degrees of freedom \cite{Hayden:2007cs,Yoshida:2021xyb,Nakayama:2023kgr}.

In the semi-classical spacetime picture, the island formula \cite{Penington:2019npb,Almheiri:2019psf,Almheiri:2019hni,Penington:2019kki,Almheiri:2019qdq} provides the realization of the embedding of black hole interior information into Hawking radiation as in the Hayden-Preskill protocol. The island formula computes the entropy of Hawking radiation, and, much like the HEE formula \cite{Ryu:2006bv,Ryu:2006ef,Hubeny:2007xt,Faulkner:2013ana,Engelhardt:2014gca} in the AdS/CFT correspondence, it determines the entanglement wedge of the Hawking radiation inside the black hole — the island region. In the semi-classical picture, the recovery of the diary in the Hayden-Preskill protocol can thus be understood as the diary falling into the island region.

The island formula is typically formulated in a setup where a non-gravitating heat bath is attached outside the black hole, serving as a reservoir for the Hawking radiation \cite{Almheiri:2019hni,Penington:2019npb}. If the black hole interior is embedded in the Hawking radiation, it might seem that operations performed on the radiation in the heat bath could modify the interior state. Since the heat bath and the black hole interior are not causally connected, however, such a possibility would imply a violation of causality\footnote{In appendix \ref{app:CausalityViolation}, we summarize the relation between the condition for the error recoverability and commutativity of operators on the code subspace and the Kraus operators for the error.}.

Kim, Tang, and Preskill  argued that operations on the Hawking radiation with complexity polynomial in the black hole entropy cannot alter the interior \cite{Kim:2020cds}. More concretely, they showed that when the interior is encoded jointly in the black hole and its radiation, it is protected against low-complexity errors acting on the radiation by showing the decoupling condition \cite{Kim:2020cds} holds. Balasubramanian et al. \cite{Balasubramanian:2022fiy} reproduced this result using a simple topological model of gravity  with end-of-the-world (EoW) branes \cite{Penington:2019kki}. Nonetheless, in the presence of operations of sufficiently high complexity the decoupling condition does not hold, therefore these operations  can indeed change the black hole interior.

These results were derived under the assumption that gravity does not act on the heat bath storing the Hawking radiation. In our actual universe, by contrast, gravity does act on the Hawking radiation once it has propagated far from the black hole. In particular, highly complex quantum operations on the radiation would require enormous energy, making their gravitational backreaction impossible to ignore. In this paper, we investigate the properties of the black hole interior as a quantum error-correcting code in situations where gravity also acts on the Hawking radiation.

We carry out this computation in two models of two-dimensional gravity. The first is the West Coast model \cite{Penington:2019kki}, which incorporates only the topological degrees of freedom of gravity. In this model, the microscopic states of the black hole are realized by EoW branes. We further construct the states of the Hawking radiation using EoW branes, and examine whether decoupling occurs. We find that even in this model, when the errors are sufficiently large, the decoupling condition breaks down in a brief interval after the Page time, as seen by computing a specific R\'enyi-2 mutual information (RMI). This happens because, in that parameter regime, the fully connected wormhole linking the boundary corresponding to the Hawking radiation, the boundary corresponding to the black hole degrees of freedom, and their replica copies becomes the dominant saddle for a certain entropy contribution to the RMI.

We then investigate the same condition in JT gravity \cite{Jackiw:1984je,Teitelboim:1983ux,Maldacena:2016upp}, which includes dynamical degrees of freedom. In this case, due to the backreaction induced by the errors, the fully connected wormhole cannot be the dominant saddle across all parameter regimes, and therefore decoupling holds.

This means that once the dynamical degrees of freedom of gravity are taken into account, no matter how large the errors are, they cannot alter the black hole interior. 
After the Page time, a spacetime configuration is realized in which the 
black hole and Hawking radiation are connected through a wormhole, and 
the code subspace lies within the entanglement wedge of the Hawking 
radiation. However, the fact that gravity acts on the Hawking radiation 
implies that the gravitational backreaction induced by errors deforms the spacetime, protecting the code subspace. The detailed mechanism of this protection remains to be fully clarified. 

The rest of this paper is organized as follows. In Section \ref{sec:QECNobackreaction}, we first introduce the models of gravity and explain the condition under which the bulk Hilbert space is encoded in the physical Hilbert space consisting of Hawking radiation and black hole microstates. We then compute the R\'enyi-2 mutual information, which characterizes the decoupling, using gravitational path integrals in the West Coast model. We show that in this topological model, decoupling fails for a brief interval after the Page time when the error is large, due to the dominance of the fully connected saddle. In Section \ref{sec:backreaction}, we compute the same R\'enyi-2 mutual information in the presence of dynamical degrees of freedom, namely in JT gravity. We show that the backreaction of the error acting on the radiation degrees of freedom makes the action of the fully connected wormhole sufficiently large to render it subdominant. As a result, the R\'enyi-2 mutual information vanishes, and the decoupling condition always holds.
In Section \ref{sec:discussion}, we conclude the paper and outline future directions.

\section{QEC in a gravitating bath in the West Coast model} \label{sec:QECNobackreaction}

\subsection{Black hole interior and QEC}

The problem we are interested in is whether the Hilbert space of the black hole interior degrees of freedom, when embedded into the Hilbert space of the bipartite system of the black hole microstates  and the Hawking radiation, functions as a quantum error-correcting code against errors acting on the Hawking radiation. In particular, \cite{Balasubramanian:2022fiy} considered this problem in the case where gravity does not act on the Hawking radiation, and concluded that sufficiently complex operations can modify the interior degrees of freedom. In this paper, we analyze how this conclusion is modified when gravity acts on the Hawking radiation.

Let us briefly recall how an evaporating black hole can be identified with a quantum error correcting code when the radiation degrees of freedom are non-gravitating. If we denote by $H_{BH}$ the Hilbert space of microstates of a black hole, and by $H_{HR}$ the Hilbert space of the Hawking radiation, then the state on the bipartite system $H_{BH} \otimes H_{HR}$ is given by
\be
|\psi \rangle = \frac{1}{\sqrt{k}} \sum_{\al=1}^{k} |\psi_{\al} \rangle_{BH} \otimes | \al \rangle_{HR}.
\ee
By increasing $k$, the entanglement between the black hole and the Hawking radiation becomes large. This means that naively $k$ corresponds to the time elapsed since the black hole began evaporating.

Here, the microscopic black hole state $|\psi_{i}\rangle_{BH}$ is realized as a state with an EoW brane inserted in the bulk \cite{Penington:2019kki}. The gravitational path integral associated with $|\psi_{\al}\rangle$ includes a sum over brane configurations in the bulk spacetime. On the other hand, the Hawking radiation part is non-gravitating and contains no nontrivial dynamics.

The Hilbert space of the bulk effective QFT on the black hole can be divided into the part corresponding to the interior degrees of freedom, $H_{in}$, and the part corresponding to the exterior degrees of freedom, $H_{ex}$. A QFT state with excitations inside the black hole, $|i\rangle_{in}$, is embedded into the full state as
\begin{equation}
    |i \rangle_{in} \;\;\rightarrow \;\; |\psi_{i} \ra_{phys}=\frac{1}{\sqrt{k}} \sum_{\al} |\psi_{\al,i} \rangle_{BH} \otimes |\al \rangle_{HR}. \label{eq;entsstate}
\end{equation}

Through this embedding, when we define a quantum error-correcting code with $H_{in}$ as the code subspace and $H_{BH} \otimes H_{HR}$ as the physical Hilbert space $H_{phys}$, we would like to understand whether it is protected against operations on the Hawking radiation $H_{HR}$, which are described by a quantum channel $\mathcal{E}$. For this purpose, in addition to $H_{phys}$ we introduce a reference system $H_{ref}$ isomorphic to the code subspace $H_{code}$, and an environment system $H_{env}$, and consider a state on $H_{ref} \otimes H_{phys} \otimes H_{env}$
\be
|\Psi \ra =\sum_{i=1}^{d_{in}}\sum_{i'=1}^{d_{ex}}\sum_{m=1}^{d_{env}} |i,i'\ra_{ref} \otimes E_{m}|\psi_{i,i'} \ra_{phys} \otimes |e_{m} \ra_{env} \label{eq;Totalstate}
\ee
where $E_{m}$ denotes the Kraus operators of $\mathcal{E}$, and the dimension of the environment system is determined by the rank of the error. States in the code subspace are labeled by two indices $i,i'$ because we also included excitations on the exterior region specified by the index $i'$.

The condition that the black hole interior states are protected against $\mathcal{E}$ is that the reduced density matrix on $H_{ref}\otimes H_{env}$ factorizes, namely
\begin{equation}
	\begin{aligned}
		\rho_{ref(in),\, ref(ex),\, E}' = \rho_{ref(in)}' \otimes \rho_{ref(ex),\, E}.
	\end{aligned}\label{eq:decoupling-condition}
\end{equation}
This is called the decoupling condition \cite{Schumacher:1996dy,Nielsen:1996pv,Hayden:2016pru,Dupuis:2010jrv,Dupuis:2014tkp}.

Our interest is in the case where gravity also acts on the Hawking radiation. The simplest way to model this gravitational effect is to describe the states of the Hawking radiation in terms of  EoW branes. This corresponds to considering two entangled black holes living in universes A and B, respectively \cite{Balasubramanian:2021wgd,Miyata:2021ncm,Anderson:2020vwi}. Now the states on the physical Hilbert space take the following form, 
\begin{equation}
	\sum_{\alpha=1}^{k} \ket{\psi^{\alpha}}_{A}^{*} \ket{\psi^{\alpha}}_{B},
\end{equation}

The new effects that appear when gravity acts on the radiation are: (1) the existence of a wormhole connecting universes A and B, which must be included in the gravitational path integral, and (2) Errors acting on the Hawking radiation play the role of  a source of  the gravitational dynamics. In this paper, we analyze how these effects influence the decoupling condition.

To check the decoupling condition, we study the following R\'enyi mutual information 
\begin{equation}
	\begin{aligned}
			I^{(2)}(ref(in),\, ref(ex)\cup E)&=S^{(2)}(\rho_{ref(in)}') + S^{(2)}(\rho_{ref(ex)\cup E}') - S^{(2)}(\rho_{ref(in),\,  ref(ex)\cup E}'),
	\end{aligned} \label{eq:renyiTwoAsymmetri}
\end{equation}
by computing gravitational path integrals (GPIs) involving EoW branes.  We will do so in two kinds of theories: (1) The first theory is topological and it is  called the West Coast model \cite{Penington:2019kki}  where each saddle contributes to the GPI through its Euler character.
 (2) The second theory  is JT gravity with a dynamical dilaton and dynamical branes.

In this section we perform the calculation of \eqref{eq:renyiTwoAsymmetri} in the topological model (the West Coast model). Since the gravitational backreactions from errors are absent in this model, we obtain a qualitatively similar result to that for the non-gravitating bath. 
In the next section \ref{sec:backreaction}, we study the same setup  in dynamical JT gravity where  the gravitational backreaction plays a very important role for the robustness of the system under the errors.

\subsection{Dynamical JT gravity with a gravitating bath}

To explain the setup,  we introduce full dynamical theory, namely the (Euclidean) two-dimensional Jackiw-Teitelboim (JT) gravity and end-of-the-world (EoW) branes with tension $\mu$ first, then make it topological to simplify the calculations.

The action of the full theory is 
\begin{equation}
	\begin{aligned}
	    I_{JT+\text{EoW}}=I_{\text{Topo. Dilaton}}+I_{\text{Dyn. Dilaton}}+I_{{ \rm EoW}}, \label{eq:westAction}
	\end{aligned}
\end{equation}
where $I_{\text{Topo. Dilaton}}$ and $I_{\text{Dyn. Dilaton}}$ are defined by
\begin{equation}
\begin{aligned}
    I_{\text{Topo. Dilaton}}=-S_{0}\chi ,
\end{aligned}\label{eq:TopoAction}
\end{equation}
\begin{equation}
   I_{\text{Dyn. Dilaton}}= - \left[ \frac{1}{2}\int_{\mathcal{M}} \phi (R+2) + \int_{\partial \mathcal{M} } \sqrt{h}\phi (K-1)  \right],\label{eq:DynamiAction}
\end{equation}
and $S_{0}$ is the extremal entropy, $\chi$ is the Euler characteristic of the (Euclidean) spacetime $\mathcal{M}$, $K$ is an extrinsic curvature of the boundary $\partial \mathcal{M}$. 
The AdS boundary conditions on the metric and the dynamical dilaton are
\begin{equation}
	ds^{2}|_{\partial \mathcal{M}} = \frac{du^{2}}{\epsilon^{2}}, \qquad \phi|_{\partial \mathcal{M}} =\phi_{b}  = \frac{\phi_{r}}{\epsilon},\label{eq:bdyMetricDilaton}
\end{equation}
where $\epsilon$ is a cutoff parameter, $u$ is a physical boundary time with Euclidean periodicity $\beta$, $\phi_{b}$ is a boundary dilaton value, and $\phi_{r}$ is a (constant) renormalized dilaton value.
We note that, for the validity of the description in JT gravity, it is necessary to impose the following condition that the topological dilaton part $I_{\text{Topo. Dilaton}}$ is sufficiently larger than the dynamical dilaton part $I_{\text{Dyn. Dilaton}}$,
\begin{equation}
\phi_{0} \gg \phi_{b}.\label{eq:TopologicalDominantCond}
\end{equation}

In addition to them,  we have the action for the EoW brane, 
\be
I_{{ \rm EoW}}= \mu \int_{\text{brane}}ds,
\ee
where the integral is defined along the  (Euclidean ) brane trajectory. The equations of motion for the brane profile
 impose boundary conditions on the EoW branes,
\begin{equation}
	\left.\partial_{n}\phi \right|_{\text{brane}} = \mu, \quad \left.K\right|_{\text{brane}}=0,
\end{equation}
where $\partial_{n}$ is an outward normal derivative to the EoW branes. 
Note that, due to the condition \eqref{eq:TopologicalDominantCond}, in the West Coast model, dominant contributions are given by the topological part of the action $I_{\text{Topo. Dilaton}}$.

\subsection{Branes describing the code excitations}

 We also need to define the code subspace,  which consists of excitations on the black hole interior in the bulk effective QFT Hilbert space. 
 
Such excitations are modeled by additional bulk branes, which we refer to as code branes with flavors on the geometries, whose action is given by 
\begin{equation}
	I_{\text{code}}=m \int_{\text{brane}'}ds, \label{eq:bulkExcitationAction}
\end{equation}
where $m$ is an energy scale of bulk excitations.
In this modeling, the possible brane configurations on background geometries  are interpreted as different trajectories of bulk excitations, and the flavors are considered as the labels of states of bulk excitations.

\subsection{Topological model (West Coast model)}\label{subsec:topolo}
Since the fully dynamical theory is a bit involved, we consider a simpler theory where we set $I_{\text{Dyn. Dilaton}} =I_{\rm EoW} = 0$ and keep the contribution of the topological part. The contributions of both types of branes are treated by the standard rule \cite{Penington:2019kki}, namely, when we connect two brane profiles with indices $i$ and $j$, we associate it with the Kronecker delta $\delta_{i,j}$\footnote{We forbid connecting two different kinds of branes.}.

\subsection{Embedding of the code subspace}

In this system described by the gravitational  action, we consider the situation that two gravitating universes $A$ and $B$ including black holes are entangled, and one of them, say, the universe $B$ is regarded as the gravitating bath system collecting Hawking quanta. This system is treated in the topological model \cite{Anderson:2020vwi}.
 This, as in \eqref{eq;entsstate}, results in the (un-normalized) entangled state between the two universes \cite{Anderson:2020vwi},
\begin{equation}
	\sum_{\alpha=1}^{k} \ket{\psi^{\alpha}}_{A}^{*} \ket{\psi^{\alpha}}_{B},
\end{equation}
where $\ket{\psi^{\alpha}}_{A}$ and $\ket{\psi^{\alpha}}_{B}$ denote gravitating states with EoW branes in state $\alpha$ on systems $A$ and $B$, respectively.  
The parameter $k$ characterizes the entanglement between the two systems, and the CPT conjugation is applied for the universe $A$ to make geometries orientable. For simplicity, we consider the micro-canonical ensemble and assume that black holes in the system $A$ and $B$ have the same black hole entropy $S_{BH}$, which is equal to the extremal entropy $S_{BH}=S_{0}$ in the current simplified situation.
 We also assume the planar limit $k,e^{S_{BH}}\gg 1$ to simplify our computations.

\begin{figure}[th]
	\centering
	\includegraphics[scale=0.8]{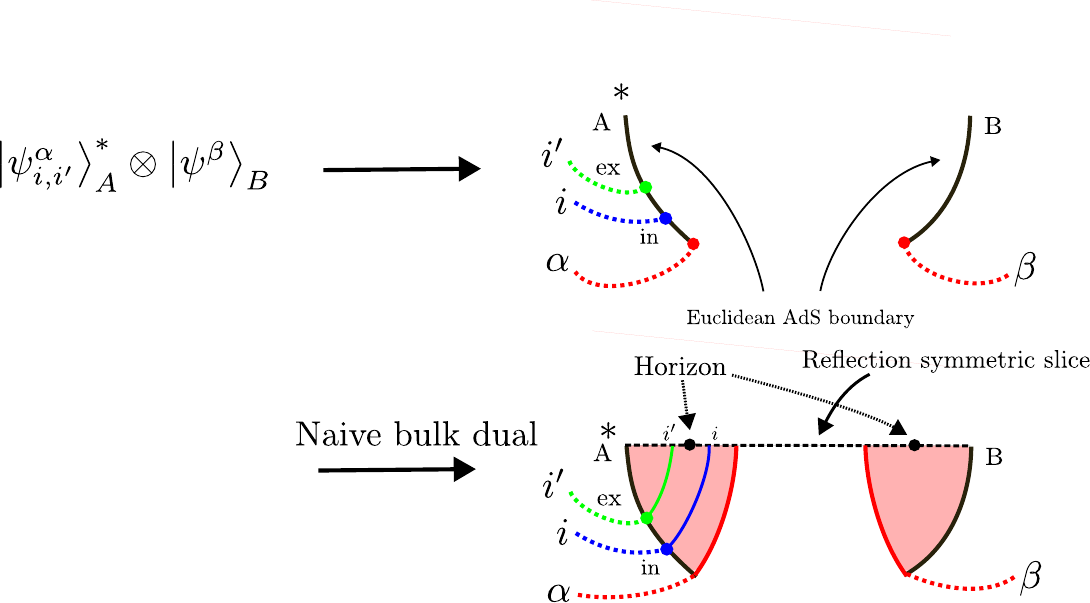}
	\caption{Diagrammatic representation of the boundary condition for the state $\left|\psi_{i, i^{\prime}}^\alpha\right\rangle_A^{*} \otimes\left|\psi^\beta\right\rangle_B$ and its naive bulk dual. The red solid lines denote the EoW branes, and the green and blue solid lines are bulk branes corresponding to exterior and interior code excitations. The red, blue and green dashed lines denote the labels of states for the EoW brane, interior and exterior code excitations respectively. The star $*$ denotes the CPT conjugation acting on the universe $A$ to keep the orientation of the total system $AB$. We refer to it as the naive bulk dual because there is another possibility for bulk geometry where the two branes are connected.}\label{fig:State-bdy_condition}
\end{figure}

On this gravitating system, we consider bulk semi-classical excitations for the black hole interior and exterior in the universe $A$.
They form a bulk Hilbert space on a semi-classical saddle geometry, $H_{code}$, and the Hilbert space consists of the interior and exterior ones, i.e., $H_{code}=H_{code,in}\otimes H_{code,ext}$. Let $\ket{i,i'}_{\text{code}} \, (i = 1,\cdots,d_{in}, i'= 1,\cdots , d_{ex})$ denote an orthonormal code state with interior and exterior excitations in the state $i$ and $i'$ respectively, and the bulk physical excitations on the system can be introduced by embedding the code states into the entangled state, in particular, the universe $A$. Then, the system state is given by
\begin{equation}
	\ket{\Psi_{i,i'}}_{phys} = V \ket{i,i'}_{\text{code}}  = \frac{1}{\sqrt{N}} \sum_{\alpha=1}^{k} \ket{\psi_{i,i'}^{\alpha}}_{A}^{*} \ket{\psi^{\alpha}}_{B}, \label{eq:phsyStateTwoGraASym1}
\end{equation}
where $V$ is an embedding map from the code Hilbert space $H_{code}$ to the physical Hilbert space $H_{phys}$, and $N$ is a normalization factor, which we will determine later. We need to clarify the precise meaning of the state $\ket{\psi_{i,i'}^{\alpha}}_{A}$ on the universe $A$ with interior and exterior code excitations in the state $i$ and $i'$ respectively.  Each of these excitations is specified by  two code branes with flavors $i$ and $i'$ respectively (the green and blue lines in figure \ref{fig:State-bdy_condition}). Through the AdS/CFT correspondence,  we can prepare these  states $\ket{\Psi_{i,i'}}_{phys} $  by appropriately inserting suitable CFT defects on the Euclidean AdS boundary \cite{Marolf:2017kvq}. The CFT defects are dual to the bulk branes with the flavor $i$ and $i'$, connecting between the defects.

To obtain consistent bulk excitations, we impose two assumptions on the CFT defect and the dual branes.
First, we assume that the number of the flavors, which corresponds to the dimensions of the interior  and exterior code Hilbert spaces $d_{in}d_{ex}$, is sufficiently smaller than the exponential of the black hole entropy, $d_{in}d_{ex} \ll e^{S_{BH}}$, which is necessary for the embedding to be isometric\footnote{One would be able to generalize our discussion to non-isometric cases \cite{Balasubramanian:2022fiy,Akers:2022qdl}.}, but larger than $1$ to simplify Wick calculus; $1\ll d_{in}, d_{ex} \ll e^{S_{BH}}$.  We note that, in the current situation the black hole entropy $S_{BH}$ is given by $S_{0}$, defined in \eqref{eq:TopoAction}.
Next, the bulk brane corresponding to interior excitations is assumed to have a configuration passing through a region corresponding to a black hole interior in order to describe the interior excitations\footnote{While other bulk brane configurations are possible, they would not correspond to interior excitations. Moreover, there are different prescriptions for setting up interior configurations, and the final results depend on which prescription is adopted. One can easily check that the final results depend on the prescription.}. This condition implies that later, when we consider gravitational path integrals to compute entropies using the replica trick, the brane configurations must be such that they pass through the black hole interior. The more precise meaning of this statement will become clear when we compute the entropies.

\begin{figure}[ht]
  \centering
  \begin{minipage}[b]{0.49\linewidth}
    \centering
    \includegraphics[scale=0.9]{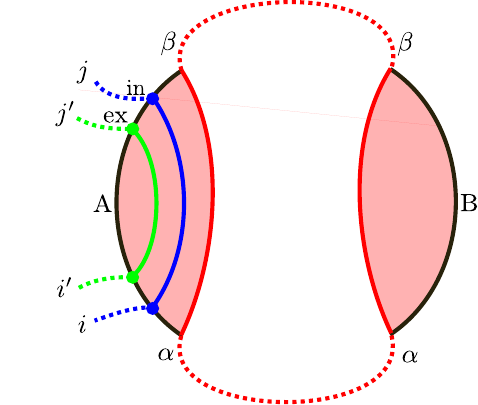}
    \subcaption{Fully disconnected saddle}\label{fig:Normalization_Haw}
  \end{minipage}
  \hfill
  \begin{minipage}[b]{0.49\linewidth}
    \centering
    \includegraphics[scale=0.9]{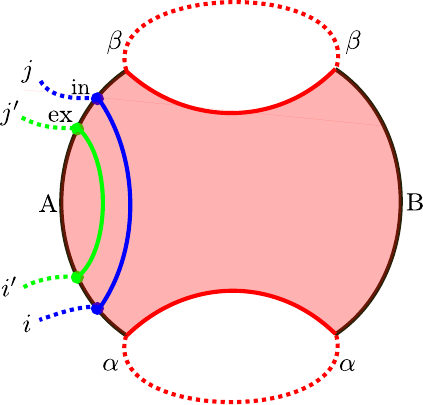}
    \subcaption{Fully connected saddle}\label{fig:Normalization_Worm}
  \end{minipage}
  \caption{Gravitational saddles appearing in the computation of the overlap used to evaluate the normalization factor $N$ in (\ref{eq:overlapPsi}). The red solid lines represent EoW branes for the black hole microstates, while the green and blue solid lines denote bulk code branes. In the topological model, these solid lines enforce Kronecker deltas between the indices associated with their endpoints. For example, the two red solid lines impose $\delta_{\alpha\beta}\delta_{\alpha\beta}$ in the left figure and $\delta_{\alpha\alpha}\delta_{\beta\beta}$ in the right figure. Dashed lines connecting the endpoints of solid lines indicate that the corresponding indices are identified and contracted. In these diagrams, the red lines form a single loop in the left panel and two loops in the right, contributing factors of $k$ and $k^{2}$, respectively, in the gravitational path integral for the overlap. Also, the number of pink regions bounded by the black and red solid lines gives the Euler characteristic of each gravitational saddle: $2$ for the left figure and $1$ for the right. These, in turn, yield factors of $e^{2S_{0}}$ for the left figure and $e^{S_{0}}$ for the right figure in the gravitational path integral for the overlap. In the following, we will omit writing the indices explicitly for notational simplicity.}
  \label{fig:our_point_HawConn}
\end{figure}
 Noting the above discussions, let us determine the normalization factor $N$ of the state \eqref{eq:phsyStateTwoGraASym1}. We first evaluate the gravitational path integral of the physical state overlap. In the topological model, we have 
\begin{equation}
	\begin{aligned}
		\overline{ \braket{\Psi_{i,i'} | \Psi_{j,j' } } } &= \frac{1}{N} \sum_{\alpha,\beta =1} ^{k}  \overline{ \braket{ \psi_{j,j'}^{\beta} | \psi_{i,i'}^{\alpha} }_{A} \braket{\psi^{\alpha} | \psi^{\beta}  }_{B}   } \\
		&= \frac{1}{N} \delta_{ij} \delta_{i'j'}\left[ k\, e^{2S_{0}} +  k^{2}\, e^{S_{0}} \right],
	\end{aligned}\label{eq:overlapPsi}
\end{equation}
where the overline means we compute the quantity by a suitable  gravitational path integral. The first term  in the second line of \eqref{eq:overlapPsi} comes from the disconnected Hawking saddle (figure \ref{fig:Normalization_Haw}), in which the two universes $A$ and $B$ are disconnected, and the second term from the connected saddle (figure \ref{fig:Normalization_Worm}), in which the two universes $A$ and $B$ are connected by a wormhole\footnote{We note that this is not a replica wormhole since it is connecting two different universes in a single replica, but not two different replicas.}. 
Thus, the normalization factor is given by
\begin{equation} 
	\begin{aligned}
		N &=  \left(k \, d_{BH} \right)^{2}  \left[ \frac{1}{k} +\frac{1}{d_{BH}} \right] \\
		&=   \left(k \,  d_{BH} \right)^{2}  \, \gamma^{2},
	\end{aligned}\label{eq:normalization}
\end{equation}
where $d_{BH}=e^{S_{0}}=e^{S_{BH}}$, and we defined 
\begin{equation}
	\begin{aligned}
		\gamma^{2} &=  \frac{1}{k} +\frac{1}{d_{BH}}. \label{eq:normaliGamma}
	\end{aligned}
\end{equation} 
Note that the normalization factor $N$ is of order  $\max\{k^{a}(d_{BH})^{b}|a+b=3, a,b\in \mathbb{Z}_{\geq 0}\}$.

Later, we use the dominant saddle approximation to compute entanglement entropies. In this  approximation, depending on the value of $k$,
 the normalization factor is given by 
 \begin{equation}
 	N \approx \begin{dcases}
 		k\, (d_{BH})^{2} & \qquad \text{ for } k < d_{BH} \\
 		k^{2}\, d_{BH}  & \qquad \text{ for }  d_{BH}  < k.
 	\end{dcases}\label{eq:normaliApprox}
 \end{equation}

\subsection{Error on the gravitating bath}

Under the above setup, we are interested in the situation that a CPTP error $\mathcal{E}$ with the Kraus representation $\{E_{m}\}_{m=1}^{d_{E}}$, which is completely-positive (CP) and trace-preserving (TP), acts on the physical state \eqref{eq:phsyStateTwoGraASym1}, in particular, the gravitating bath $B$ ,
	\begin{equation}
	\sum_{m=1}^{d_{E}} E_{m} \ket{\Psi_{i,i'}}_{phys}\otimes \ket{e_{m}}_{E} = \frac{1}{\sqrt{N}} \sum_{\alpha=1}^{k} \sum_{m=1}^{d_{E}} \ket{\psi_{i,i'}^{\alpha}}_{A}^{*} E_{m}\ket{\psi^{\alpha}}_{B} \otimes \ket{e_{m}}_{E},\label{eq:phsyStateTwoGraASymError}
\end{equation}
where $E$ denotes the environment system  implementing the error, and $\ket{e_{m}}_{E}$ is an orthonormal state in the environment system. Here, the environment system is assumed to be non-gravitating for simplicity. We note that the trace-preserving  nature of the error implies the following relation,
\begin{equation}
	\sum_{m=1}^{d_{E}}  E^{\dagger}_{m}E_{m}=I_{BH_{B}}.\label{eq:KrausTP}
\end{equation}
Here, it would be better to clarify the meaning of the error action on the gravitating bath system $B$ with the black hole.
In the above state, the Kraus operator acts on the system $B$, implying the Kraus operator acts on the black hole living in the system $B$ representing a gravitating bath storing Hawking radiation. Thus, the error acts on the Hilbert space spanned by the black hole microstates of the universe $B$ with the dimension $d_{BH}$.

\subsection{The decoupling condition}

To study whether the error $\mathcal{E}$ in the gravitating bath $B$ brings us to serious problems for the semi-classical states on $A$, we investigate the decoupling condition. If it holds, we can ``cancel" the error effect by a suitable recovery operation on the bath $B$.

 For this purpose, we introduce reference systems for the interior and exterior code  subspaces, and consider the state
\begin{equation}
	\begin{aligned}
		\ket{\Psi'} &= \frac{1}{\sqrt{N_{\Psi'}}} \sum_{i=1}^{d_{in}} \sum_{i'=1}^{d_{ex}} \sum_{m=1}^{d_{E}} \ket{i }_{ref(in)}\otimes \ket{i' }_{ref(ex)}\otimes E_{m}\ket{\Psi_{i,i'}}_{phys} \otimes\ket{e_{m}}_{E}\\
		&= \frac{1}{\sqrt{N\, N_{\Psi'} }} \sum_{i=1}^{d_{in}} \sum_{i'=1}^{d_{ex}} \sum_{\alpha=1}^{k} \sum_{m=1}^{d_{E}} \ket{i }_{ref(in)}\otimes \ket{i' }_{ref(ex)}\otimes \ket{\psi_{i,i'}^{\alpha}}_{A}^{*}\otimes (E_{m} \ket{\psi^{\alpha}}_{B})\otimes\ket{e_{m}}_{E},
	\end{aligned} \label{eq:stateBHEroGravitatingAsym}
\end{equation}
where $N_{\Psi'}$ is a normalization factor for this state. We can evaluate the norm of this state with gravitational path integral by noting the overlap \eqref{eq:overlapPsi},
\begin{equation}
	\begin{aligned}
		\overline{ \braket{\Psi' | \Psi' } }  &= \frac{1}{NN_{\Psi'}} \sum_{i=1}^{d_{in}} \sum_{i'=1}^{d_{ex}} \sum_{m=1}^{d_{E}}  \overline{\bra{\Psi_{i,i'}} E_{m}^{\dagger} E_{m}\ket{\Psi_{i,i'}}}_{phys}\\
		&= \frac{1}{NN_{\Psi'}} \sum_{i=1}^{d_{in}} \sum_{i'=1}^{d_{ex}}    \overline{\braket{\Psi_{i,i'}|\Psi_{i,i'}}}_{phys}\\
		&= \frac{1}{N_{\Psi'}} \sum_{i=1}^{d_{in}} \sum_{i'=1}^{d_{ex}}  \delta_{ij} \delta_{i'j'}\\
		&=\frac{d_{in} d_{ex}}{N_{\Psi'}},
	\end{aligned}
\end{equation}
where, in the second line, we used the trace-preserving property of the Kraus operators, \eqref{eq:KrausTP}.
 Thus, the normalization factor is given by 
\begin{equation}
	N_{\Psi'} = d_{in} d_{ex}.
\end{equation}
By using the above state, the decoupling condition is given by \eqref{eq:decoupling-condition} \footnote{In the language of Kraus operators, the decoupling condition \eqref{eq:decoupling-condition} is known to be equivalent to the Knill-Laflamme condition,
\begin{equation}
    P_{code} E_{m}^{\dagger} E_{n} P_{code} = \lambda_{mn} P_{code} \qquad m,n =1,\cdots, d_{E},\label{eq:KLcondition}
\end{equation}
where $\lambda_{mn}=\lambda_{nm}^{*}\in \mathbb{C}$, and $P_{code}$ is a projection operator onto a code subspace of the physical Hilbert space. Thus, one can investigate the Knill-Laflamme condition to study the correctability of the error $\mathcal{E}$. \label{foot:KLcondi} }. This decoupling condition is for the decoupling between the interior reference system $ref(in)$ and the subsystems $ref(ex) \cup E$ playing the role of error for the interior reference system. One can also consider other possibilities for the choice of subsystems for the decoupling condition, but we focus on the above decoupling condition \eqref{eq:decoupling-condition} for our purpose.

\begin{figure}[ht]
	\centering
	\includegraphics[scale=0.5]{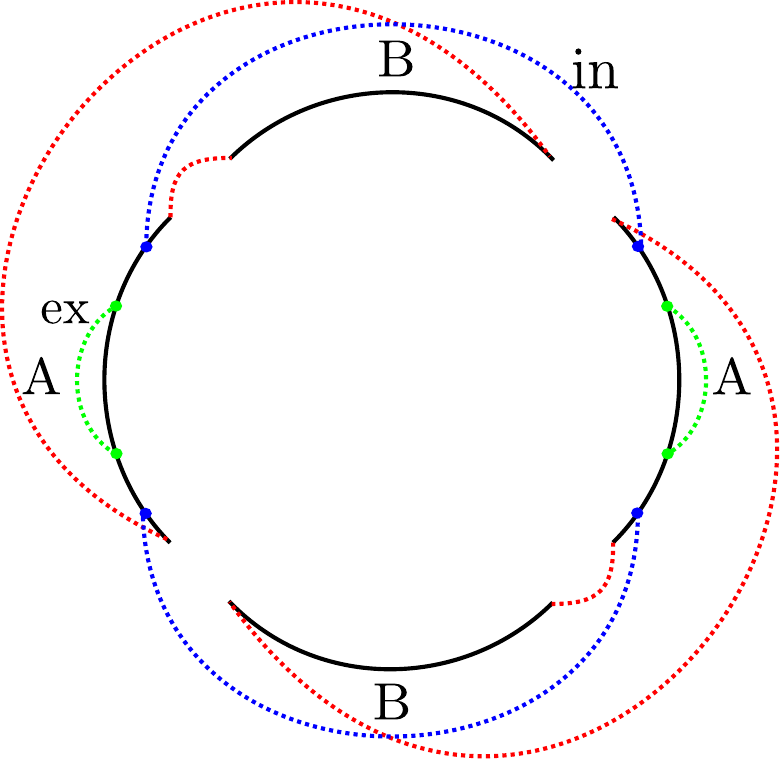}
	\caption{Diagrammatic representation of the boundary condition for $\tr \left(\rho_{ref(in)}'  \right)^{2}$, (\ref{eq:renyiTwoOnesystem}). }\label{fig:RenyiNoError}
\end{figure}

\begin{figure}[ht]
	\begin{tabular}{cc}
	\begin{minipage}[t]{0.5\hsize}
        \centering
		\includegraphics[scale=0.5]{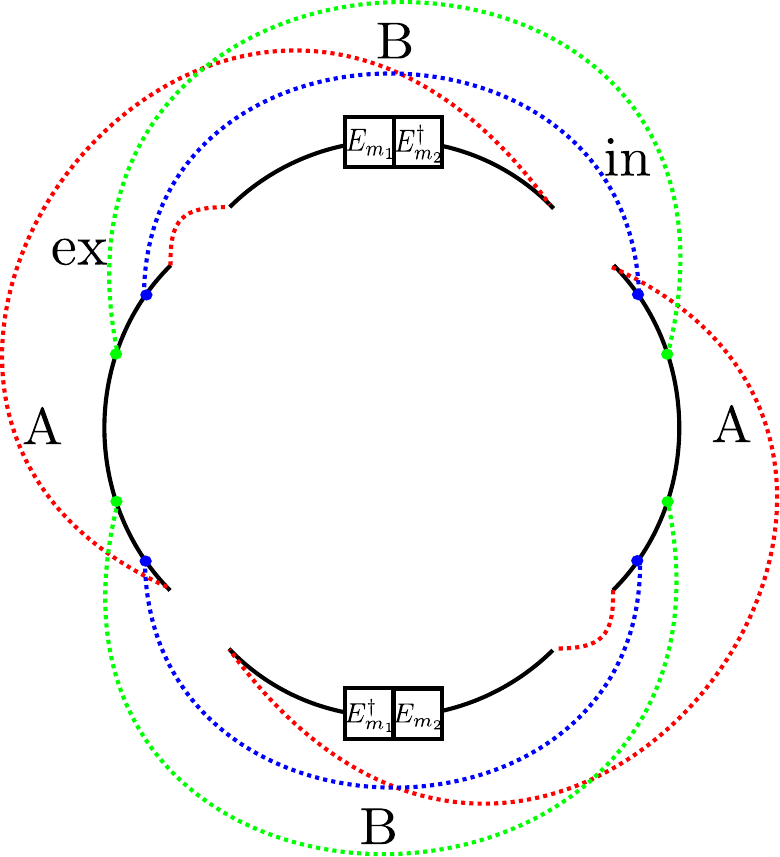}
    \subcaption{$\tr \left(\rho_{ref(in),\, ref(ex),\, E}'  \right)^{2} $}\label{fig:RenyitwoRiReEnBdyCondi}
		\end{minipage}& 
	\begin{minipage}[t]{0.5\hsize}
        \centering
		\includegraphics[scale=0.5]{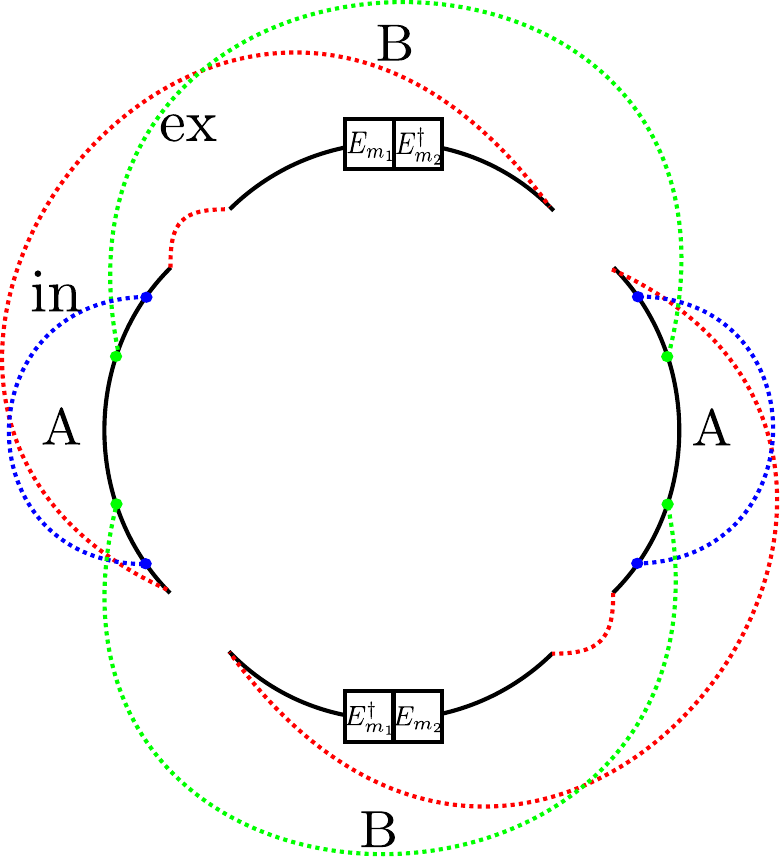}
    \subcaption{$\tr \left(\rho_{ref(ex),\, E}'  \right)^{2} $
    }\label{fig:RenyitwoReEnBdyCondi}
		\end{minipage}
	\end{tabular}
	\caption{Diagrammatic representation of boundary conditions (\ref{eq:renyiTwoThreesystem}) and (\ref{eq:renyiTwoTwosystem}), which include the Kraus operators. Only the difference between $(a)$ and $(b)$ is the way the dotted blue lines (indices of the interior code subspace) are contracted.}\label{fig:RenyiWithError}
\end{figure}

In the following, we investigate the decoupling condition by evaluating the following R\'{e}nyi-two mutual information, \eqref{eq:renyiTwoAsymmetri}.
In this case, we need to consider the gravitational path integral of the following quantities 
\begin{equation}
	\begin{aligned}
		\tr \left(\rho_{ref(in)}'  \right)^{2}&= \frac{1}{(d_{in}d_{ex} N )^{2}} \sum_{\bm{i}=1}^{d_{in}}\sum_{\bm{i}'=1}^{d_{ex}} \braket{\Psi_{i_{2},i_{1}'} | \Psi_{i_{1},i_{1}' } } \braket{\Psi_{i_{1},i_{2}'} | \Psi_{i_{2},i_{2}'} } \\
		&=\frac{1}{(d_{in} d_{ex} N )^{2}} \sum_{\bm{i}=1}^{d_{in}}\sum_{\bm{i}'=1}^{d_{ex}} \sum_{\bm{\alpha},\bm{\beta}=1}^{k} \braket{\psi_{i_{1},i_{1}'}^{\beta_{1}}|\psi_{i_{2},i_{1}'}^{\alpha_{1}} } _{A}\braket{\psi_{i_{2},i_{2}'}^{\beta_{2}}|\psi_{i_{1},i_{2}'}^{\alpha_{2}} }_{A}\braket{\psi^{\alpha_{1}}| \psi^{\beta_{1}}}_{B} \braket{\psi^{\alpha_{2}}|\psi^{\beta_{2}}}_{B},
	\end{aligned}\label{eq:renyiTwoOnesystem}
\end{equation}
\begin{equation}
	\begin{aligned}
		\tr \left(\rho_{ref(in),\, ref(ex),\, E}'  \right)^{2}&= \frac{1}{(d_{in}d_{ex} N )^{2}} \sum_{\bm{i}=1}^{d_{in}}\sum_{\bm{i}'=1}^{d_{ex}} \sum_{\bm{m}=1}^{d_{E}}\braket{\Psi_{i_{2},i_{2}'} | E_{m_{2}}^{\dagger} E_{m_{1}} | \Psi_{i_{1},i_{1}' } } \braket{\Psi_{i_{1},i_{1}'} | E_{m_{1}}^{\dagger} E_{m_{2}} | \Psi_{i_{2},i_{2}'} } \\
		 &=\frac{1}{(d_{in} d_{ex} N )^{2}} \sum_{\bm{i}=1}^{d_{in}}\sum_{\bm{i}'=1}^{d_{ex}} \sum_{\bm{\alpha},\bm{\beta}=1}^{k} \sum_{\bm{m}=1}^{d_{E}} \braket{\psi_{i_{1},i_{1}'}^{\beta_{1}}|\psi_{i_{2},i_{2}'}^{\alpha_{1}} } _{A}\braket{\psi_{i_{2},i_{2}'}^{\beta_{2}}|\psi_{i_{1},i_{1}'}^{\alpha_{2}} }_{A}\\
		& \hspace{4cm}  \times \braket{\psi^{\alpha_{1}}| E_{m_{2}}^{\dagger} E_{m_{1}} |\psi^{\beta_{1}}}_{B} \braket{\psi^{\alpha_{2}}| E_{m_{1}}^{\dagger} E_{m_{2}} |\psi^{\beta_{2}}}_{B} ,
	\end{aligned}\label{eq:renyiTwoThreesystem}
\end{equation}
and
\begin{equation}
	\begin{aligned}
		\tr \left(\rho_{ref(ex),\, E}'  \right)^{2}&= \frac{1}{(d_{in}d_{ex} N )^{2}} \sum_{\bm{i}=1}^{d_{in}}\sum_{\bm{i}'=1}^{d_{ex}} \sum_{\bm{m}=1}^{d_{E}}\braket{\Psi_{i_{1},i_{2}'} | E_{m_{2}}^{\dagger} E_{m_{1}} | \Psi_{i_{1},i_{1}' } } \braket{\Psi_{i_{2},i_{1}'} | E_{m_{1}}^{\dagger} E_{m_{2}} | \Psi_{i_{2},i_{2}'} } \\
		 &=\frac{1}{(d_{in} d_{ex} N )^{2}} \sum_{\bm{i}=1}^{d_{in}}\sum_{\bm{i}'=1}^{d_{ex}} \sum_{\bm{\alpha},\bm{\beta}=1}^{k} \sum_{\bm{m}=1}^{d_{E}} \braket{\psi_{i_{1},i_{1}'}^{\beta_{1}}|\psi_{i_{1},i_{2}'}^{\alpha_{1}} } _{A}\braket{\psi_{i_{2},i_{2}'}^{\beta_{2}}|\psi_{i_{2},i_{1}'}^{\alpha_{2}} }_{A}\\
		& \hspace{4cm}  \times \braket{\psi^{\alpha_{1}}| E_{m_{2}}^{\dagger} E_{m_{1}} |\psi^{\beta_{1}}}_{B} \braket{\psi^{\alpha_{2}}| E_{m_{1}}^{\dagger} E_{m_{2}} |\psi^{\beta_{2}}}_{B} ,
	\end{aligned}\label{eq:renyiTwoTwosystem}
\end{equation}
where the bold letters $\bm{i},\bm{i}',\bm{\alpha},\bm{m}$ denote the summation with respect to the set of indices, e.g., $ \sum_{\bm{i}=1}^{d_{in}}= \sum_{i_{1},i_{2}=1}^{d_{in}}$.
These quantities specify the boundary conditions of the gravitational path integral (see figures \ref{fig:RenyiNoError} and \ref{fig:RenyiWithError} for the diagrammatic representations of their boundary conditions), and we can evaluate their R\'{e}nyi-two geometries by  considering their gravitational path integral in a way similar to the West Coast model, albeit with modifications to treat gravitational overlaps including the Kraus operators. In appendix \ref{appsub:WCwithKraus}, we explain the explicit rules for their calculation. Below, some examples of the gravitational path integral of such overlaps are
\begin{equation}
	\overline{ \braket{ \psi^{\alpha} | E_{m}^{\dagger} E_{n} | \psi^{\beta}   } }_{B} =  \delta_{\alpha \beta} \cdot    \tr\left[ E_{m}^{\dagger} E_{n} \right],\label{eq:overlapKraus}
\end{equation}
\begin{equation}
	\begin{aligned}
		&\overline{ \braket{ \psi^{\alpha_{1}} | E_{m_{1}}^{\dagger} E_{n_{1}} | \psi^{\beta_{1}}  }_{B}   \braket{ \psi^{\alpha_{2}} | E_{m_{2}}^{\dagger} E_{n_{2}} | \psi^{\beta_{2}}   }  }_{B} \\
		 & \qquad =\delta_{\alpha_{1} \beta_{1}}  \cdot \delta_{\alpha_{2} \beta_{2}} \cdot    \tr\left[ E_{m_{1}}^{\dagger} E_{n_{1}} \right]\cdot  \tr\left[ E_{m_{2}}^{\dagger} E_{n_{2}} \right] + \delta_{\alpha_{1} \beta_{2} }  \cdot \delta_{\alpha_{2}\beta_{1}} \cdot     \tr\left[ E_{m_{1}}^{\dagger} E_{n_{1}} E_{m_{2}}^{\dagger} E_{n_{2}}  \right],	
   \end{aligned} \label{eq:overlapKraus2}
\end{equation}
\begin{equation}
	\begin{aligned}
		&\overline{ \braket{ \psi_{i,i' }^{\alpha_{1}} | \psi_{j,j'}^{\beta_{1}}  }_{A}   \braket{ \psi^{\alpha_{2}} | E_{m}^{\dagger} E_{n}  | \psi^{\beta_{2}}   }  }_{B} \\
		 &  = \delta_{ij}\delta_{i'j'}   \cdot  \delta_{\alpha_{1} \beta_{1}}  \delta_{\alpha_{2} \beta_{2}} \cdot  d_{BH} \cdot   \tr\left[ E_{m}^{\dagger} E_{n} \right]+ \delta_{ij} \delta_{i'j'}     \cdot  \delta_{\alpha_{1} \beta_{2}} \delta_{\alpha_{2} \beta_{1} }  \cdot  \tr\left[ E_{m}^{\dagger} E_{n} \right]\\
		 &= \left[\delta_{ij}\delta_{i'j'}   \cdot  \delta_{\alpha_{1} \beta_{1}}  \delta_{\alpha_{2} \beta_{2}} \cdot  (d_{BH})^{2}  + \delta_{ij} \delta_{i'j'}     \cdot  \delta_{\alpha_{1} \beta_{2}} \delta_{\alpha_{2} \beta_{1} }\cdot d_{BH}  \right]\cdot \frac{1}{d_{BH}}  \tr\left[ E_{m}^{\dagger} E_{n} \right]\\
		 & =\overline{ \braket{ \psi_{i,i' }^{\alpha_{1}} | \psi_{j,j'}^{\beta_{1}}  }_{A}   \braket{ \psi^{\alpha_{2}} |\psi^{\beta_{2}}   }  }_{B}\cdot \frac{1}{d_{BH}}  \tr\left[ E_{m}^{\dagger} E_{n} \right]
 	\end{aligned}  \label{eq:overlapKraus3}
\end{equation}
where the traces of the Kraus operators are defined on the Hilbert space spanned by the black hole microstate\footnote{For readers who are familiar with the Knill-Laflamme condition \cite{Knill:1996ny}, we give a comment on the relation between the gravitational path integrals of these overlaps and the Knill-Laflamme condition \eqref{eq:KLcondition}. As we noted in the footnote \ref{foot:KLcondi}, we can also study the Knill-Laflamme condition, instead of the decoupling condition. In that case, we can investigate whether the following equality holds or not,
\begin{equation}
	\bra{\Psi_{i}}E_{m}^{\dagger}E_{n}\ket{\Psi_{j}} \overset{?}{=} \lambda_{mn}  \delta_{ij}, \qquad \text{ for } \forall i,j, \, \forall m,n,
\end{equation}
where $\lambda_{mn}=\lambda_{nm}^{*}$ is a constant.
To check the equality, since the background geometry has a large fluctuation, we need to evaluate the gravitational path integrals of the above quantity and its variance. We can evaluate such gravitational path integrals, but in this paper, we focus only on the R\'{e}nyi-two mutual information \eqref{eq:renyiTwoAsymmetri}. }.


\subsection{R\'{e}nyi-two entropies and mutual information in the topological model}\label{subsec:RenyiTwo}

We can explicitly evaluate products of  overlaps \eqref{eq:renyiTwoOnesystem}, \eqref{eq:renyiTwoThreesystem} and \eqref{eq:renyiTwoTwosystem} by gravitational path integrals with appropriate boundary conditions, but even in the topological model their full expressions do not have simple forms. Thus, we employ  the dominant saddle approximation to simplify the expressions.
We note that in parameter regions where the dominant saddle switches, the candidate saddles contribute almost equally, making the dominant saddle approximation less accurate. However, the qualitative behavior of the results obtained using the dominant saddle approximation remains similar to that obtained by summing over all saddles. Furthermore, the result obtained by summing over all saddles can be understood as a smoothed version of the result from the dominant saddle approximation.  
Also note that in the following, when presenting the result for the R\'{e}nyi-two entropy, we use $\approx$, which represents an approximate equality based on the dominant saddle point approximation. Also, keep in mind that these results become asymptotically accurate as the difference between both sides of the associated inequalities increases.

In appendix \ref{app:renyiTwo}, we present details of the evaluation of the R\'{e}nyi-two entropies \eqref{eq:renyiTwoOnesystem}, \eqref{eq:renyiTwoTwosystem} and \eqref{eq:renyiTwoThreesystem}. 
In this subsection, we show the results of their R\'{e}nyi-two entropies after the gravitational path integral, and evaluate the R\'{e}nyi-two mutual information.

\subsubsection{R\'{e}nyi-two entropies in the topological model}

First, we give expressions of the R\'{e}nyi-two entropies,  obtained by evaluating the right hand sides of \eqref{eq:renyiTwoInFull}, \eqref{eq:renyiTwoRiReEConnectedDisco} and \eqref{eq:renyiTwoReEConnectedDisco} via gravitational path integrals in the topological model.

The dominant saddle for each entropy changes as we vary two parameters $k$ and $\;I_{c}^{(2)}\left(\frac{I_{BH}}{d_{BH}},\mathcal{E}\right)$. $k$ appears in the physical state \eqref{eq:phsyStateTwoGraASym1}, characterizing the entanglement between A and B. 

$I_{c}^{(2)}\left(\frac{1}{d_{BH}}I_{BH},\mathcal{E}\right)$ is  the R\'{e}nyi-two coherent information\footnote{The appearance of the coherent information is parallel to that for the non-gravitating bath case \cite{Balasubramanian:2022fiy}, in which the coherent information is defined with an input state given by a maximally mixed state of  Hawking radiation.} 

\begin{equation}
	I_{c}^{(2)}\left(\frac{1}{d_{BH}}I_{BH},\mathcal{E}\right)=S^{(2)}(\sigma_{\text{Bath}})-S^{(2)}(\sigma_{E}).\label{eq:coherentGraviCase}
\end{equation}

where $\sigma_{E},\sigma_{\text{Bath}}$ are defined by
\begin{equation}
		\sigma_{E}= \sum_{m,n=1}^{d_{E}} \dfrac{\tr_{BH} \left\{ E_{m}E_{n}^{\dagger} \right\} }{ d_{BH} } \ket{e_{m}}_{E} \bra{e_{n}},\label{eq:defSigmaE}
\end{equation}
\begin{equation}
	\sigma_{\text{Bath}} = \sum_{m=1}^{d_{E}} E_{m}  \left( \frac{ I_{BH} }{d_{BH}}\right) E_{m}^{\dagger}.\label{eq:defSigmaBH}
\end{equation}

Intuitively, the density matrices $\sigma_{E},\sigma_{\text{Bath}}$ correspond to coarse-grained density matrices for the environment $E$ and the gravitating bath $B$, respectively\footnote{The appearance of these coarse-grained density matrices also happens for the non-gravitating case \cite{Balasubramanian:2022fiy}. }. 

 Roughly speaking, the coherent information  quantifies the magnitude of the error acting on the radiation degrees of freedom. We will explain quantum information theoretic properties of this quantity later.

We present their results separately for early and late times, accompanied by figures of the dominant saddles.

\begin{figure}[ht]
\centering
\begin{tabular}{cc}
	  \begin{minipage}[t]{0.45\hsize}
    \centering
    \includegraphics[scale=0.5]{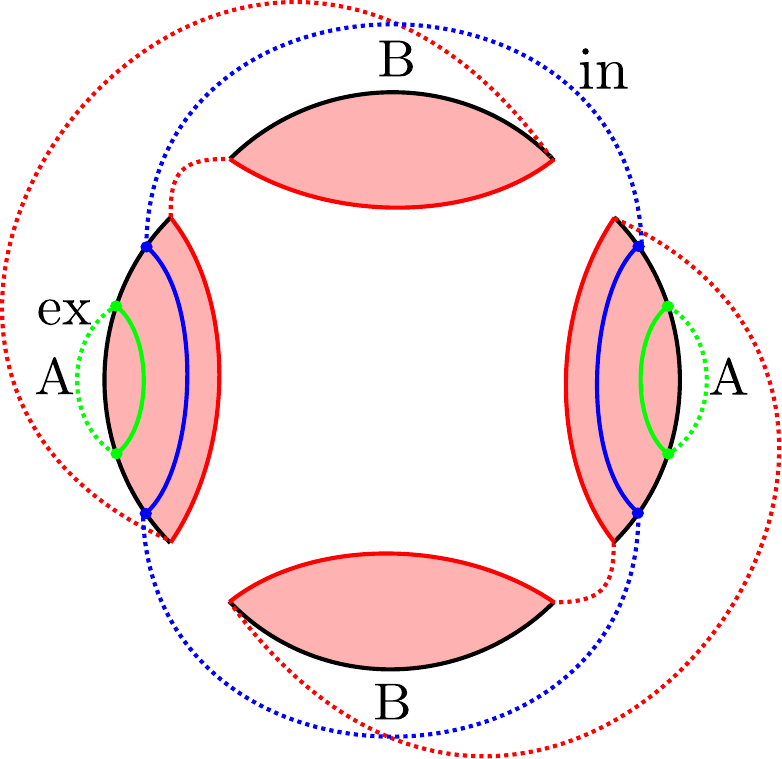}
    \subcaption{Fully disconnected saddle}\label{fig:FullDiscoMain}
  \end{minipage}
  \begin{minipage}[t]{0.45\hsize}
    \centering
    \includegraphics[scale=0.5]{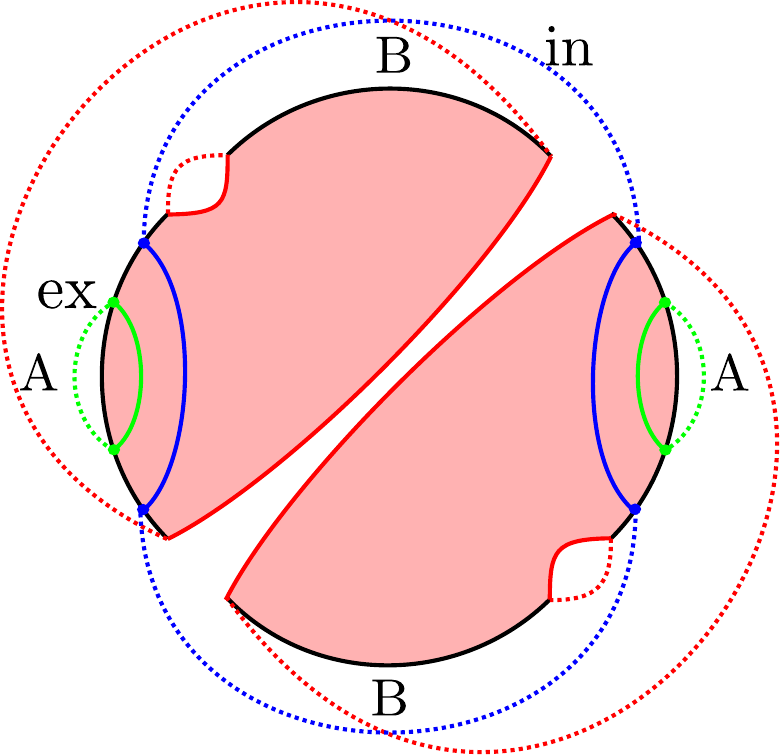}
    \subcaption{Two-$(A,B)$-wormhole saddle}\label{fig:TwoABwormWithoutErrorMain}
  \end{minipage}
\end{tabular}
  \caption{Dominant saddles for the R\'{e}nyi-two entropy, $\overline{ S^{(2)}\left(\rho_{ref(in)}'  \right) }$ . The fully disconnected saddle is dominant at early times, but as we increase the entanglement between A and B, and when  $k> d_{BH},$ wormholes connecting A and B appear.}\label{fig:RenyiRiSaddlesOtherMain}
\end{figure}

\begin{figure}[ht]
\centering
	\begin{tabular}{cc}
	\begin{minipage}[t]{0.5\hsize}
        \centering
		\includegraphics[scale=0.45]{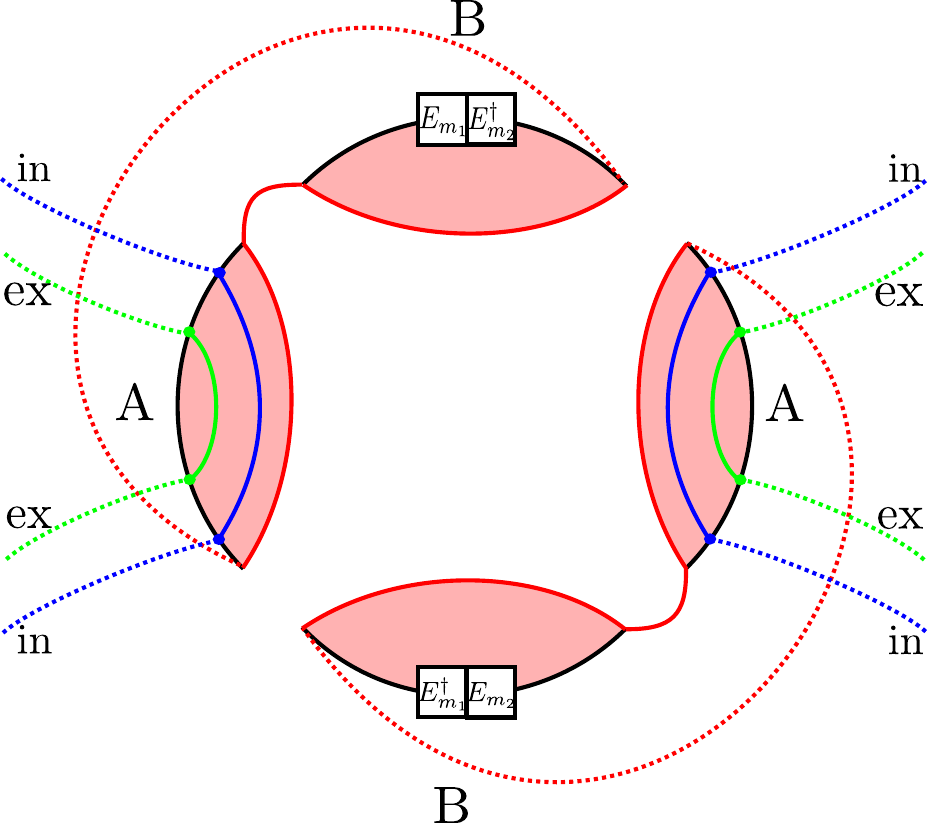}
		\subcaption{Fully disconnected saddle}
		\label{fig:FullDiscoErrorMain}
		\end{minipage}& 
	\begin{minipage}[t]{0.5\hsize}
        \centering
		\includegraphics[scale=0.45]{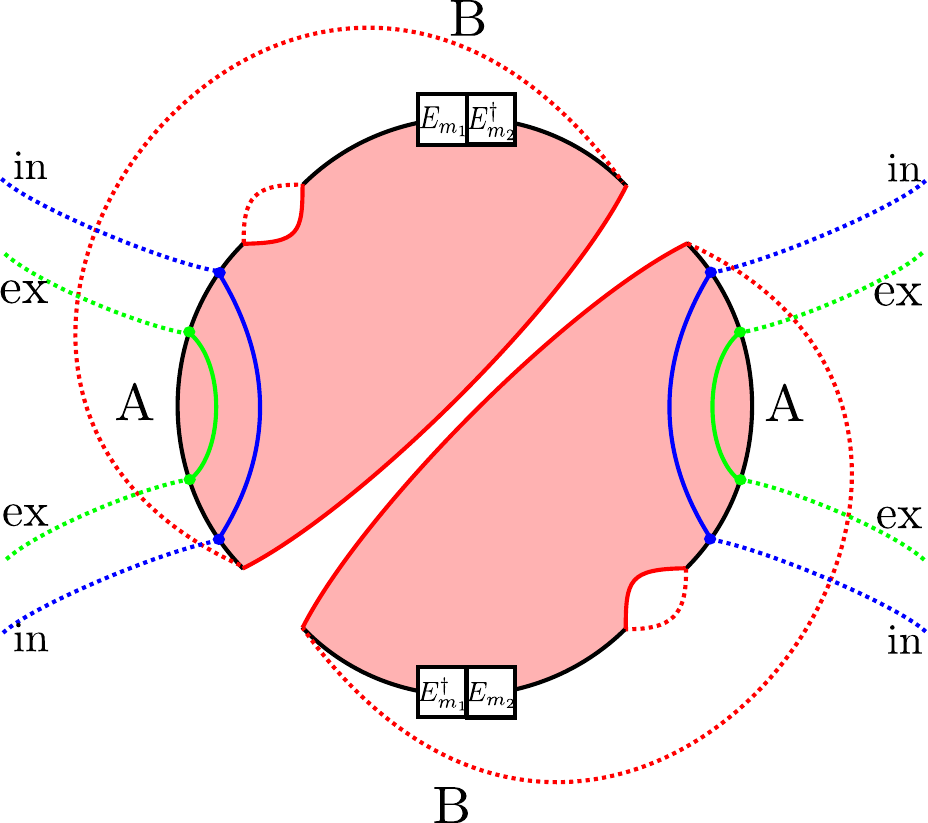}
		\subcaption{Two-$(A,B)$-wormholes saddle}
		\label{fig:TwoABWormholeErrorMain}
		\end{minipage}\\ 
	\begin{minipage}[t]{0.5\hsize}
        \centering
		\includegraphics[scale=0.45]{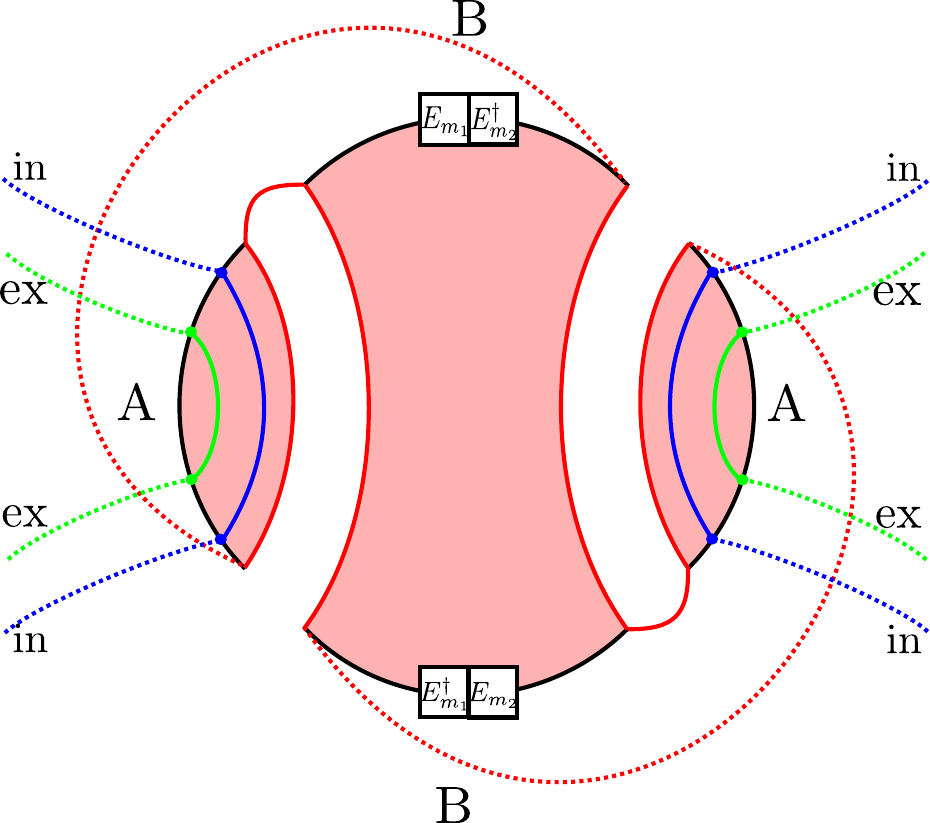}
		\subcaption{One-$(B,B)$-replica wormhole saddle}
		\label{fig:BBWormholeErrorMain}
	\end{minipage}&
	\begin{minipage}[t]{0.5\hsize}
        \centering
		\includegraphics[scale=0.45]{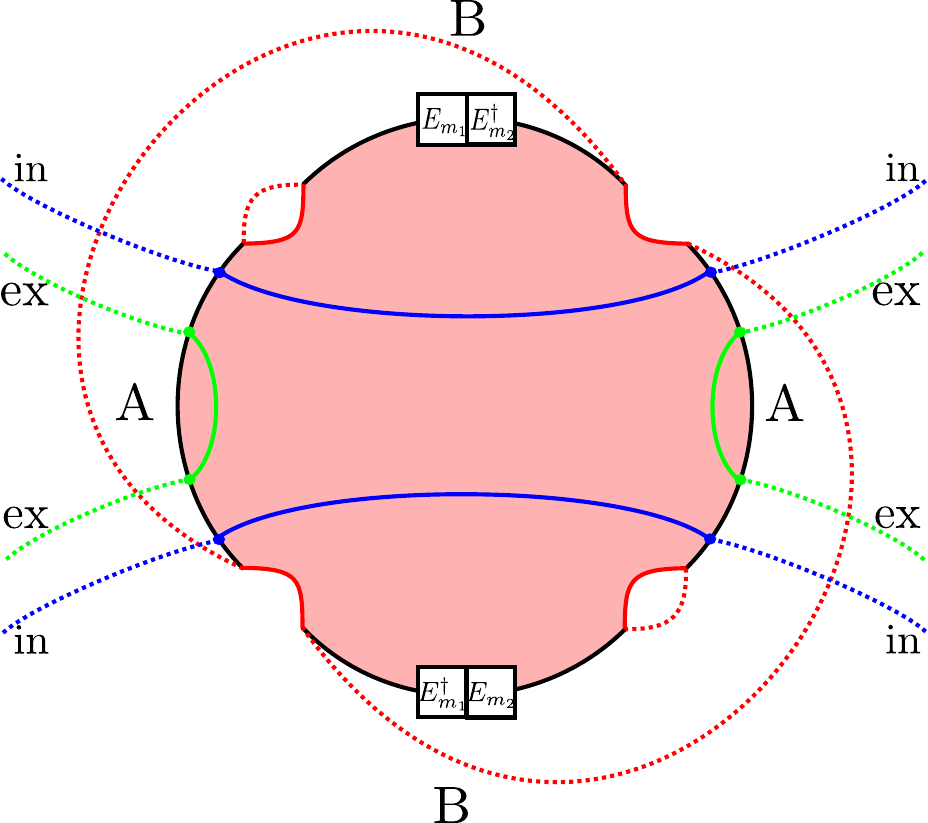}
		\subcaption{Fully connected saddle}
		\label{fig:FullConneErrorMain}
		\end{minipage}
	\end{tabular}	
	\caption{Dominant saddles for the R\'{e}nyi-two entropies, $\overline{S^{(2)}\left(\rho_{ref(in),\,ref(ex),\, E}'  \right)}$ and $\overline{S^{(2)}\left(\rho_{\,ref(ex),\, E}'  \right)}$, at early and late times. In these diagrams, we need to correctly contract blue and green dashed lines, corresponding to flavor indices of the interior and exterior excitations respectively, in a way that  the resulting contractions are consistent with the  boundary condition in figures \ref{fig:RenyitwoRiReEnBdyCondi} and \ref{fig:RenyitwoReEnBdyCondi}. Then, the resulting diagrams represent saddles to compute the R\'{e}nyi-two entropies.}
	\label{fig:DominantDiagramsForEarlyLateMain}	
\end{figure}

\paragraph{Early times $k < d_{BH}$}
For early times, the R\'{e}nyi-two entropies are given by
\begin{equation}
	\overline{ S^{(2)}\left(\rho_{ref(in)}'  \right) }  \approx \log d_{in}
 	\quad  (\text{figure } \ref{fig:FullDiscoMain}),\label{eq:renyitwoRiEarly}
\end{equation}
\begin{equation}
	\begin{aligned}
		 &\overline{S^{(2)}\left(\rho_{ref(in),\,ref(ex),\, E}'  \right)}\\
		&  \approx \begin{dcases}
			\log d_{in} +\log d_{ex} +S^{(2)}(\sigma_{E})  & \text{ for }-\log k < I_{c}^{(2)}\left(\frac{1}{d_{BH}}I_{BH},\mathcal{E}\right) \quad (\text{figure } \ref{fig:FullDiscoErrorMain}), \\
			\log d_{in} +\log d_{ex}+ \log k + S^{(2)}(\sigma_{\text{Bath}})    & \text{ for } -\log d_{BH} \leq   I_{c}^{(2)}\left(\frac{1}{d_{BH}}I_{BH},\mathcal{E}\right) <  -\log k \quad (\text{figure } \ref{fig:BBWormholeErrorMain}),		\end{dcases}
	\end{aligned} \label{eq:renyitwoRiReEnEarly}
\end{equation}
and
\begin{equation}
	\begin{aligned}
		 &\overline{S^{(2)}\left(\rho_{ref(ex),\, E}'  \right)}\\
		  & \approx \begin{dcases}
		  	\log d_{ex} + S^{(2)}(\sigma_{E})  &  \text{ for } -\log k < I_{c}^{(2)}\left(\frac{1}{d_{BH}}I_{BH},\mathcal{E}\right) \quad (\text{figure } \ref{fig:FullDiscoErrorMain}), \\
			 \log d_{ex} + \log k + S^{(2)}(\sigma_{\text{Bath}}) & \text{ for }  -\log d_{BH} \leq   I_{c}^{(2)}\left(\frac{1}{d_{BH}}I_{BH},\mathcal{E}\right) < -\log k \quad (\text{figure } \ref{fig:BBWormholeErrorMain}).\\
		  \end{dcases}
	\end{aligned}\label{eq:renyitwoReEnEarly}
\end{equation}

\paragraph{Late times $d_{BH} < k$}
Next, for late times, their R\'{e}nyi-two entropies are given by 
\begin{equation}
	\overline{ S^{(2)}\left(\rho_{ref(in)}'  \right) }  \approx \log d_{in} \quad (\text{figure } \ref{fig:TwoABwormWithoutErrorMain}),\label{eq:renyitwoRiLate}
\end{equation}
\begin{equation}
	\begin{aligned}
		 &\overline{S^{(2)}\left(\rho_{ref(in),\,ref(ex),\, E}'  \right)}\\
		&  \approx \begin{dcases}
			\log d_{in} +\log d_{ex} +S^{(2)}(\sigma_{E}) & \\
			 & \hspace{-4cm} \text{ for } \max\{-\log k + \log d_{in},-\log d_{BH}\} < I_{c}^{(2)}\left(\frac{1}{d_{BH}}I_{BH},\mathcal{E}\right) \quad (\text{figure } \ref{fig:TwoABWormholeErrorMain}),  \\
			 \log d_{ex} +\log k + S^{(2)}(\sigma_{\text{Bath}})   & \\
			 & \hspace{-4cm} \text{ for } -\log d_{BH} \leq I_{c}^{(2)}\left(\frac{1}{d_{BH}}I_{BH},\mathcal{E}\right) < \max\{-\log k + \log d_{in},-\log d_{BH}\} \quad (\text{figure } \ref{fig:FullConneErrorMain}) ,
		\end{dcases}
	\end{aligned} \label{eq:renyitwoRiReEnLate}
\end{equation}
and
\begin{equation}
	\begin{aligned}
		 &\overline{S^{(2)}\left(\rho_{ref(ex),\, E}'  \right)}\\
		  & \approx	\log d_{ex} + S^{(2)}(\sigma_{E})  & \text{ for }-\log d_{BH}  \leq I_{c}^{(2)}\left(\frac{1}{d_{BH}}I_{BH},\mathcal{E}\right) \quad (\text{figure } \ref{fig:TwoABWormholeErrorMain}).
	\end{aligned}\label{eq:renyitwoReEnLate}
\end{equation}

Using and interpolating the above results, we can write down phase diagrams of the dominant saddle for the above R\'{e}nyi-two entropies on the $\log k$ and $I_{c}^{(2)}\left(\frac{1}{d_{BH}}I_{BH},\mathcal{E}\right)\;$ plane\footnote{Near the phase boundaries where dominant saddles change, other partially connected saddles which are not appeared in the above results, can be comparable to the saddle appearing in the above results. Thus, strictly speaking, we should consider their partially connected saddles, but, here we simply ignore their contributions and write down the phase diagrams from the asymptotic results.}. Their phase diagrams are shown in figure \ref{fig:PhaseDiagramRenyiMain}. We also show the Haar random error case, discussed in appendix \ref{app:coherentHaar}, in figure \ref{fig:PhaseDiagramRenyiMain}. Since the R\'{e}nyi-two entropy $\overline{ S^{(2)}\left(\rho_{ref(in)}'  \right) }$ does not depend on $I_{c}^{(2)}\left(\frac{1}{d_{BH}}I_{BH},\mathcal{E}\right)$, we do not show the diagram.

\begin{figure}[t]
\centering
	  \begin{minipage}[t]{1\hsize}
    \centering
    \includegraphics[width=0.93\textwidth]{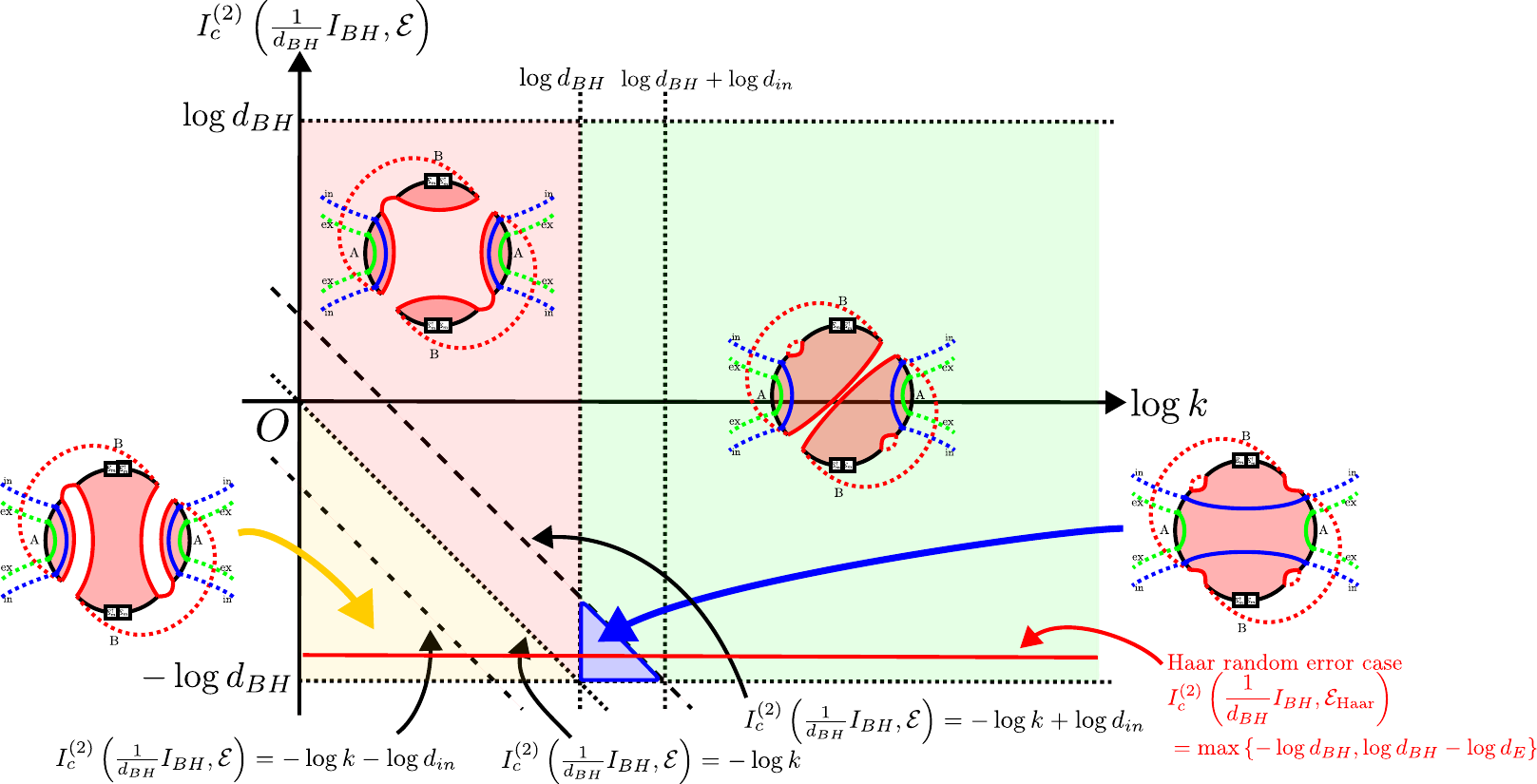}
 \subcaption{$\overline{S^{(2)}\left(\rho_{ref(in),\,ref(ex),\, E}'  \right)}$}\label{fig:PhaseDiagramRenyiRiReEnMain}
  \end{minipage}\\
  \hfil
  \begin{minipage}[t]{1\hsize}
    \centering
    \includegraphics[width=0.93\textwidth]{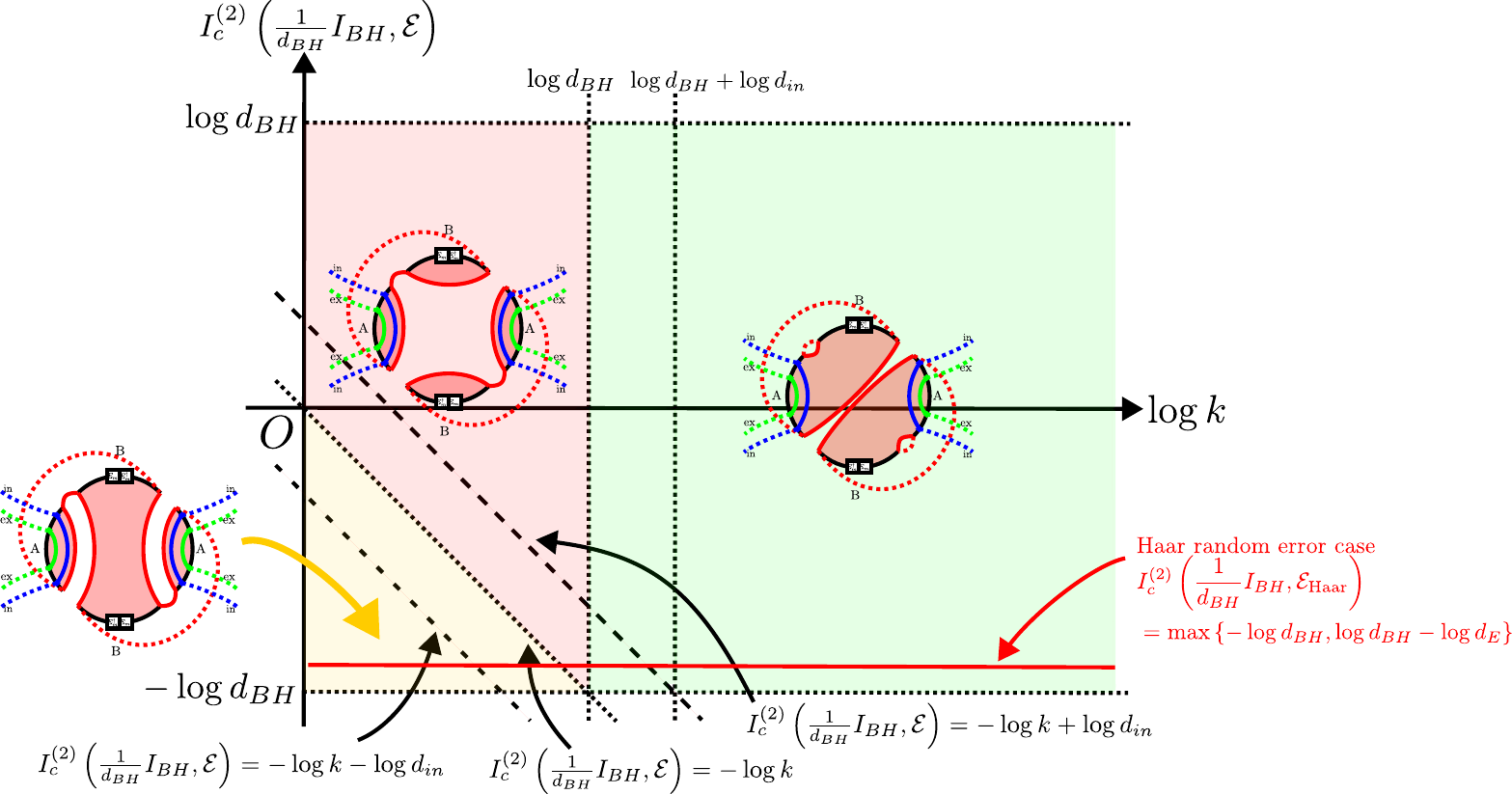}
    \subcaption{$\overline{S^{(2)}\left(\rho_{\,ref(ex),\, E}'  \right)}$}\label{fig:PhaseDiagramRenyiReEnMain}
  \end{minipage}
  \caption{Phase diagrams of dominant saddles for the R\'{e}nyi-two entropies, $\overline{S^{(2)}\left(\rho_{ref(in),\,ref(ex),\, E}'  \right)}$ and $\overline{S^{(2)}\left(\rho_{\,ref(ex),\, E}'  \right)}$ on the $\log k$ - $I_{c}^{(2)}\left(\frac{1}{d_{BH}}I_{BH},\mathcal{E}\right)$ plane. 
   The red solid lines correspond to the R\'{e}nyi-two coherent information (\ref{eq:coherentGraviCase}) for the Haar random error. 
   In the calculation of the R\'{e}nyi-two coherent information for the Haar random error, we assume $\log d_{E}\geq 2\log d_{BH} -\log d_{in} $, and in this case $I_{c}^{(2)}\left(\frac{1}{d_{BH}}I_{BH},\mathcal{E}\right)=\log d_{BH}-\log d_{E} \leq -\log d_{BH}+\log d_{in} $.
   }
  \label{fig:PhaseDiagramRenyiMain}
\end{figure}

Let us summarize the behavior of these R\'enyi-two entropies $\overline{S^{(2)}\left(\rho_{ref(in),\,ref(ex),\, E}'  \right)}$ and $\overline{S^{(2)}\left(\rho_{\,ref(ex),\, E}'  \right)}$. As shown in figure \ref{fig:PhaseDiagramRenyiMain},
before the Page time $k < d_{BH}$, there are two candidates for the dominant saddles (Fully disconnected saddle (a) and $(B,B)$-replica wormhole (c) in figure \ref{fig:DominantDiagramsForEarlyLateMain}). In this regime, there is not enough entanglement between universes A and B to create a wormhole connecting them. However, when the coherent information is sufficiently negative, i.e., the effect of the error is sufficiently large, then a wormhole connecting universes B of different replicas will be created. This is one realization of the ER=EPR \cite{Maldacena:2013xja} induced by the large entanglement of the environment system, $S^{(2)}(\sigma_{E}) \lesssim \log k+ S^{(2)}(\sigma_{\text{Bath}})$. In this case, we note that the entanglement between the universe $A$, i.e., the black hole, and the universe $B$, i.e., the Hawking radiation, is smaller than the entanglement between the environment $E$ and the universe $B$.

After the Page time $d_{BH} < k$, the above story changes. To see this, we note that the error $\mathcal{E}$ acts on the Hilbert space of the black hole of the universe $B$. This means that the effect of the error cannot be infinitely large, but rather its effect is bounded by the dimensions of the black hole on the universe $B$. For instance, the coherent information $I_{c}^{(2)}\left(\frac{1}{d_{BH}}I_{BH},\mathcal{E}\right)$  obeys the inequality \eqref{eq:coherentInfoBound}.
Thus, after the Page time $d_{BH} < k$, the effect of the entanglement between the universes $A$ and $B$ is more dominant than that of the entanglement between the environment system $E$ and the universe $B$. Around  the Page time, these two effects  are comparable to each other. As a result, 
 two-$(A,B)$-wormholes saddle that connects universes $A$ and $B$ in each replica dominates sufficiently after the Page time, and around  the Page time, the fully connected wormhole saddle can dominate in $\overline{S^{(2)}\left(\rho_{ref(in),\,ref(ex),\, E}'  \right)}$.

\subsection{Properties of the coherent information}

Now let us briefly  explain the properties of the  coherent information \cite{Schumacher:1996dy,Nielsen_Chuang_2010}. In general, the coherent information $I_{c}(\tau_{B},\mathcal{N})$, ($\tau_{B}$: quantum state, $\mathcal{N}$: quantum channel having the Kraus representation $\{K_{m}\}$), is defined by
\begin{equation}
	I_{c}(\tau_{B},\mathcal{N})=S(\mathcal{N}(\tau_{B}))-S(\mathcal{N},\tau_{B}),\label{eq:coherentInf}
\end{equation} 
where the second term $S(\mathcal{N},\tau_{B})$ is called the entropy exchange given by the von Neumann entropy of the state $\mathcal{N}^{c}(\tau_{B})$ defined by\footnote{The quantum channel $\mathcal{N}^{c}$ is called the complement channel for the quantum channel $\mathcal{N}$. We can interpret the complement channel as follows: First, we consider the Stinespring  representation of the quantum channel $\mathcal{N}$ by introducing a suitable environment system $E$ with an initial state $\ket{e_{0}}_{E}$ and a suitable unitary $U_{B,E}$ such that 
\begin{equation}
	\mathcal{N}(\tau_{B}) =\tr_{E} \left[ U_{B,E}\left(\tau_{B} \otimes \ket{e_{0}}\bra{e_{0}}  \right)U^{\dagger}_{B,E}   \right]. 
\end{equation}
Then, the complement channel is given by 
\begin{equation}
	\mathcal{N}^{c}(\tau_{B}) =\tr_{B} \left[ U_{B,E}\left(\tau_{B} \otimes \ket{e_{0}}\bra{e_{0}}  \right)U^{\dagger}_{B,E}   \right]. 
\end{equation}
  \label{foot:ComplementChannel}}
\begin{equation}
	\mathcal{N}^{c}(\tau_{B})=\sum_{m,n} \ket{e_{m}}_{E}\bra{e_{n}}\cdot \tr_{B}\left[K_{m} \tau_{B} K_{n}^{\dagger}\right].\label{eq:complementErrorChannel}
\end{equation}
Here, $ \ket{e_{m}}_{E}$ is an orthonormal basis in an environment system for the quantum channel.
In general, the coherent information can be positive or negative; it is smaller than the von Neumann entropy of the input state $S(\tau_{B})$ \cite{Schumacher:1996dy,Nielsen_Chuang_2010}, and larger than the negative of the entropy exchange.
 This coherent information quantifies the extent to which the information of the input state $\tau_{B}$ is preserved under the quantum channel $\mathcal{N}$.  In appendix \ref{app:CoherentInformation}, we give additional details regarding the coherent information. 

In our case, we have the R\'{e}nyi-two version of the coherent information \eqref{eq:coherentGraviCase}.
This R\'{e}nyi-two coherent information satisfies the following condition,
\begin{equation}
	\max\{-\log d_{BH},-\log d_{E}\} \leq I_{c}^{(2)}\left(\frac{1}{d_{BH}}I_{BH},\mathcal{E}\right) \leq \log d_{BH},\label{eq:coherentInfoBound}
\end{equation}
where the upper bound comes from the trivial case, where $\mathcal{E}(\tau)=\tau$, and the lower bound from the weak subadditivity, $-\log d_{BH} \leq S^{(2)}(\sigma_{\text{Bath}})-S^{(2)}(\sigma_{E})$, \eqref{eq:weakSubaddiExpli} and inequalities $S^{(2)}(\sigma_{E}) \leq \log d_{E}$, $0\leq S^{(2)}(\sigma_{\text{Bath}}) $. In appendix \ref{app:weakSubadd}, we give the derivation of the lower bound from the weak subadditivity.

\subsubsection{R\'{e}nyi-two mutual information}

\begin{figure}[ht]
	\centering
	\includegraphics[width=1.0\textwidth]{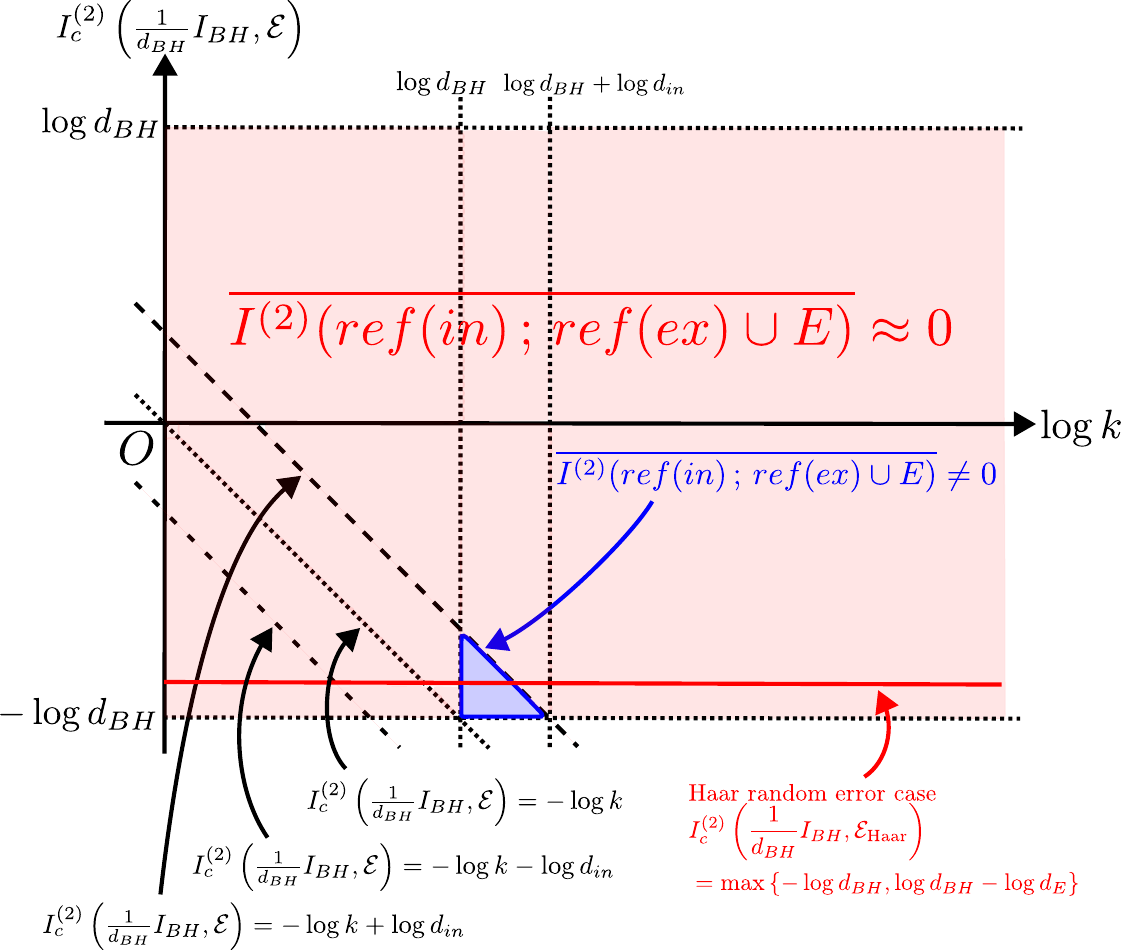}
	\caption{Phase diagram of the R\'{e}nyi-two mutual information, $\overline{I^{(2)}_{\ket{\Psi'}}(ref(in)\, ;\, ref(ex)\cup E)}$ on the $\log k$ - $I_{c}^{(2)}\left(\frac{1}{d_{BH}}I_{BH},\mathcal{E}\right)$ plane. 
	The red solid line again corresponds to the R\'{e}nyi-two coherent information (\ref{eq:coherentGraviCase}) for the Haar random error. We chose parameters such that $\log d_{E}\geq 2\log d_{BH} -\log d_{in}$.
	  }\label{fig:PhaseDiagramRenyiMutual}
\end{figure}

By assembling the above results, 
we now would like to compute the R\'{e}nyi-two mutual information \eqref{eq:renyiTwoAsymmetri} under the dominant saddle approximation. Again we discuss the details of the calculation  in appendix \ref{app:renyiTwo}.  the precise expression of the R\'{e}nyi-two mutual information is found in \eqref{eq:mutualFullEarlyLate}.  We plot its dependence on two parameters, $k$ and $I_{c}^{(2)}\left(\frac{1}{d_{BH}}I_{BH},\mathcal{E}\right)$ in figure \ref{fig:PhaseDiagramRenyiMutual}.

It is useful to explain the results by separating them into three cases, depending on the value of $k$.

\paragraph{Early times $k < d_{BH}$}
We start with the early times, where the R\'{e}nyi-two entropies are given by \eqref{eq:renyitwoRiEarly}, \eqref{eq:renyitwoRiReEnEarly} and \eqref{eq:renyitwoReEnEarly}. In these early times, the R\'{e}nyi-two mutual information vanishes regardless of the value of the R\'{e}nyi-two coherent information $I_{c}^{(2)}\left(\frac{1}{d_{BH}}I_{BH},\mathcal{E}\right)$,
\begin{equation}
	\begin{aligned}
		&\overline{I^{(2)}_{\ket{\Psi'}}(ref(in)\, ;\, ref(ex)\cup E)}\approx 0.	
	\end{aligned}\label{eq:mutualWObackreactionEarly}
\end{equation}
Here, we note that we obtained this vanishing R\'{e}nyi-two mutual information by using the dominant saddle approximation. Of course, if we include sub-dominant saddle contributions, the R\'{e}nyi-two mutual information takes a small but nonzero value.

\paragraph{Around the page time $ d_{BH} < k <d_{BH} \;d_{in}$}

Only just after the Page time, the R\'enyi mutual information can be non-vanishing. By combining the results for the R\'{e}nyi-two entropies \eqref{eq:renyitwoRiLate}, \eqref{eq:renyitwoRiReEnLate} and \eqref{eq:renyitwoReEnLate}, the R\'enyi mutual information is given by
\begin{equation}
\begin{aligned}
		&\overline{I^{(2)}_{\ket{\Psi'}}(ref(in)\, ;\, ref(ex)\cup E)}\\
		 &\approx
		 \begin{dcases}
		 	0 & \hspace{-6cm} \text{for } -\log k +\log d_{in}< I_{c}^{(2)}\left(\frac{1}{d_{BH}}I_{BH},\mathcal{E}\right)  \\
		 	   \left(-\log k + \log d_{in}\right) - I_{c}^{(2)}\left(\frac{1}{d_{BH}}I_{BH},\mathcal{E}\right) & \\
		 	  & \hspace{-6cm} \text{for }  -\log d_{BH} \leq  I_{c}^{(2)}\left(\frac{1}{d_{BH}}I_{BH},\mathcal{E}\right) < -\log k + \log d_{in}.
		 \end{dcases}
	\end{aligned}
    \label{eq:mutualWObackreactionLate}
\end{equation}

On the $\log k$ - $I_{c}^{(2)}\left(\frac{1}{d_{BH}}I_{BH},\mathcal{E}\right)$ plane, the  R\'enyi mutual information is non-vanishing  in the blue shaded region in figure 
\ref{fig:PhaseDiagramRenyiMutual}.

\paragraph{Late times $ d_{BH}\;d_{in} < k $}

In this parameter regime, again the R\'enyi-2 mutual information is vanishing.

The above results imply that, when the R\'{e}nyi-two coherent information $I_{c}^{(2)}\left(\frac{1}{d_{BH}}I_{BH},\mathcal{E}\right)=S^{(2)}(\sigma_{\text{Bath}})-S^{(2)}(\sigma_{E})$ is \textit{not} sufficiently negative, the R\'{e}nyi-two mutual information vanishes regardless of the value of $k$. On the other hand, when the R\'{e}nyi-two coherent information is sufficiently negative, the R\'{e}nyi-two mutual information  is \textit{non-vanishing} in the parameter window $d_{BH}<k< d_{BH}\; d_{in}$. This stems from the fact that the fully connected saddle (see figure \ref{fig:PhaseDiagramRenyiMain}) becomes the dominant one for $\overline{S^{(2)}\left(\rho_{ref(in),\,ref(ex),\, E}'  \right)}$.

Thus, for the error giving \textit{not} sufficiently negative R\'{e}nyi-two coherent information, the decoupling condition \eqref{eq:decoupling-condition} holds, implying the error is perfectly correctable.
However, for the error giving sufficiently negative R\'{e}nyi-two coherent information,   the decoupling condition \eqref{eq:decoupling-condition} fails to hold, implying the error is approximately correctable at best\footnote{For such an approximately correctable error, almost all of the encoded information on the universe $B$ flows into the environment system $E$ due to the interaction between the universe $B$ and the environment system $E$. Thus, if we can access the environment system $E$ in addition to the universe $B$, then we can recover code information from the universe $B$ and the environment system $E$.}.

\subsection{Comparison to the non-gravitating case}

Now let us  compare our results for the gravitating bath to those for the non-gravitating bath obtained in \cite{Balasubramanian:2022fiy}. To do so, first we show related R\'enyi-two quantities for the non-gravitating bath case. 

In this case, since the Hilbert space of the black hole microstates in universe~B is replaced by the non-gravitating radiation Hilbert space $H_{R}$, the B-boundary does not appear in the relevant gravitational path integrals. Moreover, in the expression for the coherent information that quantifies the magnitude of the error, the maximally mixed state on $I_{BH}/d_{BH}$ is replaced by that on $H_{R}$ (see appendix \ref{app:NongraRenyi}),
\begin{equation}
    I_{c}^{(2)}\!\left(\tfrac{1}{d_{BH}}I_{BH}, \mathcal{E}\right) 
    \;\;\longrightarrow\;\; 
    I_{c}^{(2)}\!\left(\tfrac{1}{k}I_{R}, \mathcal{E}'\right).
\end{equation}
We should also keep in mind that, similar to \eqref{eq:coherentInfoBound}, the new coherent information obeys the bound, 
\be
\max\left\{ -\log k ,-\log d_{E}\right\} \leq   I_{c}^{(2)} \! \left(\tfrac{1}{k}I_{R}, \mathcal{E}'\right) \leq \log k.
 \ee
In appendix \ref{app:NongraRenyi}, we give brief derivations of them with using the dominant saddle approximation again.

\begin{figure}[t]
\centering
\begin{tabular}{cc}
	  \begin{minipage}[t]{1\hsize}
    \centering
    \includegraphics[scale=0.42]{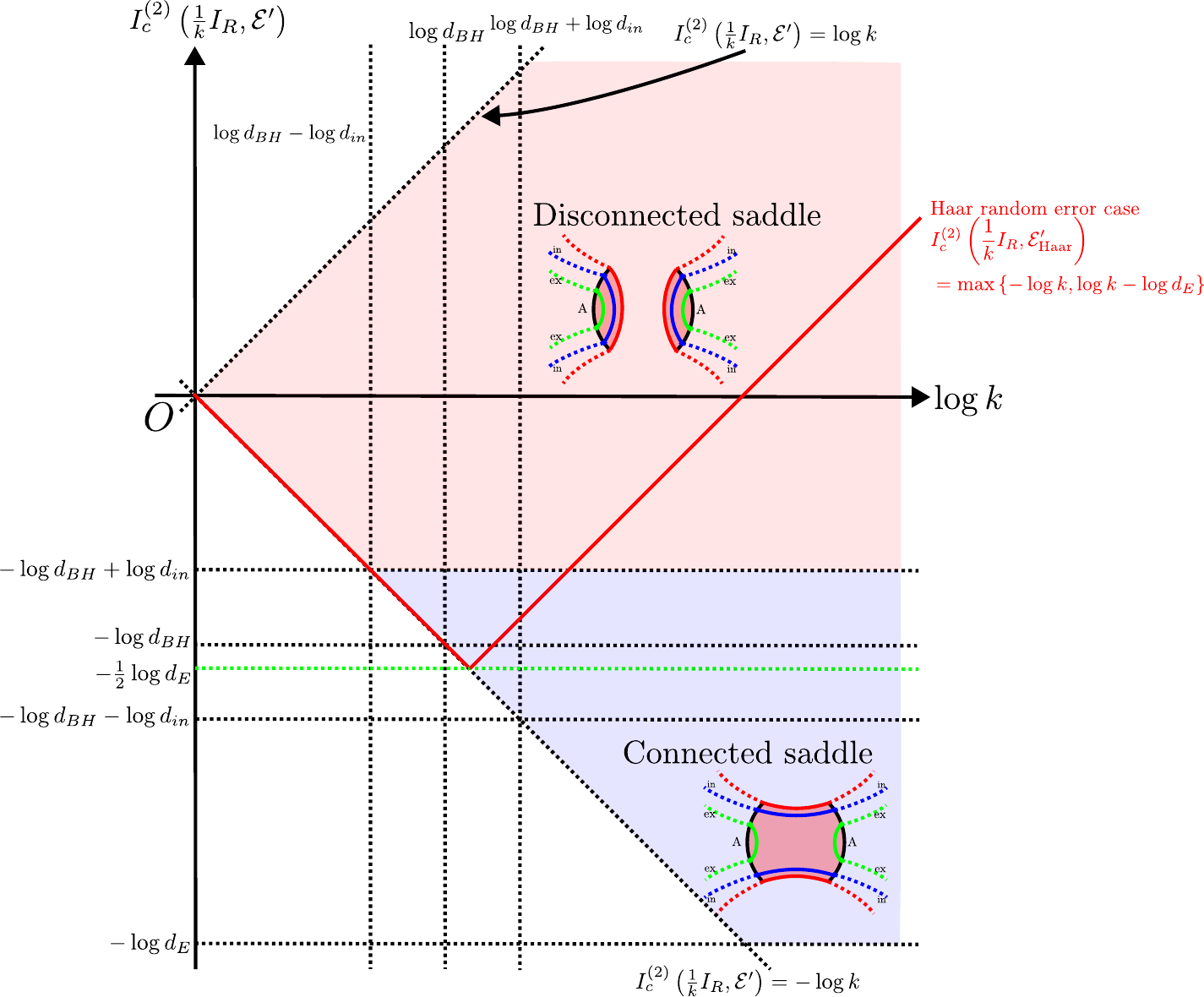}
    \subcaption{$\overline{S^{(2)}\left(\rho_{\text{NG};\,ref(in),\,ref(ex),\, E}'  \right)}$}\label{fig:PhaseDiagramNonGraRenyiRiReEn}
  \end{minipage}\\
 \hfill
  \begin{minipage}[t]{1\hsize}
    \centering
    \includegraphics[scale=0.42]{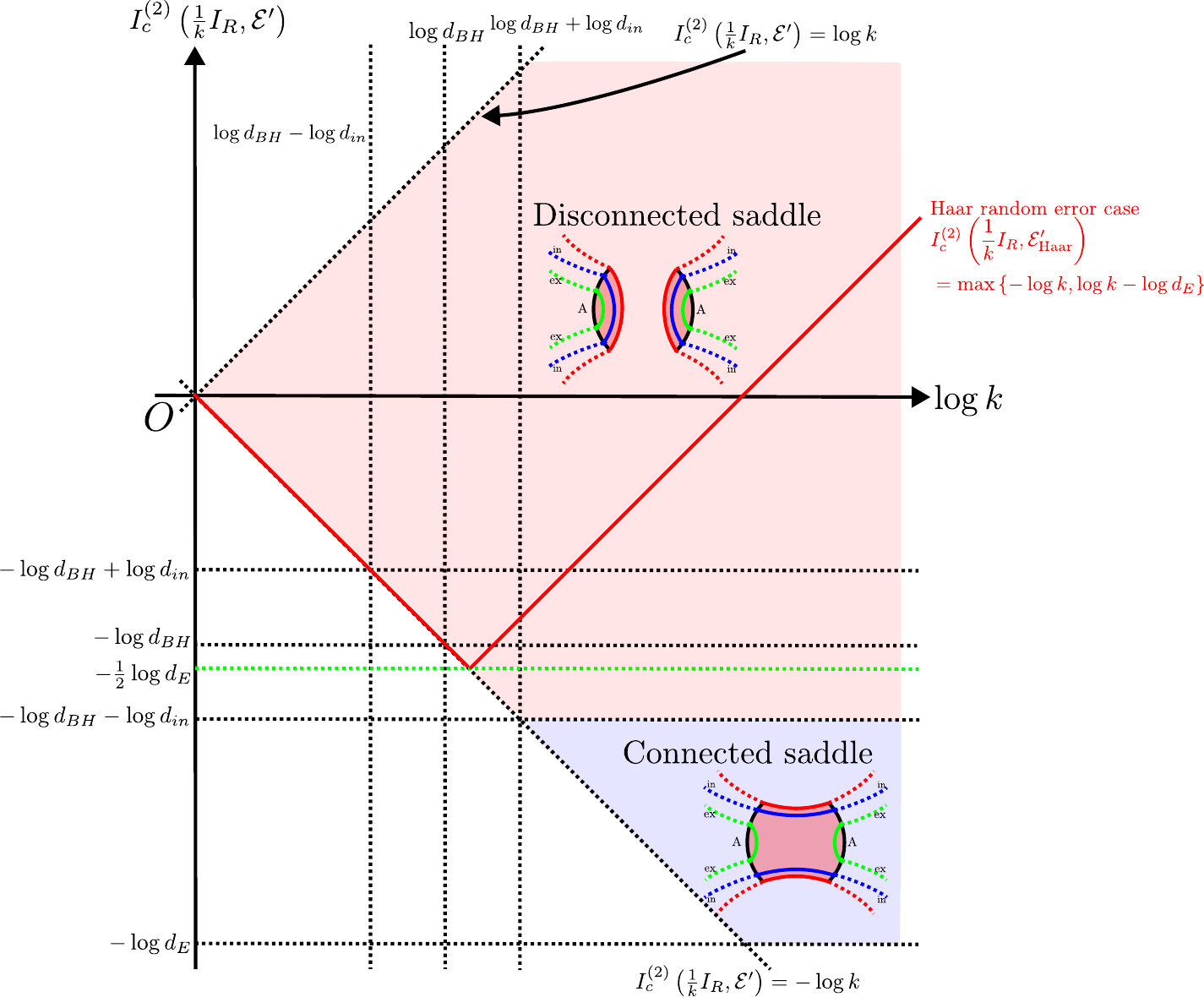}
    \subcaption{$\overline{S^{(2)}\left(\rho_{\text{NG};\,ref(ex),\, E}'  \right)}$}\label{fig:PhaseDiagramNonGraRenyiReEn}
  \end{minipage}
\end{tabular}
  \caption{Phase diagrams of dominant saddles for the R\'{e}nyi-two entropies (\ref{eq:NonGraRenyiTwoInExtEn}) and (\ref{eq:NonGraRenyiTwoExtEn}) on the $\log k$ - $ I_{c}^{(2)}\left(\frac{1}{k}I_{R},\mathcal{E}'\right)$ plane. We chose parameters such that $\frac{1}{2}\log d_{E}\geq \log d_{BH} $.
   The red solid lines correspond to the R\'{e}nyi-two coherent information (\ref{eq:coherentGraviCase}) for Haar random error. }
  \label{fig:PhaseDiagramNonGraRenyi}
\end{figure}

The R\'enyi-two entropies are given by \eqref{eq:NonGraRenyiTwoIn}, \eqref{eq:NonGraRenyiTwoInExtEn} and \eqref{eq:NonGraRenyiTwoExtEn},
\begin{equation}
	\begin{aligned}
		\overline{S^{(2)} \left(\rho_{\text{NG};\,ref(in)}'  \right)} \approx  \log d_{in} \qquad (\text{figure } \ref{fig:NonGraFullDiscoError}),
	\end{aligned}\label{eq:NonGraRenyiTwoInMain}
\end{equation}
\begin{equation}
	\begin{aligned}
		&\overline{S^{(2)}  \left(\rho_{\text{NG};\,ref(in),\,ref(ex),\, E}'  \right)}\\
		&\approx \begin{dcases}
			\log d_{in} +\log d_{ex} + S^{(2)}\left(\tau_{E}\right) & \text{ for } -\log d_{BH}+\log d_{in} < I_{c}^{(2)}\left(\frac{1}{k}I_{R},\mathcal{E}'\right) \quad   (\text{figure } \ref{fig:NonGraFullDiscoError}), \\
			\log d_{ex}+\log d_{BH}  + S^{(2)}\left(\tau_{\text{Bath}}\right) & \text{ for }  I_{c}^{(2)}\left(\frac{1}{k}I_{R},\mathcal{E}'\right) < -\log d_{BH}+\log d_{in} \quad  (\text{figure } \ref{fig:NonGraFullConneError}),
		\end{dcases}
	\end{aligned}\label{eq:NonGraRenyiTwoInExtEnMain}
\end{equation}
and
\begin{equation}
	\begin{aligned}
		&\overline{S^{(2)}  \left(\rho_{\text{NG};\,ref(ex),\, E}'  \right)}\\
		&\approx \begin{dcases}
			\log d_{ex} + S^{(2)}\left(\tau_{E}\right) &  \hspace{-3cm} \text{ for } -\log d_{BH}-\log d_{in} < I_{c}^{(2)}\left(\frac{1}{k}I_{R},\mathcal{E}'\right) \quad   (\text{figure } \ref{fig:NonGraFullDiscoError}), \\
			\log d_{in} +\log d_{ex}+\log d_{BH}  + S^{(2)}\left(\tau_{\text{Bath}}\right)& \\
			& \hspace{-3cm} \text{ for }  I_{c}^{(2)}\left(\frac{1}{k}I_{R},\mathcal{E}'\right) < -\log d_{BH}-\log d_{in} \quad  (\text{figure } \ref{fig:NonGraFullConneError}),
		\end{dcases}
	\end{aligned}\label{eq:NonGraRenyiTwoExtEnMain}
\end{equation}
where the subscript ``NG" means that they are quantities for the non-gravitating bath case, and the figures in the above expressions refer to the dominant saddles for their cases. 
 In figure \ref{fig:PhaseDiagramNonGraRenyi}, we give the phase diagrams of the dominant saddles for the R\'enyi-two entropies \eqref{eq:NonGraRenyiTwoInExtEnMain} and \eqref{eq:NonGraRenyiTwoExtEnMain} with noting the bound \eqref{eq:NonGracoherentInfoBound}.

Combining these results, we obtain the expression of the R\'enyi-two mutual information $\overline{I^{(2)}_{\ket{\Psi'}_{\text{NG}}}(ref(in)\, ;\, ref(ex)\cup E)}$
for non-gravitating bath. Again it is convenient to present the results in three separate cases.

\paragraph{Early times $k < d_{BH}/d_{in}$}

In this case, the RMI is vanishing for any value of the coherent information $I_{c}^{(2)}\left(\frac{1}{k}I_{R},\mathcal{E}'\right)$. 

\paragraph{Late times $k >d_{BH}/d_{in}$}
 In this case, the RMI is non vanishing when the coherent information is sufficiently negative, 
\begin{equation}
	\begin{aligned}
		&\overline{I^{(2)}_{\ket{\Psi'}_{\text{NG}}}(ref(in)\, ;\, ref(ex)\cup E)}\\
		 &\approx
		 \begin{dcases}
		 	0 & \hspace{-5cm} \text{ for } -\log d_{BH}+\log d_{in} < I_{c}^{(2)}\left(\frac{1}{k}I_{R},\mathcal{E}'\right),\\
		 	  \left(-\log d_{BH} + \log d_{in}\right) - I_{c}^{(2)}\left(\frac{1}{k}I_{R},\mathcal{E}'\right) & \\
		 	  & \hspace{-5cm} \text{ for } -\log d_{BH}-\log d_{in} < I_{c}^{(2)}\left(\frac{1}{k}I_{R},\mathcal{E}'\right) < -\log d_{BH}+\log d_{in}, \\
		 	2\log d_{in} & \hspace{-5cm} \text{ for }  I_{c}^{(2)}\left(\frac{1}{k}I_{R},\mathcal{E}'\right) < -\log d_{BH}-\log d_{in}.
		 \end{dcases}
	\end{aligned} \label{eq:NonGramutualFullMain}
\end{equation}

\begin{figure}[ht]
	\centering
	\includegraphics[scale=0.7]{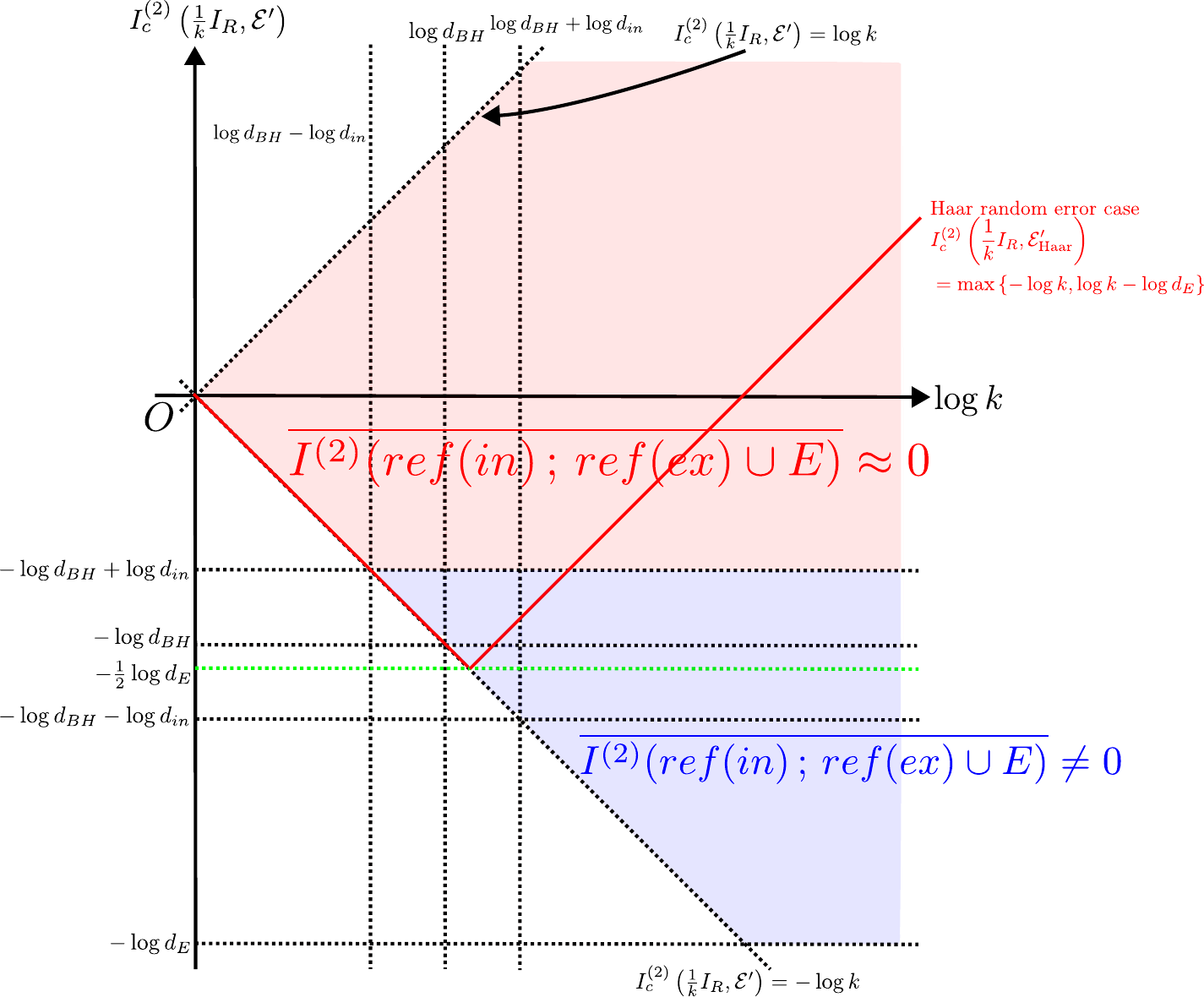}
	\caption{Phase diagram of the R\'{e}nyi-two mutual information  (\ref{eq:NonGramutualFull}) on the $\log k$- $I_{c}^{(2)}\left(\frac{1}{k}I_{R},\mathcal{E}'\right)$ plane. The parameters are equal to those in figure \ref{fig:PhaseDiagramNonGraRenyi}. The red line again corresponds to the R\'{e}nyi-two coherent information (\ref{eq:coherentGraviCase}) for the Haar random error.}\label{fig:PhaseDiagramNonGraRenyiMutual}
\end{figure}

The phase diagram of the value of the R\'enyi-two mutual information are given in figure \ref{fig:PhaseDiagramNonGraRenyiMutual}.

Now that we have presented  the R\'enyi-two entropies and mutual information for the non-gravitating bath case, we compare our gravitating bath results to them.
The R\'enyi-two mutual information exhibits quantitatively different behaviors for the gravitating and non-gravitating cases, but there is an important similarity between them; when the coherent information is sufficiently negative, the R\'enyi-two mutual information does not vanish. This implies that at the parameter regions, the decoupling condition does not hold, and thus, the errors acting on the gravitating and non-gravitating baths $B$, i.e., the Hawking radiation, can change the information of interior semi-classical excitations on the black holes $A$. In other words, the black hole interior is protected against error whose coherent information is \textit{not} sufficiently negative, but not for those with sufficiently negative coherent information\footnote{This result is similar to that in \cite{Kim:2020cds}, where errors acting on a bath are classified in terms of complexity in a quantum information theoretic setup, i.e., a black hole interior is protected against low-complexity errors on the bath, but not against high-complexity ones. It is tempting to argue that an error with very negative coherent information is highly complex.  However, we note that there is no exact correspondence between the complexity argument and the coherent information argument \cite{Balasubramanian:2022fiy}. Indeed, in the paper, the authors provide an example of an error that has sufficiently negative coherent information but is not high-complexity. From the example, one may generally argue that high-complexity errors have sufficiently negative coherent information, and as its contraposition, errors that do not have sufficiently negative coherent information are low-complexity.
}.

From the above calculations 
we conclude that the violation of the decoupling condition observed in non-gravitating bath case also happens when the bath is  gravitating. 
However, let us point out that  in our computations so far, we have been treating effects of gravity only at the level of topological theory. In the presence of dynamical gravity, we expect an error causes  gravitational backreactions to the saddle point geometry. This was absent in the non-gravitating bath case. In the next section, we will reexamine the arguments presented in this chapter while taking these gravitational backreactions into account.

\section{QEC in a gravitating bath including gravitational backreactions}\label{sec:backreaction}

In the previous section, we argued that when we compute $\overline{S^{(2)}\left(\rho_{ref(in),\,ref(ex),\, E}'  \right)}$ in the topological model,  there is a parameter region where the fully connected wormhole dominates, when the error becomes large (figure \ref{fig:PhaseDiagramRenyiMain}). Because of this, in this regime, the R\'enyi-two mutual information \eqref{eq:mutualWObackreactionLate} takes a nonzero value. In this case, the result we have obtained is almost the same as in cases where the dynamics are governed by the Haar random unitaries \cite{Kim:2020cds}.

Once gravity is made dynamical, however, the error induces a gravitational backreaction on the spacetime. A large error should therefore correspond to a significant backreaction that cannot be ignored.  

In this section, we study this effect in the simplest dynamical gravity setup, namely JT gravity. In this theory, dynamical properties of gravity are carried by the dynamical dilaton action \eqref{eq:DynamiAction}, $I_{\text{Dyn. Dilaton}}$, and by considering the contribution in the total action
\begin{equation}
	I=I_{\text{Topo. Dilaton}}+I_{\text{Dyn. Dilaton}}+ I_{\text{code}} +I_{\text{error}},
    \label{eq;totalaction}
\end{equation}

Our focus is on how the phase boundary between the fully connected wormhole dominant phase (the blue region in figure \ref{fig:PhaseDiagramRenyiRiReEnMain}) and the two (A,B)-wormhole saddle dominant phase (the green  region in the same figure)  of the R\'enyi-2 entropy  is shifted 
when such backreaction effects are included.

We now give a brief explanation of how the gravitational back reaction from the error alters the nature of the phase transition. From now on, we model the error by inserting operators at the boundary. On the bulk side, this corresponds to introducing a new type of brane, which we call the \emph{error brane}, with the action $I_{\text{error}}$ defined 
by
\begin{equation}
	I_{\text{error}}=\Delta \int_{\text{Error brane}} ds. \label{eq:bulkMassive}
\end{equation}

 The error brane modifies the dilaton configuration, and in the fully connected phase, it generates  cusps at the AdS boundary. More precisely, we regard the Kraus operators as local scaling operators $\mathcal{O}_{\Delta}$ with some scaling dimension $\Delta$, which would be determined by the details of the error. 
Then the gravitational backreaction from the dual error brane is represented by gluing together two black hole spacetimes along the brane \cite{Goel:2018ubv,Bulycheva:2019naf}. 
When the error is large, the contribution from these cusps  to the gravitational action \cite{Hayward:1993my} becomes significant. In contrast, the (A,B) wormhole has no such cusp contribution, and its action remains unaffected. Thus, for large errors, the contribution from the fully connected wormhole saddle is expected to be suppressed relative to that of the (A,B) wormhole saddle. As a result, the fully connected wormhole dominant phase should shrink, or possibly disappear altogether. In the following, we demonstrate that this is indeed the case.

\subsection{Gravitational backreaction from error as Massive brane in Post-Page times}\label{subsec:GraBackPostPage}
 
\begin{figure}[t]
\centering
	  \begin{minipage}[t]{1\hsize}
    \centering
    \includegraphics[width=0.93\textwidth]{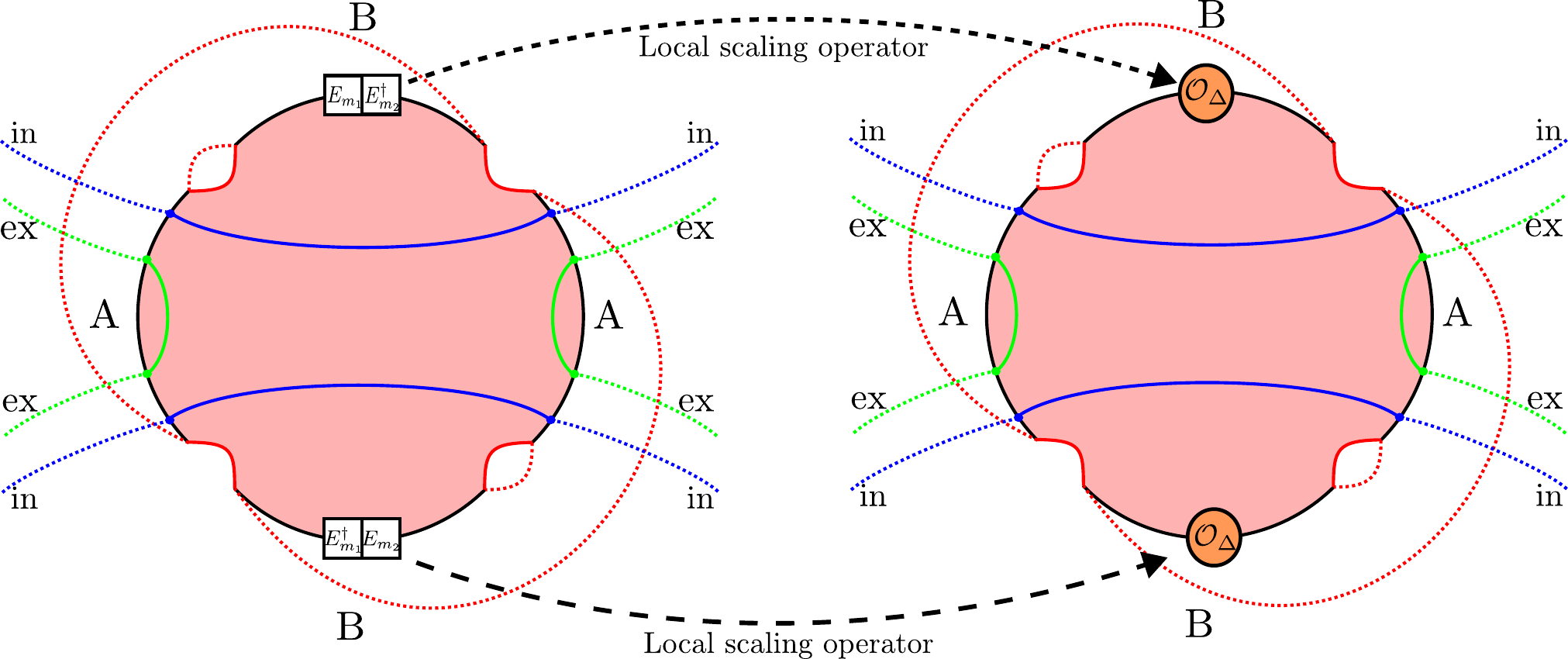}
 \subcaption{Fully connected saddle}\label{fig:Operator_replacing_1}
  \end{minipage}\\
  \hfil
  \begin{minipage}[t]{1\hsize}
    \centering
    \includegraphics[width=0.93\textwidth]{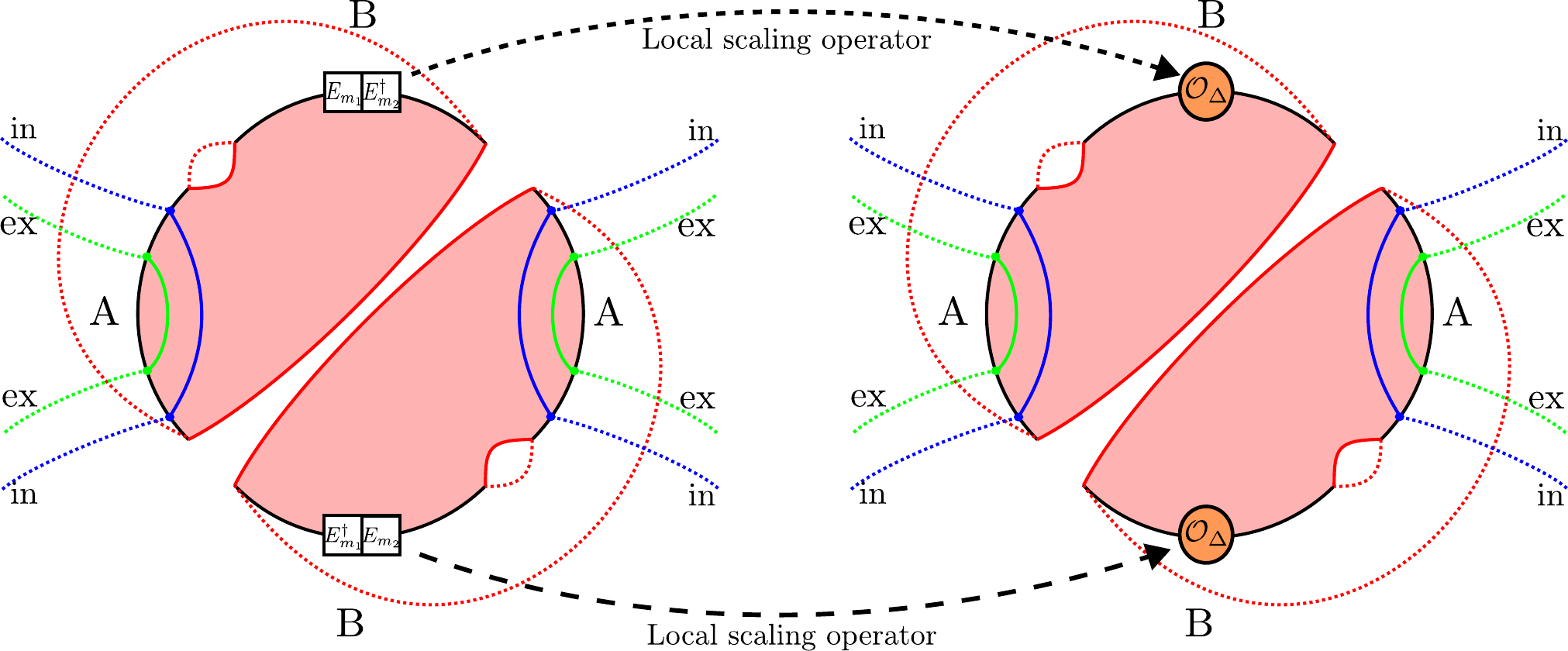}
    \subcaption{Two-$(A,B)$-wormholes saddle}\label{fig:Operator_replacing_2}
  \end{minipage}
  \caption{Two saddles for $\overline{S^{(2)}\left(\rho_{ref(in),\,ref(ex),\, E}'  \right)}$ in (dynamical) JT gravity .  The Kraus operators are treated as local scaling operators $\mathcal{O}_{\Delta}$. The fully connected saddle induces a two-point function of the local scaling operators. On the other hand, the two-$(A,B)$-wormholes saddle induces one-point functions for each replica.}
  \label{fig:Operator_replacing}
\end{figure}

In the above mentioned  description, the fully connected saddle (figure \ref{fig:FullConneErrorMain}) induces a two-point function of the local scaling operators $\mathcal{O}_{\Delta}$ living on the asymptotic AdS boundaries, while the two-(A,B) wormhole saddle (figure \ref{fig:TwoABWormholeErrorMain}) induces a one-point function for each replica as manifested in figure \ref{fig:Operator_replacing}.

 We evaluate the two-point function in the bulk  by considering the brane action $I_{\text{error}}$.
 
\begin{figure}[ht]
	\centering
	\includegraphics[scale=0.6]{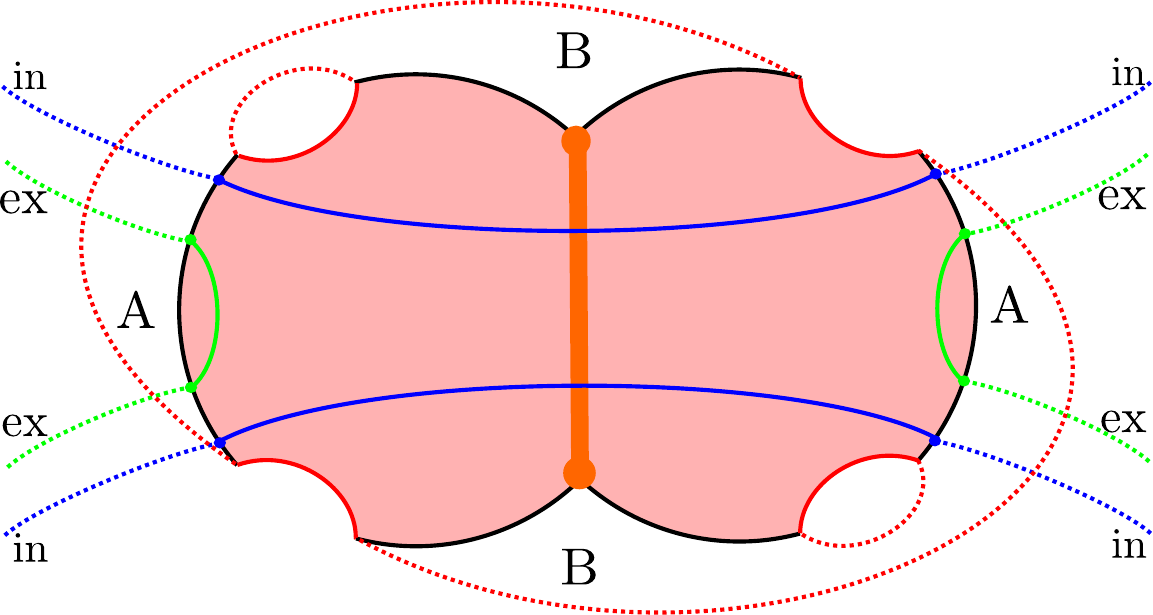}
	\caption{Backreacted fully connected saddle by the Kraus operators. The end points of the thick orange line correspond to the Kraus operators, and the orange line ending at those points correspond to the brane described by the brane action (\ref{eq:bulkMassive}).}\label{fig:saddle_cusp}
\end{figure}

 When the brane tension is large\footnote{The ``large" means that it is comparable to or larger than $\phi_{r}$ \cite{Bulycheva:2019naf}.}, the bulk massive particle backreacts on the geometry, and  the resulting geometry is given by gluing together two black hole spacetimes along the brane as explained in \cite{Bulycheva:2019naf}.
 The resulting geometry has cusps as depicted in figure \ref{fig:saddle_cusp}. Since the AdS boundary is no longer smooth, we need to introduce its contribution to the boundary Gibbons Hawking action in \eqref{eq;totalaction}, called the Hayward term   \cite{Arias:2021ilh}\footnote{See also \cite{Takayanagi:2019tvn} for the related discussion in general AdS/CFT setups.}.

We note that, for the two-$(A,B)$ wormhole saddle, there is no such worldline action since each of the two local scaling operators lives on different replicas, which are not connected by a (replica) wormhole. Thus, it is sufficient to consider the modification for the fully connected saddle.

In this case, we need to evaluate the on-shell action \eqref{eq;totalaction} of the fully connected wormhole  with backreaction from the error (figure \ref{fig:saddle_cusp}). Here, we note that the topological contribution is not affected by the backreaction\footnote{If the error effect is too large, there is a possibility that the topology is changed by the backreaction. Here, we do not focus on such a case, and leave it as a future work.}. Thus, let us focus on the dynamical part of the action, that is, the dynamical dilaton part plus the brane action coming from the error. As discussed in \cite{Bulycheva:2019naf}, their on-shell action can be divided into two parts, which come from the AdS boundaries and the cusps,
\begin{equation}
	\begin{aligned}
		\left.I_{\text{Dyn. Dilaton}}+I_{\text{error}}\right|_{\text{on-shell}} =\left.I_{\text{Dyn. Dilaton, seg}}\right|_{\text{on-shell}} + \left.I_{\text{Dyn. Dilaton, cusp}}\right|_{\text{on-shell}},
	\end{aligned}
\end{equation}
where the first term of the right hand side $I_{\text{Dyn. Dilaton, seg}}$ comes from the regular part of the AdS boundaries, and the second term $I_{\text{Dyn. Dilaton, cusp}}$ from the cusps on the same boundary \footnote{We note that the brane action $I_{\text{error}}$ does not appear explicitly in the right hand side since the brane contribution is incorporated as the gluing operation of two black hole spacetimes.}.
 The first term, $I_{\text{Dyn. Dilaton, seg}}$, is almost the same as that without backreactions.
   On the other hand, the on-shell cusp contribution depends on the backreaction, since the cusp angle becomes large as the backreaction becomes larger. Let us briefly evaluate the on-shell cusp contribution following \cite{Bulycheva:2019naf}. First, to evaluate the cusp contribution, we need to find a cusp angle $\theta$. The angle is related to the mass of the bulk massive particle, $\Delta$, by the relation in our convention\footnote{In appendix \ref{app:UpperBoundScaling}, we discuss the meaning of an upper bound of $\Delta$ coming from the relation \eqref{eq:angleDeltaPhib}.  }
 \begin{equation}
 	\sin \theta = \frac{\Delta}{2\phi_{b}},\label{eq:angleDeltaPhib}
 \end{equation}
 where $\phi_{b}$ is the boundary dilaton value, \eqref{eq:bdyMetricDilaton}. 
  This relation follows from the fact that the shift of the horizon induced by the backreaction, which is related to the mass $\Delta$, can be equivalently characterized by the cusp angle. 

Since the boundary dilaton can be generally assumed to be large $\phi_{b} \gg 1, \Delta$ implying $\theta$ is small, and the above relation reduces to
  \begin{equation}
 	 \theta \approx \frac{\Delta}{2\phi_{b}}.\label{eq:cuspAngleApproximated}
 \end{equation}
We now evaluate the on-shell cusp contribution by regularizing the cusps with arcs with an infinitesimally small radius $r$ spanning twice the cusp angle, $2\theta$,
\begin{equation}
	\begin{aligned}
		 \left.I_{\text{Dyn. Dilaton, cusp}}\right|_{\text{on-shell}}&= -\phi_{b}  \int_{\text{cusps}} \sqrt{h} K\\
		 &=-\phi_{b} \, \lim_{r\to +0}  \int_{\text{arcs with radius $r$ and angle $2\theta$}} \sqrt{h} K\\
		 &=2\cdot\phi_{b}\cdot  2\theta\\
		 &\approx 2\Delta, \label{eq:cuspOnShellaction}
	\end{aligned}
\end{equation}
where, in the third line, the additional factor $2$ comes from the fact that there are two cusps (figure \ref{fig:saddle_cusp}), and in the final line, we used the cusp angle \eqref{eq:cuspAngleApproximated}.
Thus, this new contributions implies that the R\'{e}nyi-two entropy \eqref{eq:renyitwoRiReEnLate} for late times should be modified such that it includes the factor $2\Delta$.

Having evaluated the contribution from the cusps, 
now let us go back to the calculation of the gravitation path integral of  the R\'{e}nyi-two entropy \eqref{eq:renyitwoRiReEnLate}. The fully connected saddle of our interest can potentially be the dominant one when the coherent information becomes sufficiently negative and $k \sim d_{BH}$. In  this case, \eqref{eq:PurityConnectedRiReEApprox} for late times $d_{BH}< k$ should be modified as follows
\begin{equation}
	\begin{aligned}
		&\left. \overline{\tr \left(\rho_{ref(in),\,ref(ex),\, E}'  \right)^{2}} \right|_{\text{Connected}} \approx
			\frac{1}{d_{ex}}   \cdot \frac{1}{k}\cdot \tr \left[(\sigma_{\text{Bath}})^{2}\right]\cdot e^{2\Delta}   & \text{ for } d_{BH} < k,
	\end{aligned}\label{eq:PurityConnectedRiReEApproxBackreaction}
\end{equation}
leading to the modification of the candidate contribution of the R\'{e}nyi-two entropy \eqref{eq:renyiTwoRiReEConnected} coming from the fully connected saddle (figure \ref{fig:FullConneErrorMain}),
\begin{equation}
	\begin{aligned}
		&\left.  \overline{S^{(2)}\left(\rho_{ref(in),\,ref(ex),\, E}'  \right)} \right|_{\text{Connected}} \approx 
			 \log d_{ex} + \log k + S^{(2)}(\sigma_{\text{Bath}}) +2\Delta  & \text{ for }  d_{BH} < k.
	\end{aligned}\label{eq:renyiTwoRiReEConnectedBackreact}
\end{equation}
On the other hand, the contribution from the two $(A,B)$ wormhole saddle $\left.  \overline{S^{(2)}\left(\rho_{ref(in),\,ref(ex),\, E}'  \right)} \right|_{\text{Disconnected}}  $, \eqref{eq:RenyiTwoDisconnectedRiReE} is not affected by the backreaction of the error. 

Thus, comparing the candidate contributions of the R\'{e}nyi-two entropy, \eqref{eq:RenyiTwoDisconnectedRiReE}  and \eqref{eq:renyiTwoRiReEConnectedBackreact}, we get the R\'{e}nyi-two entropy,
\begin{equation}
	\begin{aligned}
		 &\overline{S^{(2)}\left(\rho_{ref(in),\,ref(ex),\, E}'  \right)}\\
		 &= \min \left\{  \left.  \overline{S^{(2)}\left(\rho_{ref(in),\,ref(ex),\, E}'  \right)} \right|_{\text{Disconnected}}   ,  \left.  \overline{S^{(2)}\left(\rho_{ref(in),\,ref(ex),\, E}'  \right)} \right|_{\text{Connected}} \right\}\\
		&  \approx \min\left\{ \log d_{in} +\log d_{ex} +S^{(2)}(\sigma_{E}), \log d_{ex}+ \log k + S^{(2)}(\sigma_{\text{Bath}})+ 2\Delta \right\}\\
		 &=\log d_{ex} + \log d_{in} +S^{(2)}(\sigma_{E}) +  \min\left\{ 0, \log k + I_{c}^{(2)}\left(\frac{1}{d_{BH}}I_{BH},\mathcal{E}\right) -\log d_{in} + 2\Delta \right\}\\
		 & \hspace{12cm} \text{ for } d_{BH} < k.
	\end{aligned} \label{eq:BackreactedRenyitwoRiReEnLate}
\end{equation}
where, in the second equality, the first argument comes from the two-$(A,B)$ wormhole saddle (figure \ref{fig:TwoABWormholeErrorMain}), the second argument from the backreacted fully connected saddle (figure \ref{fig:saddle_cusp}), and in the second equality, we used the definition of the coherent information \eqref{eq:coherentGraviCase}.

From this result \eqref{eq:BackreactedRenyitwoRiReEnLate},
the phase boundary between the fully connected wormhole dominant phase  and the two (A,B)-wormhole saddle dominant phase of the R\'enyi-2 entropy (in figure \ref{fig:PhaseDiagramRenyiMain}) is shifted as follows. First,  if $2\Delta$ is not so large, then we get the similar result as \eqref{eq:renyitwoRiReEnLate} with replacing $S^{(2)}(\sigma_{\text{Bath}})$ by $S^{(2)}(\sigma_{\text{Bath}})+ 2\Delta$. As a result, in the phase diagram (figure \ref{fig:PhaseDiagramRenyiMutual}), the triangle region, where the mutual information does not vanish, becomes smaller since the position of the top vertex of the triangle region is at $\left(\log k,I_{c}^{(2)}\left(\frac{1}{d_{BH}}I_{BH},\mathcal{E}\right)\right)=(\log d_{BH},-\log d_{BH}+\log d_{in} -2\Delta)$ and the lower bound (the vertical axis) is at  $I_{c}^{(2)}\left(\frac{1}{d_{BH}}I_{BH},\mathcal{E}\right)= -\log d_{BH}$, which is fixed by  the weak subadditivity \eqref{eq:coherentInfoBound}.
  However, if $\Delta > \Delta_{\text{Criti.} }= \frac{1}{2} \log d_{in}$ in $d_{BH} < k$\footnote{More precisely, if 
  \begin{equation}
  	2\Delta > \log d_{in}+ \log d_{BH} -\log k  \quad \text{ for } \log d_{BH} \leq \log k \leq  \log d_{BH} + \log d_{in},
  \end{equation}
 then the triangle region can no longer appear in the phase diagram. For simplicity, we focus on the  upper value of the right hand side,
 \begin{equation}
 	 \log d_{in} \geq  \log d_{in}+ \log d_{BH} -\log k  \quad \text{ for } \log d_{BH} \leq \log k \leq  \log d_{BH} + \log d_{in}.
 \end{equation}
 }, then the triangle region no longer appears in the phase diagram, and the mutual information identically vanishes.

  Let us summarize the result.
\paragraph{Case 1: $\Delta \leq \Delta_{\text{Criti.} }= \frac{1}{2} \log d_{in}$.}
The R\'{e}nyi-two entropy for late times $d_{BH} <  k$, \eqref{eq:renyitwoRiReEnLate}, becomes 
\begin{equation}
	\begin{aligned}
		 &\overline{S^{(2)}\left(\rho_{ref(in),\,ref(ex),\, E}'  \right)}\\
		&  \approx \begin{dcases}
			\log d_{in} +\log d_{ex} +S^{(2)}(\sigma_{E}) & \\
			 & \hspace{-4cm} \text{ for } \max\{-\log k + \log d_{in}-2\Delta,-\log d_{BH}\} < I_{c}^{(2)}\left(\frac{1}{d_{BH}}I_{BH},\mathcal{E}\right) ,  \\
			 \log d_{ex} +\log k + S^{(2)}(\sigma_{\text{Bath}}) +2\Delta   & \\
			 & \hspace{-4cm} \text{ for } -\log d_{BH} \leq I_{c}^{(2)}\left(\frac{1}{d_{BH}}I_{BH},\mathcal{E}\right) < \max\{-\log k + \log d_{in}-2\Delta ,-\log d_{BH}\}.
		\end{dcases}
	\end{aligned} 
\end{equation}
Here, we note that, in this case, depending on the parameter region, either the two-(A,B) wormhole saddle (figure \ref{fig:TwoABWormholeErrorMain}) or the fully connected saddle (figure \ref{fig:FullConneErrorMain}) can be dominant.

From the above result, the R\'{e}nyi-two mutual information for the late times \eqref{eq:mutualWObackreactionLate} becomes
\begin{equation}
	\begin{aligned}
		&\overline{I^{(2)}_{\ket{\Psi'}}(ref(in)\, ;\, ref(ex)\cup E)}\\
		 &\approx
		 \begin{dcases}
		 	0 & \hspace{-7cm} \text{for } \max\{-\log k + \log d_{in}-\Delta,-\log d_{BH}\} < I_{c}^{(2)}\left(\frac{1}{d_{BH}}I_{BH},\mathcal{E}\right)  \\
		 	   \left(-\log k + \log d_{in}-2\Delta\right) - I_{c}^{(2)}\left(\frac{1}{d_{BH}}I_{BH},\mathcal{E}\right) & \\
		 	  & \hspace{-7cm} \text{for }  -\log d_{BH} \leq  I_{c}^{(2)}\left(\frac{1}{d_{BH}}I_{BH},\mathcal{E}\right)< \max\{-\log k + \log d_{in}-2\Delta,-\log d_{BH}\}.
		 \end{dcases}
	\end{aligned}
\end{equation}
Thus, the R\'{e}nyi-two mutual information still has a non-vanishing parameter region.

\paragraph{Case 2: $\Delta > \Delta_{\text{Criti.} }= \frac{1}{2}\log d_{in}$.}
For the late times, the R\'{e}nyi-two entropy, \eqref{eq:renyitwoRiReEnLate}, becomes 
\begin{equation}
	\begin{aligned}
		 &\overline{S^{(2)}\left(\rho_{ref(in),\,ref(ex),\, E}'  \right)} \approx \log d_{in} +\log d_{ex} +S^{(2)}(\sigma_{E}).
	\end{aligned} 
\end{equation}
Unlike the case 1, $2\Delta < \log d_{in}$, the two-(A,B) wormhole saddle (figure \ref{fig:TwoABWormholeErrorMain}) is the only dominant one.

\begin{figure}[ht]
	\centering
	\includegraphics[width=1.0\textwidth]{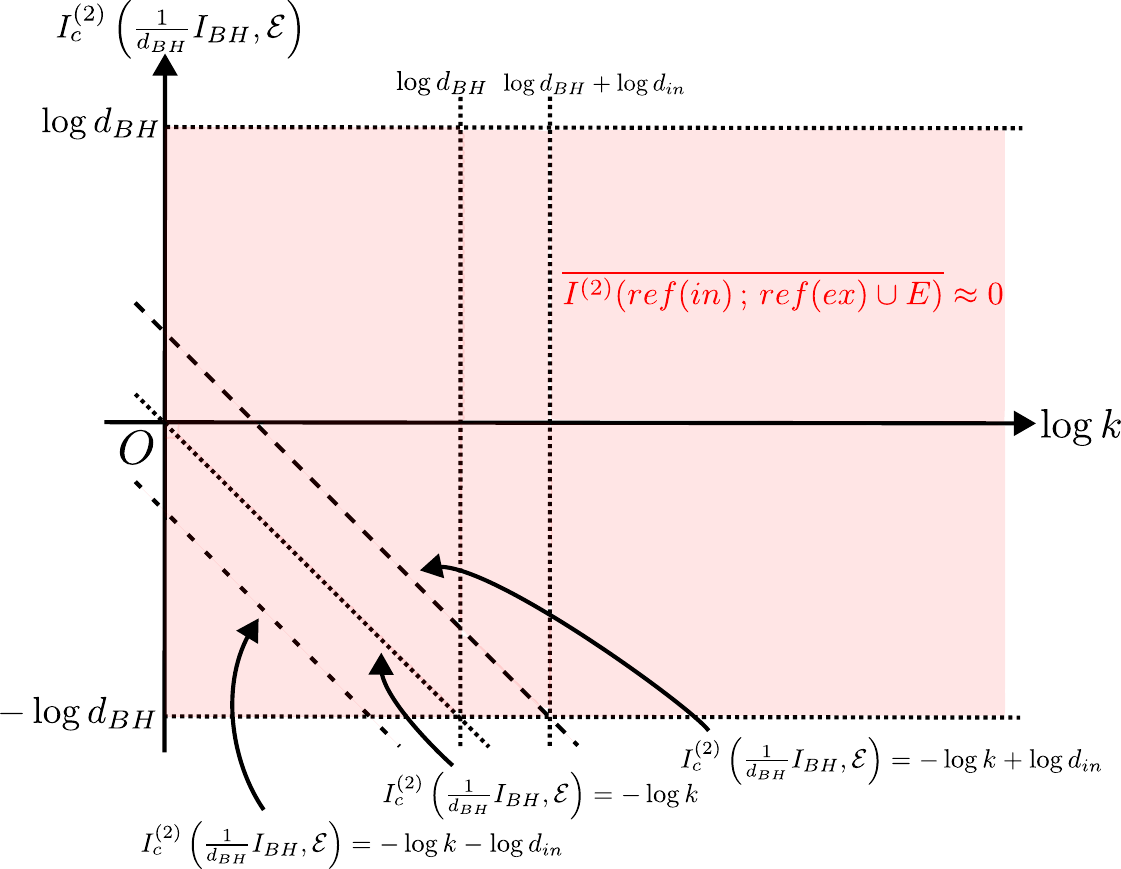}
	\caption{Phase diagram of the R\'{e}nyi-two mutual information, $\overline{I^{(2)}_{\ket{\Psi'}}(ref(in)\, ;\, ref(ex)\cup E)}$, including the large backreaction, on the $\log k$ - $I_{c}^{(2)}\left(\frac{1}{d_{BH}}I_{BH},\mathcal{E}\right)$ plane. As discussed in section \ref{subsec:gravitationa-backreaction-PrePage} later, the R\'{e}nyi-two mutual information does not change even after including the gravitational backreaction.
	  }\label{fig:PhaseDiagramRenyiMutual_Backreaction}
\end{figure}

Then, the R\'{e}nyi-two mutual information for the late times \eqref{eq:mutualWObackreactionLate} becomes 
\begin{equation}
		\overline{I^{(2)}_{\ket{\Psi'}}(ref(in)\, ;\, ref(ex)\cup E)} \approx 0 \qquad  \text{ for } d_{BH} < k.
\end{equation}
Therefore, the R\'{e}nyi-two mutual information is always vanishing regardless of the value of $I_{c}^{(2)}\left(\frac{1}{d_{BH}}I_{BH},\mathcal{E}\right) $. See figure \ref{fig:PhaseDiagramRenyiMutual_Backreaction}.

\subsection{Example of the relation of the error and the local scaling operators}

So far, we have not given the relation between $\Delta$ and the parameters of the Kraus operator, e.g., the dimension of the environment system $d_{E}$. It is a logical possibility that the two parameters $d_{E}, \Delta$ are independent; however, on general grounds, we expect that these two are related since when the error is large (the rank of the error $d_{E}$ is large), it is natural that the scaling dimension $\Delta$ is also large.
As an example of the relation, let us give one possibility of the relation and discuss the consequence.

 To this end, we assume that the environment system is given by a two-dimensional holographic CFT with central charge $c$\footnote{This assumption implies that the dynamics of the environment system is sufficiently chaotic.}, and the system is in a thermal equilibrium state. We further assume that, in the micro-canonical description, the thermal state of the system is described in terms of $d_{E}$ micro-states within an energy window $[\mathcal{E},\mathcal{E}+\delta\mathcal{E}]$. In this case, the thermal entropy of the environment system is given by
\begin{equation}
	S_{\text{Env}}^{\text{Thermal}}(\mathcal{E})= \log d_{E}.
\end{equation}
On the other hand, from Cardy's discussion, which also hold for finite temperature in holographic CFTs \cite{Hartman:2014oaa}, one can derive the following expression,
\begin{equation}
	S_{\text{Env}}^{\text{Thermal}}(\mathcal{E})\sim 2\pi\sqrt{\frac{c}{3}\mathcal{E}} \qquad \left( \mathcal{E} > \frac{c}{12} \right).
\end{equation}
Thus, by comparing the two expressions, we obtain the relation
\begin{equation}
	\mathcal{E}\sim\frac{3}{c}\left(\frac{1}{2\pi}\log d_{E} \right)^{2}.
\end{equation}
Next, let us identify the energy $\mathcal{E}$ with the brane tension $\Delta$,
\begin{equation}
	\Delta = \mathcal{E}.
\end{equation}
Combining the two expressions, we have 
\begin{equation}
	\Delta \sim \frac{3}{c}\left(\frac{1}{2\pi}\log d_{E} \right)^{2}.
    \label{eq;reldeandde}
\end{equation}
This is one example of the relation between $\Delta$ and the parameters of the Kraus operator.

If the relation \eqref{eq;reldeandde} holds, it is easy to achieve the condition
\begin{equation}
	\Delta =\frac{6}{c}\left(\frac{1}{2\pi}\log d_{E} \right)^{2} > \log d_{in},
\end{equation}
when the error is large. 
In this case, the R\'{e}nyi-two mutual information is identically vanishing, and this happens for the sufficiently large $d_{E}$. This means that the effect of gravitational backreaction is always non-negligible when the error is sufficiently large $d_{E}\gg 1$.

\subsection{Gravitational backreaction from error as Massive brane in Pre-Page times}\label{subsec:gravitationa-backreaction-PrePage}

In the previous subsection \ref{subsec:GraBackPostPage}, we have seen that, when we take the gravitational backreaction into account, the R\'enyi-two mutual information after the Page time vanishes identically depending on the scaling dimension $\Delta$. In this subsection, we investigate the effects of the gravitational backreaction on the R\'enyi-two entropies and mutual information before the Page time. 

\begin{figure}[ht]
	\centering
	\includegraphics[scale=0.5]{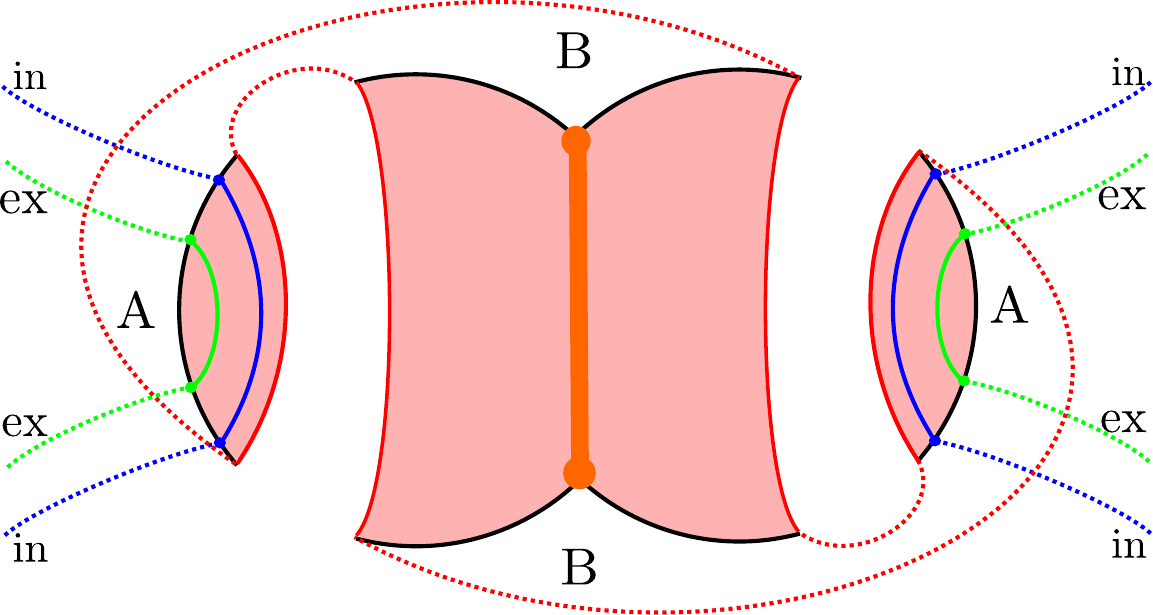}
	\caption{Backreacted one-($B,B$)-replica wormhole saddle by the Kraus operators. The blue dots denote the Kraus operators.  The end points of the thick orange line correspond to the Kraus operators, and the orange line ending at those points corresponds to the brane described by the brane action (\ref{eq:bulkMassive}).}\label{fig:saddle_cusp_BBreplica}
\end{figure}

As in the previous subsection \ref{subsec:GraBackPostPage}, the gravitational backreaction changes the geometry of the saddle having a replica wormhole connecting copies of the  universe $B$, which is what we call ($B,B$)-replica wormhole saddle (figure \ref{fig:BBWormholeErrorMain}) appearing at the computations of the R\'enyi-two entropies $\overline{S^{(2)}\left(\rho_{ref(in),\,ref(ex),\, E}'  \right)}$ and $\overline{S^{(2)}\left(\rho_{ref(ex),\, E}'  \right)}$. 
The gravitational backreaction is again introduced by putting a massive brane with tension $\Delta$ bridging the two Kraus operators as in figure \ref{fig:saddle_cusp_BBreplica}, and the geometry starts having the cusps at the endpoints of the brane.  

The evaluation of the  modified gravitational actions  for the R\'enyi-two entropies  \eqref{eq:renyitwoRiReEnEarly},  \eqref{eq:renyitwoReEnEarly} is completely parallel to that of the previous subsection \ref{subsec:GraBackPostPage}. 
For instance it is sufficient to add the cusp contribution $2\Delta$ into the expressions corresponding to  the one-(B,B)-replica wormhole saddles  in  the R\'enyi-two entropies \eqref{eq:renyitwoRiReEnEarly},  \eqref{eq:renyitwoReEnEarly}.
Therefore, we get the modified R\'enyi-two entropies  before the Page time, $k< d_{BH}$,
\begin{equation}
	\begin{aligned}
		 &\overline{S^{(2)}\left(\rho_{ref(in),\,ref(ex),\, E}'  \right)}\\
		&  \approx \begin{dcases}
			&\log d_{in} +\log d_{ex} +S^{(2)}(\sigma_{E})  \\
			  & \hspace{2cm} \text{ for }\max\{-\log k -2\Delta,-\log d_{BH}\} < I_{c}^{(2)}\left(\frac{1}{d_{BH}}I_{BH},\mathcal{E}\right), \\
			&\log d_{in} +\log d_{ex}+ \log k + S^{(2)}(\sigma_{\text{Bath}})+2\Delta  \\
			   & \hspace{2cm} \text{ for } -\log d_{BH} \leq   I_{c}^{(2)}\left(\frac{1}{d_{BH}}I_{BH},\mathcal{E}\right) <  \max\{-\log k -2\Delta,-\log d_{BH}\} ,		\end{dcases}
	\end{aligned} \label{eq:renyitwoRiReEnEarly-GraBack}
\end{equation}
and
\begin{equation}
	\begin{aligned}
		 &\overline{S^{(2)}\left(\rho_{ref(ex),\, E}'  \right)}\\
		&  \approx \begin{dcases}
			&\log d_{ex} +S^{(2)}(\sigma_{E})  \\
			  & \hspace{2cm} \text{ for }\max\{-\log k -2\Delta,-\log d_{BH}\} < I_{c}^{(2)}\left(\frac{1}{d_{BH}}I_{BH},\mathcal{E}\right), \\
			&\log d_{ex}+ \log k + S^{(2)}(\sigma_{\text{Bath}})+2\Delta  \\
			   & \hspace{2cm} \text{ for } -\log d_{BH} \leq  I_{c}^{(2)}\left(\frac{1}{d_{BH}}I_{BH},\mathcal{E}\right) <  \max\{-\log k -2\Delta,-\log d_{BH}\} ,		\end{dcases}
	\end{aligned}\label{eq:renyitwoReEnEarly-GraBack}
\end{equation}
where the first cases in \eqref{eq:renyitwoRiReEnEarly-GraBack} and \eqref{eq:renyitwoReEnEarly-GraBack} come from the fully disconnected saddle (figure \ref{fig:FullDiscoErrorMain}), and the second cases from the  backreacted one-$(B,B)$-replica wormhole saddle (figure \ref{fig:saddle_cusp_BBreplica}).

As we discussed in appendix \ref{app:UpperBoundScaling}, the upper bound of $\Delta$ is given by $\Delta_{\max}=2\phi_{b}$. Due to this  bound,  we can not eliminate one of the two cases in \eqref{eq:renyitwoRiReEnEarly-GraBack} in general, unlike the late time case \eqref{eq:BackreactedRenyitwoRiReEnLate}. This is also true for \eqref{eq:renyitwoReEnEarly-GraBack}. 
Thus, for pre-Page times, the one-$(B,B)$-replica wormhole saddle (figure \ref{fig:saddle_cusp_BBreplica}) always appears in the phase diagram even when $\Delta$ takes the maximal value $\Delta=\Delta_{\max}=2\phi_{b} \geq \frac{1}{2}\log d_{in}$.

Although the backreacted one-$(B,B)$-replica wormhole saddle can contribute to the R\'{e}nyi-two entropies $\overline{S^{(2)}\left(\rho_{ref(in),\,ref(ex),\, E}'  \right)}$ and $\overline{S^{(2)}\left(\rho_{\,ref(ex),\, E}'  \right)}$, the difference between their R\'{e}nyi-two entropies at pre-Page times does not change,
\begin{equation}
	\begin{aligned}
		\overline{S^{(2)}\left(\rho_{ref(in),\,ref(ex),\, E}'  \right)}-\overline{S^{(2)}\left(\rho_{ref(ex),\, E}'  \right)} \approx  \log d_{in} \qquad \text{ for } k<  d_{BH}.
	\end{aligned}
\end{equation}
This implies that the R\'{e}nyi-two mutual information for pre-Page times, $k < d_{BH}$, still vanishes independently of the value $\Delta (\leq \Delta_{\max})$;
\begin{equation}
		\overline{I^{(2)}_{\ket{\Psi'}}(ref(in)\, ;\, ref(ex)\cup E)} \approx 0 \qquad  \text{ for } k< d_{BH}.
\end{equation}

\subsection{Interpretation of the vanishing R\'{e}nyi-two mutual information}

 In the previous subsections, we observed that the R\'{e}nyi-two mutual information can always vanish depending on the scaling dimension $\Delta$. This implies that, unlike the previous section~\ref{sec:QECNobackreaction}, the black hole interior is protected against errors that have sufficiently negative coherent information and a relatively large scaling dimension $2\Delta > \log d_{\mathrm{in}}$ regardless of how large they are.

Let us give a physical interpretation of the above results in terms of the relationship between entanglement in the system and possible replica wormholes. For convenience, we focus on the late time regime, $d_{BH} < k$. As we have seen in the previous section~\ref{sec:QECNobackreaction}, the non-vanishing R\'{e}nyi-two mutual information originates from the existence of the fully connected saddle (figure~\ref{fig:FullConneErrorMain}), which includes the replica wormhole connecting two replicas. If we do not consider gravitational backreaction, the fully connected saddle is dominant compared to the two-$(A,B)$ wormhole saddle (figure~\ref{fig:TwoABWormholeErrorMain}). Essentially, the existence of the replica wormhole in the fully connected saddle corresponds to strong entanglement between the environment system and the  system consisting of the universe A and B, described by the physical Hilbert space.

However, if we take into account the gravitational backreaction on the fully connected saddle, the replica wormhole connecting the two replicas becomes ``longer". This means that the gravitational backreaction from the error reduces the entanglement between the environment and the other systems. Geometrically, the reduced entanglement is manifested as a longer replica wormhole\footnote{See, e.g.,~\cite{Goel:2018ubv} for a discussion on how the insertion of a heavy local operator creates a less entangled state.}. Eventually, the gravitational backreaction becomes sufficiently strong that the fully connected saddle is no longer dominant over the two-$(A,B)$ wormhole saddle. In this sense, the gravitational backreaction from the error on the gravitating bath effectively ``cuts'' the replica wormhole. This situation is reminiscent of the AMPS firewall argument~\cite{Almheiri:2012rt} and its resolution by shockwaves, e.g.,~\cite{Yoshida:2019qqw}. One might find a more precise relation between our statement and such proposals.

For early times, $k < d_{BH}$, the entanglement structure is slightly different from that after the Page time. From the viewpoint of entanglement monogamy, after the Page time, universe $B$ is almost maximally entangled with universe $A$, leaving little room for entanglement with the environment $E$. Consequently, the entanglement between $B$ and $E$ is weak and fragile. The saddle corresponding to this entanglement structure includes a replica wormhole that can be easily \textit{cut} by gravitational backreaction.

In contrast, before the Page time, universe $B$ is weakly entangled with both environments $A$ and $E$, especially when the coherent information is not sufficiently negative. However, when the coherent information becomes sufficiently negative, it indicates that $B$ is almost maximally entangled with environment $E$, leaving little room for entanglement with $A$. In this case, the entanglement between $B$ and $E$ is stronger and more resilient than that after the Page time. The corresponding saddle includes a replica wormhole (one-$(B,B)$-replica wormhole) that is \textit{not} easily cut by gravitational backreaction. Indeed, as discussed in subsection~\ref{subsec:gravitationa-backreaction-PrePage}, even after including the gravitational backreaction, the replica wormhole saddle continues to appear in the phase diagram before the Page time, while it disappears after the Page time.

Next, intuitively there is a relation between sufficiently negative coherent information and relatively large scaling dimension $2\Delta > \log d_{in}$. Here let us explain why we expect there is a relation between them. First, to implement an error with less negative coherent information, we only need a small number of combinations of (simple) fundamental operations. In our modeling, such errors lead to light branes in the bulk, because it would not cause large gravitational backreactions.
 However, to implement a large error having sufficiently negative coherent information, we would need many of such simple fundamental operations. Since this would lead to a large gravitational backreaction, it corresponds to an operator with a  large scaling dimension $\Delta$, dual to a  heavy brane in the  bulk, which can change background geometries.

\section{Discussion}\label{sec:discussion}

In this paper, we studied   an evaporating black hole entangled with a gravitating bath described by another black hole. We introduced an error represented by a general quantum channel acting on the gravitating bath system, and studied the quantum error correction properties of the black hole interior against this error. We did this by evaluating an appropriate R\'enyi-two mutual information.

In section \ref{sec:QECNobackreaction}, we focused on  the doubled PSSY (West Coast) model \cite{Anderson:2020vwi}. 
We found that in this model, the R\'enyi-two mutual information exhibits qualitatively similar behavior to the case where the bath is non-gravitating \cite{Balasubramanian:2022fiy}. That is, the R\'enyi-two mutual information takes non-zero values in a certain parameter region where the error effect is large. This implies that the black hole interior is not protected against a  large error whose coherent information is almost less than  the negative of the original black hole entropy.

However, once the gravitational backreaction from the error is included, the story completely changes. 
To see this, we model a  Kraus operator for the error channel as a local scaling operator with a certain scaling dimension, then its gravity dual is a massive brane with the tension given by the scaling dimension.  
The gravitational backreaction of the massive brane introduces a cusp in the bulk geometry which is a saddle of the  gravitational path integrals of our interest. The resulting  R\'enyi-two mutual information  vanishes, when the scaling dimension is greater than the critical value given by the size of the code subspace.

These results imply that the black hole interior might be disturbed by an error having sufficiently negative coherent information and a small scaling dimension below the critical value. However, the two conditions are difficult to make compatible with each other. This is because the implementation of an error with sufficiently negative coherent information is typically expected to require a large number of simple fundamental operations, and the resulting gravitational backreaction would easily exceed the critical value of the scaling dimension. Thus, we expect that in practice the black hole interior is typically protected against most errors that produce gravitational backreactions.

\subsection*{Future directions}
We conclude this paper by discussing several possible future directions.

\begin{itemize}

	 \item 
	 	Our semi-classical calculation demonstrates that gravitational backreaction ensures vanishing R\'enyi-two mutual information for $\Delta > \Delta_{\text{Criti.}} = \frac{1}{2}\log d_{\text{in}}$. However, the physical mechanism can be interpreted in at least two ways:

\paragraph{Interpretation 1 (Entanglement wedge transition):} 
The gravitational backreaction might effectively return the island to the black hole's entanglement wedge in an operational sense. In this case, the interior information does not belong to the entanglement wedge of the bath system, thus any physical operation acting on that system  alone cannot alter the interior information, leading to the restoration of semi-classical causality.

Testing this interpretation would ideally require explicit calculations of the island formula (or quantum extremal surfaces) on the backreacted geometry to determine the entanglement wedge assignment. However, such calculations are technically challenging in our setup. Instead, one could employ a simpler criterion: examining the decoupling condition between the interior reference system and the combined system of bath, exterior reference, and environment. Specifically, one could compute the R\'enyi-two mutual information $I^{(2)}({ref}({in}) \, ; \, B \cup {ref}({ex}) \cup E)$, where $B$ denotes the bath system.

If this mutual information vanishes (indicating decoupling), it would suggest that the interior reference system decouples from the combined system $B \cup {ref}({ex}) \cup E$, which could be interpreted as the interior belonging to the black hole's entanglement wedge rather than the bath's. Conversely, if it remains nonzero, the interior would still be entangled with the bath system, consistent with remaining in the radiation's entanglement wedge.\footnote{At the current stage, carrying out this computation is difficult due to the lack of detailed specifications of the underlying microscopic model. However, for the decoupling condition we focus on in the main body of this paper, such details are not required as extensively.} Investigating this direction would provide a concrete test of interpretation 1.

\paragraph{Interpretation 2 (Information encoding dilution):} 
Alternatively, the island may remain geometrically inside the entanglement wedge of the Hawking radiation (the bath system), but the gravitational backreaction effectively increases the negative coherent information required for errors acting on the bath to disturb the interior information, and consequently pushes it beyond the coherent-information threshold \eqref{eq:coherentInfoBound}, as discussed in section \ref{subsec:GraBackPostPage}.

In this situation, the gravitational backreaction from the error semi-classically enlarges the effective dimension of the bath system. Indeed, from \eqref{eq:BackreactedRenyitwoRiReEnLate}, the gravitational backreaction induces an increase of the semi-classical entropy of the bath,
\begin{equation}
    S^{(2)}(\sigma_{\text{Bath}})
    \quad\Longrightarrow\quad
    S^{(2)}(\sigma_{\text{Bath}}) + 2\Delta.
\end{equation}
On the other hand, the dimension of the environment system remains unchanged by the backreaction. Combining these observations with the definition of the coherent information \eqref{eq:coherentGraviCase}, we see that it becomes more difficult to reach the parameter region in which the R\'enyi-two mutual information takes a non-zero value than in the case without the gravitational backreaction. To reach such a parameter region, one must consider a stronger error compared to the non-backreacting case, but such a stronger error induces a larger gravitational backreaction. This larger backreaction further enlarges the effective dimension of the bath system. Iterating this logic, the parameter region where the R\'enyi-two mutual information takes non-zero values is pushed beyond the coherent-information threshold \eqref{eq:coherentInfoBound}, and thus can never be reached, as discussed in section \ref{subsec:GraBackPostPage}. The important aspect of this phenomenon is the effective increase of the bath-system dimension caused by the gravitational backreaction induced by the error.

One may interpret this situation as indicating that the gravitational backreaction ``dilutes" how interior information is encoded across the gravitating bath, since the effective dimension of the bath system in which the interior information is encoded becomes larger. While the total encoded degrees of freedom of the interior code subspace remain fixed, each bath degree of freedom carries less information about the interior code subspace. The error channel $\mathcal{E}$ therefore extracts insufficient information to disturb the interior information, leading to its protection. To investigate this interpretation further, one would need a more microscopic understanding of the gravitational backreaction and its effect on the structure of information encoding.

Clarifying which interpretation is correct remains an important direction for future work. Regardless of interpretation, our central finding is robust: gravitational backreaction ensures the vanishing mutual information $I^{(2)}_{\ket{\Psi'}}({ref}({in})\, ;\, {ref}({ex})\cup E) = 0$ for $\Delta > \Delta_{\text{Criti.}}$, protecting interior information in a manner qualitatively different from non-gravitating systems \cite{Balasubramanian:2022fiy}.

	\item Although we treat a quantum channel as an error, a recovery operation to extract interior information from the Hawking radiation is also implemented by a quantum channel. To extract interior information from the Hawking radiation, the required recovery operation is very complex and will lead to large gravitational backreaction. Thus, naively our findings imply that the recovery operation itself cannot access the interior information. However, this argument requires careful examination. The errors we focus on in this paper are those that do not depend on black hole microstates. If an error depends on black hole microstates, as recovery operations typically do, we cannot apply our findings directly and need to recompute the relevant quantities carefully. We leave this important problem for future work.

	\item We study the quantum error correction properties in the presence of gravitational effects on the bath system within the doubled PSSY (West Coast) model. This setup corresponds to a two-dimensional gravitational model of an evaporating black hole, which is simple enough to allow for analytical computations of many quantities. However, to establish the generality of our arguments, it is important to investigate quantum error correction properties in other setups. For example, one natural extension would be to consider doubly holographic setups, which would allow us to explore higher-dimensional cases and provide additional insights into the interplay between gravity and quantum error correction.

\end{itemize}

\section*{Acknowledgements}
This work was  supported in part by MEXT KAKENHI Grant-in-Aid for Transformative Research Areas A “Extreme Universe” No. 21H05184. T.U was also supported in part by MEXT KAKENHI Grant Number JP25K00997 and 25K07289.

\appendix

\section{Evaluation of the R\'{e}nyi-two entropies in the gravitating bath }\label{app:renyiTwo}

In this appendix, we evaluate the R\'{e}nyi-two entropies appearing in section \ref{sec:QECNobackreaction} by using the dominant saddle approximation. As we mentioned in the beginning of subsection \ref{subsec:RenyiTwo}, when presenting the result for the R\'{e}nyi-two quantities, we use $\approx$, which represents an approximate equality based on the dominant saddle point approximation. 

 We perform most of those calculations using the methods used in the West Coast model \cite{Penington:2019npb,Anderson:2020vwi}. Due to the appearance of the Kraus operators, we first need to fix the rule to compute the overlaps including the Kraus operators.

\subsection{West Coast model including the Kraus operators}\label{appsub:WCwithKraus}

In this subsection, we explain how we can include operators, e.g., Kraus operators, acting on black hole microstates in the West Coast model. 
We first consider the effect of these Kraus operators  in the probe limit without gravitational back reaction.
This allows us to start with the geometry without EoW branes, then we determine the configuration of those EoW branes later in this fixed geometry with operator insertions.

In this case, we can use the expansion of the gravitating state $\ket{\psi^{\alpha}}$ in terms of eigenvalues of the boundary Hamiltonian as discussed in appendix E of \cite{Penington:2019kki} and \cite{Kar:2022qkf,Blommaert:2021etf}. For the micro-canonical case, the expansion can be expressed as
\begin{equation}
	\ket{\psi^{\alpha}}  = \sum_{a=1}^{d_{BH}} C_{\alpha a} \ket{a},\label{eq:EigenstateExpansion}
\end{equation}
where $C_{\alpha a}$ is a complex Gaussian random variable with the normalization $\overline{C_{\alpha a}C_{\beta a}^{*}}=\delta_{ab}\delta_{\alpha}$ under the Gaussian random average denoted by the overline, and $\ket{a}$ is an orthonormal eigenstate of the boundary Hamiltonian. Here, the Gaussian random average corresponds to gravitational path integral. Using this expansion, we can evaluate the gravitational path integral of products of expectation values $ \braket{ \psi^{\alpha} | \mathcal{O} | \psi^{\beta}}$. 
For example, let us consider the following products of expectation values 
\begin{equation}
	\braket{ \psi^{\alpha_{1}} | \mathcal{O}_{1}  | \psi^{\beta_{1}}  }   \braket{ \psi^{\alpha_{2}} | \mathcal{O}_{2} | \psi^{\beta_{2}}   }=\sum_{a_{1},a_{2},b_{1},b_{2}=1}^{d_{BH}} C_{\alpha_{1}a_{1}}^{*}C_{\beta_{1}b_{1}}C_{\alpha_{2}a_{2}}^{*}C_{\beta_{2}b_{2}} \cdot \braket{a_{1}|\mathcal{O}_{1}|b_{1}}\braket{a_{2}|\mathcal{O}_{2}|b_{2}}.
\end{equation}
By considering the Gaussian random average of this quantity, we have
\begin{equation}
	\begin{aligned}
		&\overline{\braket{ \psi^{\alpha_{1}} | \mathcal{O}_{1}  | \psi^{\beta_{1}}  }   \braket{ \psi^{\alpha_{2}} | \mathcal{O}_{2} | \psi^{\beta_{2}}   }}\\
		&=\sum_{a_{1},a_{2},b_{1},b_{2}=1}^{d_{BH}} \overline{C_{\alpha_{1}a_{1}}^{*}C_{\beta_{1}b_{1}}C_{\alpha_{2}a_{2}}^{*}C_{\beta_{2}b_{2}}} \cdot \braket{a_{1}|\mathcal{O}_{1}|b_{1}}\braket{a_{2}|\mathcal{O}_{2}|b_{2}}\\
		& =\sum_{a_{1},a_{2},b_{1},b_{2}=1}^{d_{BH}} \left[ \delta_{\alpha_{1}\beta_{1}}\delta_{a_{1}b_{1}} \delta_{\alpha_{2}\beta_{2}}\delta_{a_{2}b_{2}}+  \delta_{\alpha_{1}\beta_{2}}\delta_{a_{1}b_{2}} \delta_{\alpha_{2}\beta_{1}}\delta_{a_{2}b_{1}}\right] \cdot \braket{a_{1}|\mathcal{O}_{1}|b_{1}}\braket{a_{2}|\mathcal{O}_{2}|b_{2}}\\
		&=\delta_{\alpha_{1} \beta_{1}}  \cdot \delta_{\alpha_{2} \beta_{2}} \cdot    \tr_{BH}\left[ \mathcal{O}_{1} \right]\cdot  \tr_{BH}\left[ \mathcal{O}_{2} \right] +  \delta_{\alpha_{1} \beta_{2} }  \delta_{\alpha_{2} \beta_{1} } \cdot     \tr_{BH}\left[ \mathcal{O}_{1}\mathcal{O}_{2} \right]\\
		&= (d_{BH})^{2} \cdot  \delta_{\alpha_{1} \beta_{1}}  \cdot \delta_{\alpha_{2} \beta_{2}} \cdot   \frac{1}{d_{BH}} \tr_{BH}\left[ \mathcal{O}_{1} \right]\cdot    \frac{1}{d_{BH}}\tr_{BH}\left[ \mathcal{O}_{2} \right] +  d_{BH} \cdot  \delta_{\alpha_{1} \beta_{2} }  \delta_{\alpha_{2} \beta_{1} } \cdot     \frac{1}{d_{BH}}  \tr_{BH}\left[ \mathcal{O}_{1}\mathcal{O}_{2} \right].
	\end{aligned}
\end{equation}
If we set $\mathcal{O}_{2}$ to be the identity operator, then the above expression reduces to 
\begin{equation}
	\begin{aligned}
		&\overline{\braket{ \psi^{\alpha_{1}} | \psi^{\beta_{1}}  }   \braket{ \psi^{\alpha_{2}} | \mathcal{O}_{2} | \psi^{\beta_{2}}   }}\\
		&= (d_{BH})^{2} \cdot  \delta_{\alpha_{1} \beta_{1}}  \cdot \delta_{\alpha_{2} \beta_{2}} \cdot    \frac{1}{d_{BH}}\tr_{BH}\left[ \mathcal{O}_{2} \right] +  d_{BH} \cdot  \delta_{\alpha_{1} \beta_{2} }  \delta_{\alpha_{2} \beta_{1} } \cdot     \frac{1}{d_{BH}}  \tr_{BH}\left[ \mathcal{O}_{2} \right]\\
		&=\overline{\braket{ \psi^{\alpha_{1}} | \psi^{\beta_{1}}  }   \braket{ \psi^{\alpha_{2}}  | \psi^{\beta_{2}}   }} \cdot \frac{1}{d_{BH}}  \tr_{BH}\left[ \mathcal{O}_{2} \right].
	\end{aligned}
\end{equation}

By generalizing these computations, we can find the following rule in the West Coast model including operators acting on the state $\ket{\psi^{\alpha}}$. 
\begin{enumerate}
	\item If an operator is inserted in the gravitational overlap (see e.g., figure \ref{fig:RenyiWithError}),
add a box representing the operator in the middle of the AdS boundary corresponding to the bracket.
	\item As in the usual West Coast model, we consider possible configurations of EoW branes to cap off given AdS boundaries with or without operators.
    \item If no operator is inserted in a spacetime capped off by AdS boundaries and EoW branes, then assign $d_{BH}$ to that spacetime. If there is more than one spacetime with no operator inserted, then $d_{BH}$ is assigned to each of them. 
    Otherwise, if operators are inserted into a spacetime capped off by AdS boundaries and EoW branes, then we align the operators along the orientation of the spacetime boundary and take the trace in the black hole Hilbert space in which the operators act. If there are other spacetimes with operators, perform the above operation for each of them.
    
	\item Finally, contributions from Hawking radiation can be calculated as in the usual West Coast model, by appropriately accounting for the Kronecker deltas arising from the EoW branes.
\end{enumerate}

\subsection{R\'{e}nyi-two entropies}

Now that we have given the treatment which enables us to evaluate gravitational overlaps including the Kraus operators, let us evaluate  the R\'{e}nyi-two entropies \eqref{eq:renyiTwoOnesystem}, \eqref{eq:renyiTwoThreesystem} and \eqref{eq:renyiTwoTwosystem}. 

In computing these R\'{e}nyi-two entropies, it is convenient to classify the saddles into two families; disconnected family (e.g., figure \ref{fig:RenyitwoRiDiscon}), which do not include (replica wormholes) saddles connecting two different replicas, and connected family (e.g., figure \ref{fig:RenyitwoRiConnec}), which include them. The disconnected family has four saddles, and the connected family has ten saddles.

As one can check explicitly, the purities \eqref{eq:renyiTwoOnesystem}, \eqref{eq:renyiTwoThreesystem} and \eqref{eq:renyiTwoTwosystem} for the saddles, belonging to the disconnected family, have contributions of order  $\max\{k^{a}(d_{BH})^{b}|a+b=6, a,b\in \mathbb{Z}_{\geq 0}\}$ at most. On the other hand, the purities for the connected family, are of order $\max\{k^{a}(d_{BH})^{b}|a+b=4, a,b\in \mathbb{Z}_{\geq 0}\}$ at most. We also note that the normalization factor $N$ is of order $\max\{k^{a}(d_{BH})^{b}|a+b=3, a,b\in \mathbb{Z}_{\geq 0}\}$.  Thus, if we have no contributions from the Kraus operators, R\'{e}nyi-two entropies associated with the disconnected family would be dominant ones as long as we consider the planar limit $k,e^{S_{BH}}\gg 1$. However, this is not the whole story, i.e., we have contributions from the Kraus operators for \eqref{eq:renyiTwoThreesystem} and \eqref{eq:renyiTwoTwosystem}.
 In this case, there are contributions from the Kraus operators, implying that the purities \eqref{eq:renyiTwoThreesystem} and \eqref{eq:renyiTwoTwosystem} for the connected family  are of order  $\max\{k^{a}(d_{BH})^{b} (d_{E})^2|a+b=4, a,b\in \mathbb{Z}_{\geq 0}\}$, roughly speaking. Due to the underlying structures of the purities, the purities for the disconnected family can not be modified by the contributions from the Kraus operators, thus they are still of order $\max\{k^{a}(d_{BH})^{b}|a+b=6, a,b\in \mathbb{Z}_{\geq 0}\}$. Therefore, if the Kraus operator contributions are of the order comparable to the dimensions of the Hawking radiation or the black hole, i.e.,  $d_{E}\sim k,d_{BH}$, the connected families can give dominant contributions.

\subsubsection{R\'{e}nyi-two entropy for the interior reference system}\label{appsubsub:RenyiInt}

First, we start with the evaluation of the R\'{e}nyi-two entropy, \eqref{eq:renyiTwoOnesystem}, not including the Kraus operators $E_{m}$. In this case, the competition between the disconnected and connected families are simply determined by the dimensions of the Hawking radiation and the black hole.

\begin{figure}[ht]
\centering
\begin{tabular}{cc}
	\begin{minipage}[t]{0.5\hsize}
    \centering
    \includegraphics[width=1\linewidth]{figures/GraPathRiHawking.pdf}
    \subcaption{Fully disconnected saddle}\label{fig:RenyitwoRiDiscon}
  \end{minipage}&
  \begin{minipage}[t]{0.5\hsize}
    \centering
    \includegraphics[width=1\linewidth]{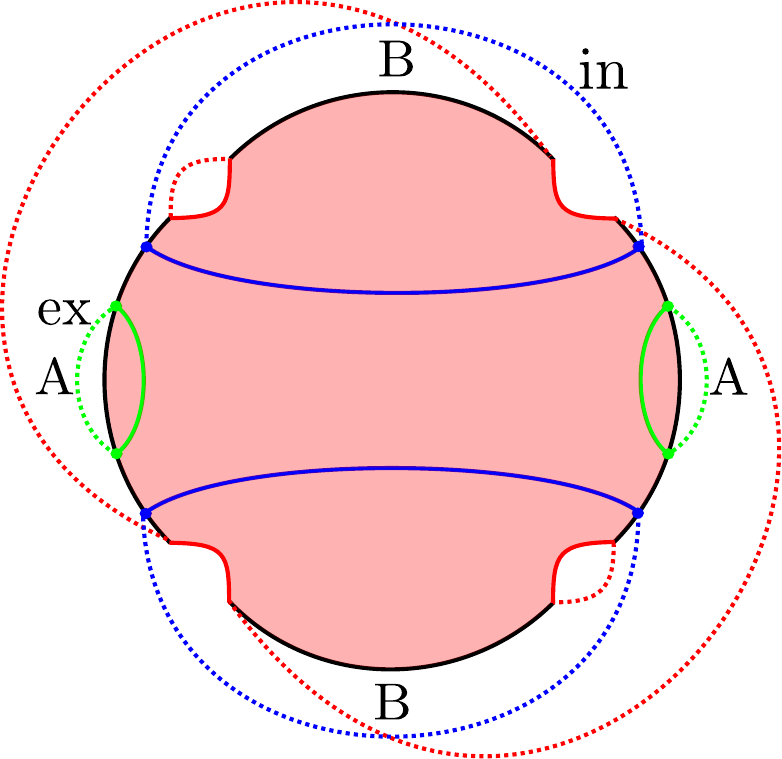}
    \subcaption{Fully connected saddle}\label{fig:RenyitwoRiConnec}
  \end{minipage}
\end{tabular}
  \caption{Two candidate saddles for the gravitational path integral of the R\'{e}nyi-two entropy, (\ref{eq:renyiTwoOnesystem}), under the boundary condition given in figure \ref{fig:RenyiNoError}. There are still nine partially connected saddles.}\label{fig:RenyiRiSaddles}
\end{figure}

Since we have given the boundary conditions for the gravitational path integral as in figures  \ref{fig:RenyiNoError}, we can do the path integral diagrammatically under the boundary condition. Two of the gravitational path integrals of the quantity \eqref{eq:renyiTwoOnesystem} are given diagrammatically by figure \ref{fig:RenyiRiSaddles}. There are still twelve partially connected diagrams.

\paragraph{Disconnected family}
Noting that the disconnected family does not include saddles having the replica wormhole connecting different replicas by definition,
their contributions to  $\tr \left(\rho_{ref(in)}'  \right)^{2}$ can be written as the purity of the gravitational path integral of the density matrix,
\begin{equation}
	\begin{aligned}
		\left. \overline{\tr \left(\rho_{ref(in)}'  \right)^{2}} \right|_{\text{Disconnected}} &= \tr \left( \overline{\rho_{ref(in)}'}  \right)^{2}    \\
		& = \frac{1}{(d_{in} d_{ex} N )^{2}} \sum_{\bm{i}=1}^{d_{in}} \sum_{\bm{i}'=1}^{d_{ex}} \overline{\braket{\Psi_{i_{2},i_{1}'} | \Psi_{i_{1},i_{1}' } } } \cdot \overline{\braket{\Psi_{i_{1},i_{2}'} | \Psi_{i_{2},i_{2}'} }}\\
		&= \frac{1}{(d_{in} d_{ex} N )^{2}} \sum_{\bm{i}=1}^{d_{in}} \sum_{\bm{i}'=1}^{d_{ex}} N^{2} \delta_{i_{2}i_{1}} \delta_{i_{1}i_{2}}\cdot \delta_{i_{1}i_{1}} \delta_{i_{2}i_{2}}\\
		&= \frac{1}{d_{in}},
	\end{aligned}\label{eq:purityRiFullDiscoFamily}
\end{equation}
where, in the third line, we used the result \eqref{eq:overlapPsi}, \eqref{eq:normalization}. 
We note that this purity is of order one, i.e., $k^{0}e^{0 \cdot  S_{BH}}$, and this is consistent with the estimation which we gave at the beginning of this subsection,
\begin{equation}
	\begin{aligned}
		\left. \overline{\tr \left(\rho_{ref(in)}'  \right)^{2}} \right|_{\text{Disconnected}} &\sim \frac{ \max\{k^{a}(d_{BH})^{b}|a+b=6, a,b\in \mathbb{Z}_{\geq 0}\} }{\left(\max\{k^{a}(d_{BH})^{b}|a+b=3, a,b\in \mathbb{Z}_{\geq 0}\}\right)^{2}}\\
		&\sim \frac{ \max\{k^{a}(d_{BH})^{b}|a+b=6, a,b\in \mathbb{Z}_{\geq 0}\} }{\max\{k^{a}(d_{BH})^{b}|a+b=6, a,b\in \mathbb{Z}_{\geq 0}\}}\\
		&\sim 1.
	\end{aligned}
\end{equation}

Thus, this expression gives the R\'{e}nyi-two entropy for the disconnected family,
\begin{equation}
	\left. \overline{S^{(2)}\left( \rho_{ref(in)}'  \right)}\right|_{\text{Disconnected}} =   \log d_{in}.
\end{equation}

\begin{figure}[ht]
\centering
\begin{tabular}{cc}
	  \begin{minipage}[t]{0.45\hsize}
    \centering
    \includegraphics[width=1\linewidth]{figures/ABwormhole.pdf}
    \subcaption{two-$(A,B)$-wormholes saddle}\label{fig:ABwormhole}
  \end{minipage}
  \begin{minipage}[t]{0.45\hsize}
    \centering
    \includegraphics[width=1\linewidth]{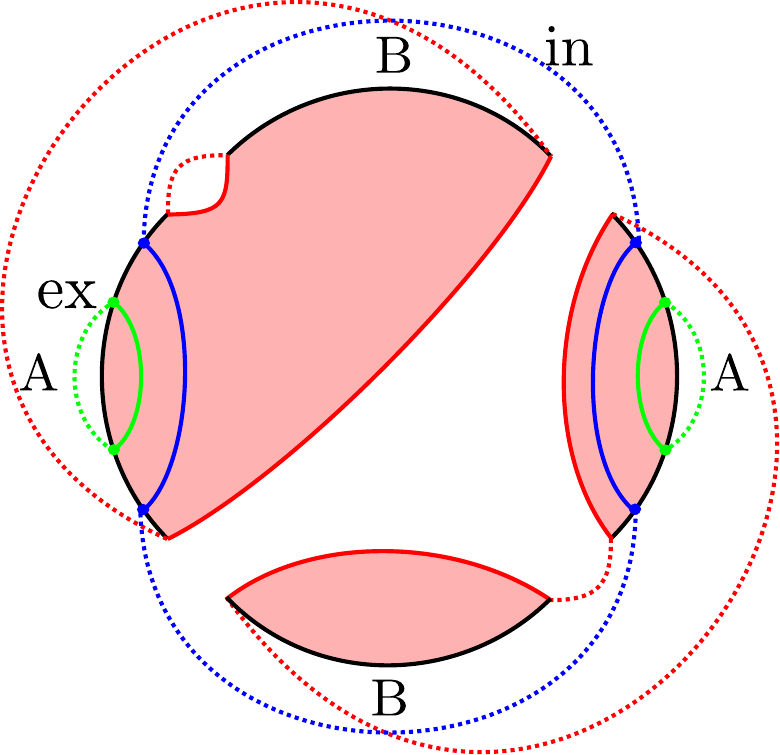}
    \subcaption{one-$(A,B)$-wormholes}\label{fig:PartiallyABworm}
  \end{minipage}
\end{tabular}
  \caption{Other candidate saddles, belonging to the disconnected family, for the gravitational path integral of the R\'{e}nyi-two entropy, (\ref{eq:renyiTwoOnesystem}), under the boundary condition given in figure \ref{fig:RenyiNoError}. }\label{fig:RenyiRiSaddlesOther}
\end{figure}

Before discussing the connected family, let us clarify the components of the disconnected family, which consists of four saddles. Although we did not explicitly consider those four saddles in deriving the above entropy for the disconnected family, we can explicitly evaluate each contribution. Their explicit evaluation gives 
\begin{equation}
	\begin{aligned}
		&\left. \overline{\tr \left(\rho_{ref(in)}'  \right)^{2}} \right|_{\text{Disconnected}}\\
		&=   \frac{1}{(d_{in} d_{ex} N )^{2}} \big[ d_{in}d_{ex}\cdot  k^{2}(d_{BH})^{4}+d_{in}d_{ex}\cdot  k^{4}(d_{BH})^{2} +2d_{in}d_{ex}\cdot  k^{3}(d_{BH})^{3} \big].
	\end{aligned}
\end{equation}
The first term originates from the fully disconnected saddle (figure \ref{fig:RenyitwoRiDiscon}), while the second term comes from the two-$(A,B)$-wormholes saddle (figure \ref{fig:ABwormhole}), which includes two wormholes connecting the universes $A$ and $B$ in each replica. The third term arises from the one-$(A,B)$-wormholes saddles (figure \ref{fig:PartiallyABworm}), including a single wormhole that connects universes $A$ and $B$ in one of the two replicas. Thus, from the power of $k$ and $d_{BH}$, at early times $k< d_{BH}$, the first term, corresponding to the fully connected saddle, is dominant, and at late times $d_{BH} < k$, the second term, corresponding to the two-$(A,B)$-wormholes saddle, is dominant;
\begin{equation}
\begin{aligned}
	\left. \overline{\tr \left(\rho_{ref(in)}'  \right)^{2}} \right|_{\text{Disconnected}}&\approx
	\begin{dcases}
		\frac{1}{(d_{in} d_{ex} N )^{2}} \cdot  d_{in}d_{ex}\cdot  k^{2}(d_{BH})^{4} &  \text{ for } k< d_{BH}\\
		\frac{1}{(d_{in} d_{ex} N )^{2}}  d_{in}d_{ex}\cdot  k^{4}(d_{BH})^{2} &  \text{ for }  d_{BH} < k
	\end{dcases}\\
	& \approx \frac{1}{d_{in}} ,
\end{aligned}
\end{equation}
where in the final line, we used the approximation for the normalization factor \eqref{eq:normaliApprox}. This reproduces the earlier result \eqref{eq:purityRiFullDiscoFamily} for early and late times.

\begin{figure}[ht]
	\centering
	\includegraphics[width=0.4\linewidth]{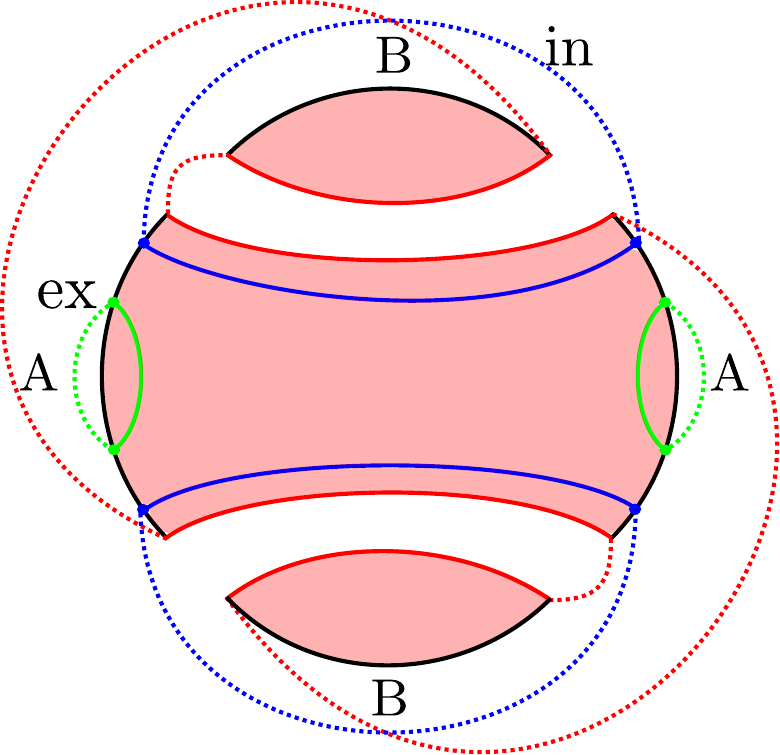}
	\caption{ The diagram of $(A,A)$-wormhole saddle. This saddle includes the replica wormhole connecting between two replicated universes $A$.}\label{fig:AAwormhole}
\end{figure}

\paragraph{Connected family}
Next, we focus on the connected family, which includes replica wormhole saddles connecting different replicas. In this case, we can not simply evaluate the purity unlike the disconnected case, \eqref{eq:purityRiFullDiscoFamily}, and we need to evaluate it for all possible saddles, belonging to the connected family, directly. We can evaluate them diagrammatically by writing down possible ten diagrams including replica wormholes with noting that the branes, corresponding to the interior excitations, must have configurations passing through cuts. This results in the following purity,
\begin{equation}
	\begin{aligned}
		&\left. \overline{\tr \left(\rho_{ref(in)}'  \right)^{2}} \right|_{\text{Connected}}\\
		&=   \frac{1}{(d_{in} d_{ex} N )^{2}} \big[ (d_{in})^{2}(d_{ex})^{2}\cdot  k^{3}d_{BH}  + (d_{in})^{2}(d_{ex})^{2}\cdot  k(d_{BH})^{3}\\
		& \hspace{4cm} + 3 d_{in}(d_{ex})^{2}\cdot  k^{2}(d_{BH})^{2} + 2 (d_{in})^{2}(d_{ex})^{2}\cdot  k^{2}(d_{BH})^{2} + 3 d_{in}(d_{ex})^{2}\cdot  k(d_{BH})^{3} \big],
	\end{aligned}\label{eq:RenyiTWoOneSystemConneFull}
\end{equation}
where the first term comes from the fully connected saddle (figure \ref{fig:RenyitwoRiConnec}), the second term from $(A,A)$-wormhole saddle (figure \ref{fig:AAwormhole}) including replica wormhole connecting between two replicated universes $A$, and the remaining terms from other partially connected replica wormhole saddles, which are not essential for our discussions.
As we noted in the beginning of this subsection, these terms without the normalization factor are of order of $\{k^{a}(d_{BH})^{b}|a+b=4, a,b\in \mathbb{Z}_{\geq 0}\}$.

At early times $k< d_{BH}$ and late times $d_{BH}< k$, this purity can be approximated as
\begin{equation}
	\begin{aligned}
		&\left. \overline{\tr \left(\rho_{ref(in)}'  \right)^{2}} \right|_{\text{Connected}}\\
		&\approx  \begin{dcases}
			\frac{1}{(d_{in} d_{ex} N )^{2}}  \cdot (d_{in})^{2}(d_{ex})^{2}\cdot  k(d_{BH})^{3}  &  \text{ for } k< d_{BH}\\
			\frac{1}{(d_{in} d_{ex} N )^{2}}  \cdot (d_{in})^{2}(d_{ex})^{2}\cdot  k^{3}d_{BH} &  \text{ for } d_{BH} < k
		\end{dcases}\\
		& \approx \frac{1}{kd_{BH}} ,
	\end{aligned}
\end{equation}
where, in the first approximation, we used $d_{in}\gg 1$, and in the second approximation, we used \eqref{eq:normaliApprox}.
Thus, we get the R\'{e}nyi-two entropy for the connected family,
\begin{equation}
	\left.  \overline{S^{(2)}\left(\rho_{ref(in)}'  \right)} \right|_{\text{Connected}} =  \log k + \log d_{BH} .
\end{equation}

\paragraph{Disconnected and connected families}

Combining the R\'{e}nyi-two entropy for the disconnected and connected families, we get the R\'{e}nyi-two entropy for early times $k< d_{BH}$ and late times $d_{BH} < k$,
\begin{equation}
	\begin{aligned}
		&\overline{S^{(2)}\left( \rho_{ref(in)}'  \right)}\\
		& = \min \left[\left. \overline{S^{(2)}\left( \rho_{ref(in)}'  \right)}\right|_{\text{Disconnected}}, \, \left.  \overline{S^{(2)}\left(\rho_{ref(in)}'  \right)} \right|_{\text{Connected}}\right]  \\
		& \approx \min \left[\log d_{in}, \log k + \log d_{BH}\right]\\
		& =    \log d_{in} ,\label{eq:renyiTwoInFull}
	\end{aligned}
\end{equation}
where, in the last line, we used $d_{in} \ll  d_{BH}$. Thus, in this R\'{e}nyi-two entropy, the disconnected family is always dominant for early times $k< d_{BH}$ and late times $d_{BH}< k$.

\subsubsection{R\'{e}nyi-two entropy for the reference systems and the environment system}
\label{appsubsub:RenyiRefEnv}

Next, we focus on the R\'{e}nyi-two entropies, \eqref{eq:renyiTwoThreesystem} and \eqref{eq:renyiTwoTwosystem}, and evaluate them for the disconnected and connected families. Unlike the R\'{e}nyi-two entropy of the interior reference system, the contributions from the Kraus operators come into the evaluation of the gravitational path integral of \eqref{eq:renyiTwoThreesystem} and \eqref{eq:renyiTwoTwosystem}.

Since we have given the boundary conditions for the gravitational path integral as in figures  \ref{fig:RenyitwoRiReEnBdyCondi} and \ref{fig:RenyitwoReEnBdyCondi}, we can again do the path integral diagrammatically under the boundary conditions.

\paragraph{Disconnected family}

As in the previous case, we can simply evaluate the R\'{e}nyi-two entropies, \eqref{eq:renyiTwoThreesystem} and \eqref{eq:renyiTwoTwosystem} for the disconnected family as follows,
\begin{equation}
	\begin{aligned}
		&\left. \overline{\tr \left(\rho_{ref(in),\,ref(ex),\, E}'  \right)^{2}} \right|_{\text{Disconnected}}\\
		&=  \tr \left(\overline{\rho_{ref(in),\,ref(ex),\, E}' } \right)^{2}\\
		&= \frac{1}{(d_{in}d_{ex} N )^{2}} \sum_{\bm{i}=1}^{d_{in}}\sum_{\bm{i}'=1}^{d_{ex}} \sum_{\bm{m}=1}^{d_{E}} \overline{\braket{\Psi_{i_{2},i_{2}'} | E_{m_{2}}^{\dagger} E_{m_{1}} | \Psi_{i_{1},i_{1}' } }  }\cdot  \overline{ \braket{\Psi_{i_{1},i_{1}'} | E_{m_{1}}^{\dagger} E_{m_{2}} | \Psi_{i_{2},i_{2}'} } }\\
		&= \frac{1}{(d_{in}d_{ex} N )^{2}} \sum_{\bm{i}=1}^{d_{in}}\sum_{\bm{i}'=1}^{d_{ex}} \sum_{\bm{m}=1}^{d_{E}} \delta_{i_{2}i_{1}}\delta_{i_{2}'i_{1}'}\, N\cdot \frac{1}{d_{BH}} \tr\left[E_{m_{2}}^{\dagger} E_{m_{1}} \right] \cdot  \delta_{i_{1}i_{2}}\delta_{i_{1}'i_{2}'}\, N \cdot \frac{1}{d_{BH}} \tr\left[E_{m_{1}}^{\dagger} E_{m_{2}} \right]\\
		 &= \frac{1}{d_{in}d_{ex}} \cdot \frac{1}{(d_{BH})^{2}} \sum_{\bm{m}=1}^{d_{E}} \tr\left[E_{m_{2}}^{\dagger} E_{m_{1}} \right]  \tr\left[E_{m_{1}}^{\dagger} E_{m_{2}} \right],
	\end{aligned}
\end{equation}
and 
\begin{equation}
	\begin{aligned}
		&\left. \overline{\tr \left(\rho_{ref(ex),\, E}'  \right)^{2}} \right|_{\text{Disconnected}}\\
		&=  \tr \left(\overline{\rho_{ref(ex),\, E}' } \right)^{2}\\
		&= \frac{1}{(d_{in}d_{ex} N )^{2}} \sum_{\bm{i}=1}^{d_{in}}\sum_{\bm{i}'=1}^{d_{ex}} \sum_{\bm{m}=1}^{d_{E}} \overline{\braket{\Psi_{i_{1},i_{2}'} | E_{m_{2}}^{\dagger} E_{m_{1}} | \Psi_{i_{1},i_{1}' } }  }\cdot  \overline{ \braket{\Psi_{i_{2},i_{1}'} | E_{m_{1}}^{\dagger} E_{m_{2}} | \Psi_{i_{2},i_{2}'} } }\\
		&= \frac{1}{(d_{in}d_{ex} N )^{2}} \sum_{\bm{i}=1}^{d_{in}}\sum_{\bm{i}'=1}^{d_{ex}} \sum_{\bm{m}=1}^{d_{E}} \delta_{i_{1}i_{1}}\delta_{i_{2}'i_{1}'}\, N \cdot \frac{1}{d_{BH}} \tr\left[E_{m_{2}}^{\dagger} E_{m_{1}} \right] \cdot  \delta_{i_{2}i_{2}}\delta_{i_{1}'i_{2}'}\, N \cdot \frac{1}{d_{BH}}\tr\left[E_{m_{1}}^{\dagger} E_{m_{2}} \right]\\
		 &= \frac{1}{d_{ex}} \cdot  \frac{1}{(d_{BH})^{2}} \sum_{\bm{m}=1}^{d_{E}} \tr\left[E_{m_{2}}^{\dagger} E_{m_{1}} \right]  \tr\left[E_{m_{1}}^{\dagger} E_{m_{2}} \right].
	\end{aligned}
\end{equation}
Here, we used the following property, which follows from \eqref{eq:overlapKraus3},
\begin{equation}
	 \overline{\braket{\Psi_{i_{1},i_{1}'} | \mathcal{O}| \Psi_{i_{2},i_{2}' } }  } =\overline{\braket{\Psi_{i_{1},i_{1}'} |\Psi_{i_{2},i_{2}' } }  }\cdot \frac{1}{d_{BH}}\tr \left[\mathcal{O}\right].
\end{equation}
By defining $\sigma_{E}$ by \eqref{eq:defSigmaE},
\begin{equation*}
		\sigma_{E}= \sum_{m,n=1}^{d_{E}} \dfrac{\tr \left\{ E_{m}E_{n}^{\dagger} \right\} }{ d_{BH} } \ket{e_{m}}_{E} \bra{e_{n}},
\end{equation*}
the above expressions can be simply written as
\begin{equation}
	\left. \overline{\tr \left(\rho_{ref(in),\,ref(ex),\, E}'  \right)^{2}} \right|_{\text{Disconnected}}=\frac{1}{d_{in}d_{ex}} \cdot\tr\left[(\sigma_{E})^{2}\right],
\end{equation}
and
\begin{equation}
	\left. \overline{\tr \left(\rho_{ref(ex),\, E}'  \right)^{2}} \right|_{\text{Disconnected}}=\frac{1}{d_{ex}} \cdot\tr\left[(\sigma_{E})^{2}\right],
\end{equation}

Thus, the R\'{e}nyi-two entropies for the disconnected family,
\begin{equation}
	\left. \overline{ S^{(2)}\left(\rho_{ref(in),\, ref(ex),\, E}'  \right) } \right|_{\text{Disconnected}} \approx  \log d_{in} +\log d_{ex} +S^{(2)}(\sigma_{E}),\label{eq:RenyiTwoDisconnectedRiReE}
\end{equation}
and 
\begin{equation}
	\left. \overline{ S^{(2)}\left(\rho_{ref(ex),\, E}'  \right) } \right|_{\text{Disconnected}} \approx \log d_{ex} +S^{(2)}(\sigma_{E}).
\end{equation}
This result exhibits the entanglement structure that the two reference systems and the environment systems are entangled with other systems, not with themselves; in other words, the entanglement wedge of their subsystems are not connected to each other.  
\begin{figure}[ht]
	\begin{tabular}{cc}
	\begin{minipage}[t]{0.5\hsize}
        \centering
		\includegraphics[width=1\linewidth]{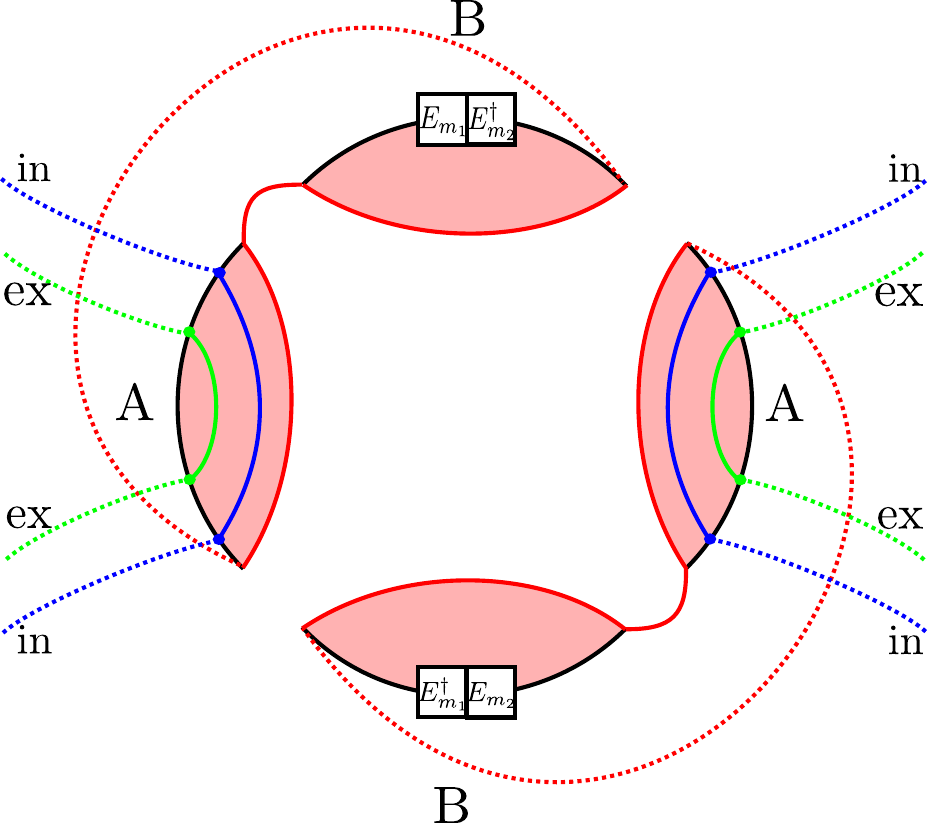}
		\subcaption{Fully disconnected saddle}
		\label{fig:FullDiscoError}
		\end{minipage}& 
	\begin{minipage}[t]{0.5\hsize}
        \centering
		\includegraphics[width=1\linewidth]{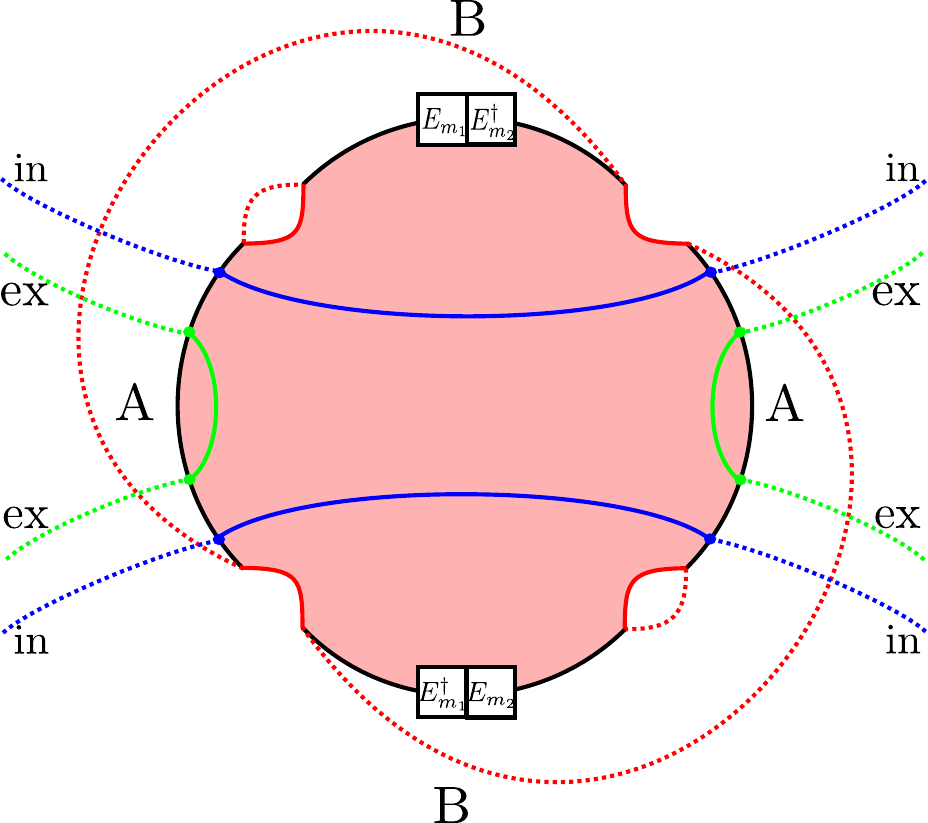}
		\subcaption{Fully connected saddle}
		\label{fig:FullConneError}
		\end{minipage}
	\end{tabular}
	\caption{Fully disconnected and connected diagrams for the gravitational path integral of the R\'{e}nyi-two entropies, (\ref{eq:renyiTwoThreesystem}) and (\ref{eq:renyiTwoTwosystem}), under the boundary condition given in figures \ref{fig:RenyitwoRiReEnBdyCondi} and \ref{fig:RenyitwoReEnBdyCondi}. In these diagrams, we need to correctly contract code indices (blue and green dashed lines) so that the resulting contractions are consistent with the  boundary condition in figures \ref{fig:RenyitwoRiReEnBdyCondi} and \ref{fig:RenyitwoReEnBdyCondi}.  }
	\label{fig:FullDisconeConneError}
\end{figure}

\begin{figure}[t]
\centering
	\begin{tabular}{cc}
	\begin{minipage}[t]{0.5\hsize}
        \centering
		\includegraphics[width=0.77\linewidth]{figures/TwoABWormholes.pdf}
		\subcaption{Two-$(A,B)$-wormholes}
		\label{fig:TwoABWormholesError}
		\end{minipage}& 
	\begin{minipage}[t]{0.5\hsize}
        \centering
		\includegraphics[width=0.77\linewidth]{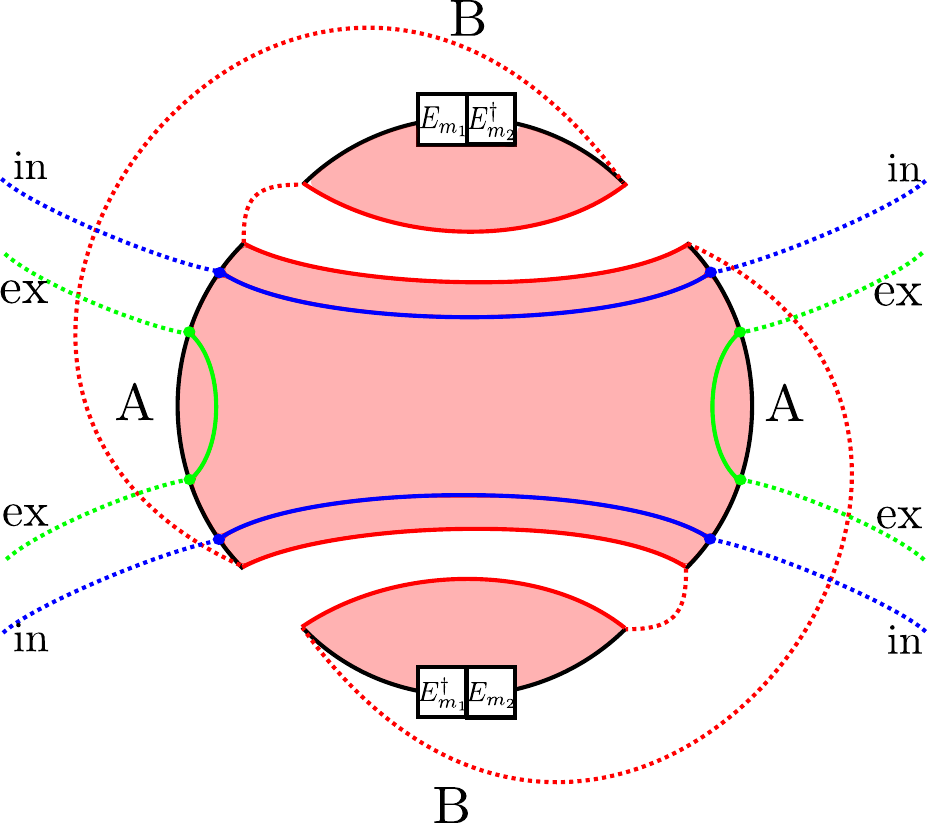}
		\subcaption{One-$(A,A)$-replica wormhole}
		\label{fig:AAWormholeError}
		\end{minipage}\\ 
	\begin{minipage}[t]{0.5\hsize}
        \centering
		\includegraphics[width=0.77\linewidth]{figures/BBWormhole.pdf}
		\subcaption{One-$(B,B)$-replica wormhole}
		\label{fig:BBWormholeError}
	\end{minipage}&
	\begin{minipage}[t]{0.5\hsize}
        \centering
		\includegraphics[width=0.77\linewidth]{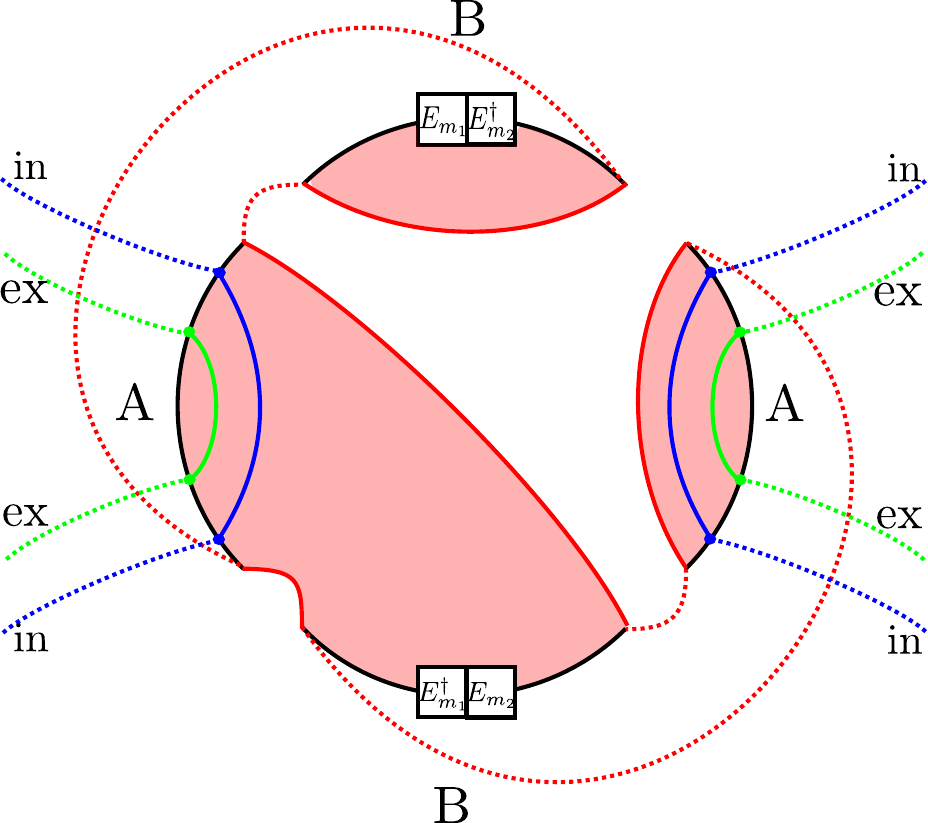}
		\subcaption{One-$(A,B)$-replica wormhole}
		\label{fig:OneAB'WormholesError}
		\end{minipage}
	\end{tabular}	
	\centering
		\begin{minipage}[t]{0.5\hsize}
        \centering
		\includegraphics[width=0.77\linewidth]{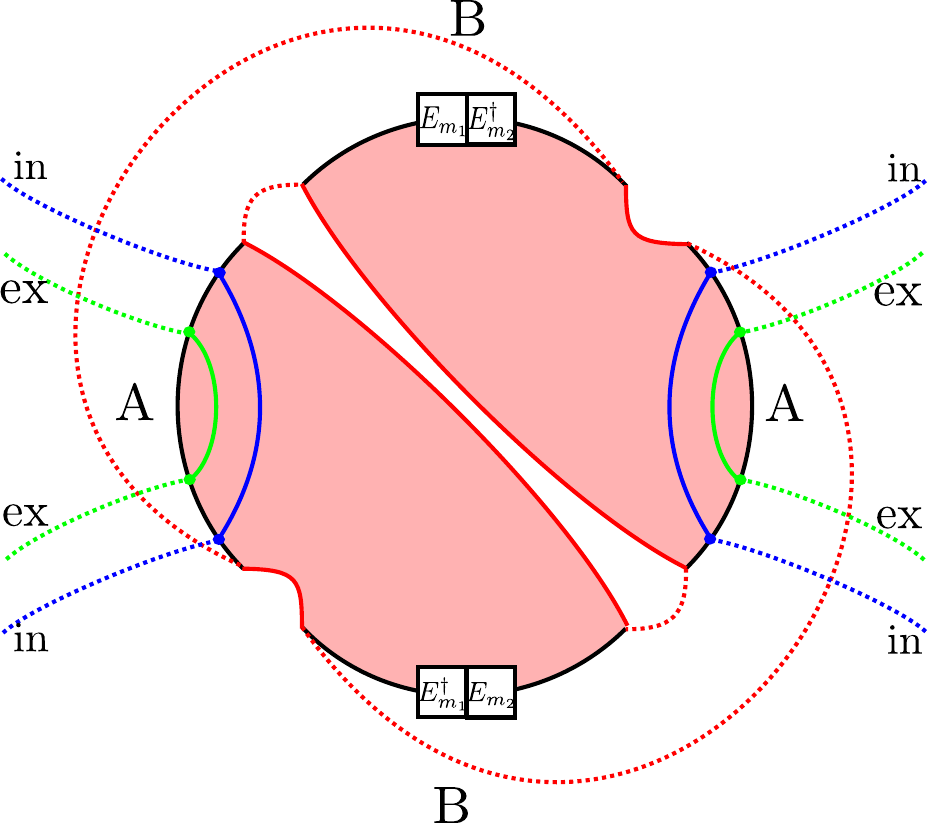}
		\subcaption{Two-$(A,B)$-replica wormholes}
		\label{fig:TwoABreplicaWormholesError}
		\end{minipage}
	\caption{Partially connected diagrams. In these diagrams, we again need to correctly contract code indices so that the resulting contractions are consistent with the  boundary condition in figures \ref{fig:RenyitwoRiReEnBdyCondi} and \ref{fig:RenyitwoReEnBdyCondi}. }
	\label{fig:partialConnected}	
\end{figure}

We note that, dominant geometries within the disconnected family at early times $k< d_{BH}$ and late times $k>  d_{BH}$, are again given by the fully disconnected saddle (figure \ref{fig:FullDiscoError}) and the two-$(A,B)$-wormholes saddle (figure \ref{fig:TwoABWormholesError}). Also, note that, unlike the previous case without the Kraus operators, the current saddles include the Kraus operators on the AdS boundaries of the universes $B$.

\paragraph{Connected family}

Next, we consider the R\'{e}nyi-two entropies for the connected family. Again, in this case, we can not simply evaluate them, but we can evaluate them individually. These brute-force computations give the following results
\begin{equation}
	\begin{aligned}
		&\left. \overline{\tr \left(\rho_{ref(in),\,ref(ex),\, E}'  \right)^{2}} \right|_{\text{Connected}}\\
		&= \frac{1}{(d_{in}d_{ex} N )^{2}}   \cdot d_{ex} \cdot \tr[(\sigma_{\text{Bath}})^{2}]\cdot   \big\{ (d_{in})^{2}\cdot k^{3} (d_{BH})^{2} + d_{in} \cdot k (d_{BH})^{4} +2d_{in}\cdot k^{2}(d_{BH})^{3} \big\}  \\
		& \quad +\frac{1}{(d_{in}d_{ex} N )^{2}} \cdot d_{ex} \cdot  \tr\left[ (\sigma_{E})^{2} \right] \cdot  \big\{ 2(d_{in})^{2}\cdot k^{2}(d_{BH})^{2} + (d_{in})^{2}\cdot k(d_{BH})^{3} \\
		&\hspace{10cm} + d_{in}\cdot k^{2}(d_{BH})^{2} + 2d_{in}\cdot k(d_{BH})^{3}  \big\}
	\end{aligned}
\end{equation}
and 
\begin{equation}
	\begin{aligned}
		&\left. \overline{\tr \left(\rho_{ref(ex),\, E}'  \right)^{2}} \right|_{\text{Connected}}\\
		&= \frac{1}{(d_{in}d_{ex} N )^{2}}   \cdot d_{ex} \cdot  \tr[(\sigma_{\text{Bath}})^{2}]\cdot   \big\{ d_{in}\cdot k^{3} (d_{BH})^{2} + (d_{in})^{2} \cdot k (d_{BH})^{4} +2(d_{in})^{2}\cdot k^{2}(d_{BH})^{3} \big\}  \\
		& \quad +\frac{1}{(d_{in}d_{ex} N )^{2}} \cdot  d_{ex} \cdot  \tr\left[ (\sigma_{E})^{2} \right] \cdot  \big\{ 2d_{in}\cdot k^{2}(d_{BH})^{2} + d_{in}\cdot k(d_{BH})^{3} \\
		&\hspace{10cm} + (d_{in})^{2}\cdot k^{2}(d_{BH})^{2} + 2(d_{in})^{2}\cdot k(d_{BH})^{3}  \big\}.
	\end{aligned}
\end{equation}
Here, $\sigma_{\text{Bath}}$ is given  by  \eqref{eq:defSigmaBH},
\begin{equation*}
	\sigma_{\text{Bath}} = \sum_{m=1}^{d_{E}} E_{m}  \left( \frac{ I_{BH} }{d_{BH}}\right) E_{m}^{\dagger}.
\end{equation*}

To pick up dominant saddle contributions, we again focus on early times $k< d_{BH}$ and late times $d_{BH} < k$. In these time regimes, $\left. \overline{\tr \left(\rho_{ref(in),\,ref(ex),\, E}'  \right)^{2}} \right|_{\text{Connected}}$ can be approximated by
\begin{equation}
	\begin{aligned}
		&\left. \overline{\tr \left(\rho_{ref(in),\,ref(ex),\, E}'  \right)^{2}} \right|_{\text{Connected}}\\
		& \approx \begin{dcases}
			\frac{1}{(d_{in}d_{ex} N )^{2}}   \cdot d_{ex} \left\{ \tr[(\sigma_{\text{Bath}})^{2}]\cdot   d_{in} \cdot k (d_{BH})^{4} +\tr\left[ (\sigma_{E})^{2} \right] \cdot (d_{in})^{2}\cdot k(d_{BH})^{3} \right\}   & \text{ for } k < d_{BH}\\
			\frac{1}{(d_{in}d_{ex} N )^{2}}   \cdot d_{ex} \cdot  \tr[(\sigma_{\text{Bath}})^{2}]\cdot (d_{in})^{2}\cdot k^{3} (d_{BH})^{2}  & \text{ for }  d_{BH} < k
		\end{dcases}\\
		& \approx \begin{dcases}
			\frac{1}{(d_{in}d_{ex} N )^{2}}  \cdot d_{ex} \cdot  \tr\left[ (\sigma_{E})^{2} \right] \cdot (d_{in})^{2}\cdot k(d_{BH})^{3}   & \text{ for } k < d_{BH}\\
			& \qquad \text{ with }S^{(2)}(\sigma_{E}) +\log d_{BH}-S^{(2)}(\sigma_{\text{Bath}}) < \log d_{in}\\
			\frac{1}{(d_{in}d_{ex} N )^{2}}   \cdot d_{ex}  \cdot  \tr[(\sigma_{\text{Bath}})^{2}]\cdot   d_{in} \cdot k (d_{BH})^{4}    & \text{ for }k < d_{BH} \\
			& \qquad \text{ with }S^{(2)}(\sigma_{E}) +\log d_{BH}-S^{(2)}(\sigma_{\text{Bath}}) > \log d_{in}\\
			\frac{1}{(d_{in}d_{ex} N )^{2}}   \cdot d_{ex} \cdot  \tr[(\sigma_{\text{Bath}})^{2}]\cdot (d_{in})^{2}\cdot k^{3} (d_{BH})^{2}  & \text{ for }  d_{BH} < k,
		\end{dcases}
	\end{aligned}\label{eq:PurityRiReEConn}
\end{equation}
where, in obtaining the second case of the first approximate equality, we used the relation\footnote{One can derive this relation from the weak subadditivity \eqref{eq:WeakSubAdditivityBound}.}
\begin{equation}
	d_{BH} \cdot  \tr[(\sigma_{\text{Bath}})^{2}]\geq \tr[(\sigma_{E})^{2}].
\end{equation}
Here, the first case comes from the $(A,A)$-wormhole saddle (figure \ref{fig:AAWormholeError}), which includes a replica wormhole connecting between two replicated universes $A$, the second case from  the $(B,B)$-wormhole saddle (figure \ref{fig:BBWormholeError}), including a replica wormhole connecting between two replicated universes $B$, and the third case from the fully connected saddle (figure \ref{fig:FullConneError}).
Next, $\left. \overline{\tr \left(\rho_{ref(ex),\, E}'  \right)^{2}} \right|_{\text{Connected}}$ can be approximated in the following form
\begin{equation}
	\begin{aligned}
		&\left. \overline{\tr \left(\rho_{ref(ex),\, E}'  \right)^{2}} \right|_{\text{Connected}}\\
		& \approx \begin{dcases}
			\frac{1}{(d_{in}d_{ex} N )^{2}}   \cdot d_{ex} \left\{ \tr[(\sigma_{\text{Bath}})^{2}]\cdot   (d_{in})^{2} \cdot k (d_{BH})^{4} +2\tr\left[ (\sigma_{E})^{2} \right] \cdot (d_{in})^{2}\cdot k(d_{BH})^{3} \right\}   & \text{ for } k < d_{BH}\\
			\frac{1}{(d_{in}d_{ex} N )^{2}}   \cdot d_{ex} \{  \tr[(\sigma_{\text{Bath}})^{2}]\cdot d_{in}\cdot k^{3} (d_{BH})^{2}+ \tr[(\sigma_{E})^{2}]\cdot (d_{in})^{2}\cdot k^{2} (d_{BH})^{2} \}  & \text{ for } d_{BH}  < k
		\end{dcases}\\
		& \approx \begin{dcases}
			\frac{1}{(d_{in}d_{ex} N )^{2}}   \cdot d_{ex} \cdot 2\tr\left[ (\sigma_{E})^{2} \right] \cdot (d_{in})^{2}\cdot k(d_{BH})^{3}   & \text{ for } k < d_{BH}\\
			& \quad \text{ with }S^{(2)}(\sigma_{E}) +\log d_{BH}-S^{(2)}(\sigma_{\text{Bath}}) < \log 2\\
			\frac{1}{(d_{in}d_{ex} N )^{2}}   \cdot d_{ex}  \cdot  \tr[(\sigma_{\text{Bath}})^{2}]\cdot   (d_{in})^{2} \cdot k (d_{BH})^{4}   & \text{ for } k < d_{BH}\\
			& \qquad \text{ with }S^{(2)}(\sigma_{E}) +\log d_{BH}-S^{(2)}(\sigma_{\text{Bath}}) > \log 2\\
			\frac{1}{(d_{in}d_{ex} N )^{2}}   \cdot d_{ex} \cdot  \tr[(\sigma_{E})^{2}]\cdot (d_{in})^{2}\cdot k^{2} (d_{BH})^{2}  & \text{ for } d_{BH}  < k \\
			& \quad \text{ with }S^{(2)}(\sigma_{E}) +\log k-S^{(2)}(\sigma_{\text{Bath}})< \log d_{in}\\
			\frac{1}{(d_{in}d_{ex} N )^{2}}   \cdot d_{ex} \cdot  \tr[(\sigma_{\text{Bath}})^{2}]\cdot d_{in}\cdot k^{3} (d_{BH})^{2}  & \text{ for } d_{BH}  < k \\
			& \quad \text{ with }S^{(2)}(\sigma_{E}) +\log k-S^{(2)}(\sigma_{\text{Bath}}) > \log d_{in}. \\
		\end{dcases}
	\end{aligned}\label{eq:PurityReEConn}
\end{equation}
The first case comes from the one $(A,B)$-replica wormhole saddle (figure \ref{fig:OneAB'WormholesError}), which includes a replica wormhole  connecting a universe $A$ of one replica with a universe $B$ of another replica, the second case from the $(B,B)$-replica wormhole saddle (figure \ref{fig:BBWormholeError}), including a replica wormhole connecting between two replicated universes $B$, the third case from two $(A,B)$-replica wormholes (figure \ref{fig:TwoABreplicaWormholesError}), which include two replica wormholes connecting two different replicas,  and the fourth case from the fully connected saddle (figure \ref{fig:FullConneError}).
To simplify the above expressions, we note that, since we have the inequality $\log d_{BH}\geq S^{(2)}(\sigma_{\text{Bath}})$, the condition 
\begin{equation}
	S^{(2)}(\sigma_{E}) +\log d_{BH}-S^{(2)}(\sigma_{\text{Bath}}) < \log 2
\end{equation}
forces the following equalities,
\begin{equation}
	S^{(2)}(\sigma_{E}) \approx 0, \quad S^{(2)}(\sigma_{\text{Bath}}) \approx \log d_{BH},
\end{equation}
implying the trivial error case.
With this observation, by combining these results with \eqref{eq:normaliApprox}, we get the following purities,
\begin{equation}
	\begin{aligned}
		&\left. \overline{\tr \left(\rho_{ref(in),\,ref(ex),\, E}'  \right)^{2}} \right|_{\text{Connected}}\\
		& \approx \begin{dcases}
			 \frac{1}{d_{ex}}   \cdot \frac{1}{k\, d_{BH}}\cdot \tr \left[(\sigma_{E})^{2}\right]  & \text{ for } k < d_{BH}\\
			& \qquad \text{ with }S^{(2)}(\sigma_{E}) +\log d_{BH}-S^{(2)}(\sigma_{\text{Bath}}) < \log d_{in}\\
			 \frac{1}{d_{in}\, d_{ex}}   \cdot \frac{1}{k}\cdot \tr \left[(\sigma_{\text{Bath}})^{2}\right]     & \text{ for } k < d_{BH} \\
			& \qquad \text{ with }S^{(2)}(\sigma_{E}) +\log d_{BH}-S^{(2)}(\sigma_{\text{Bath}}) > \log d_{in}\\
			\frac{1}{d_{ex}}   \cdot \frac{1}{k}\cdot \tr \left[(\sigma_{\text{Bath}})^{2}\right]   & \text{ for } d_{BH} < k,
		\end{dcases}
	\end{aligned}\label{eq:PurityConnectedRiReEApprox}
\end{equation}
and 
\begin{equation}
	\begin{aligned}
		&\left. \overline{\tr \left(\rho_{ref(ex),\, E}'  \right)^{2}} \right|_{\text{Connected}}\\
		& \approx \begin{dcases}
			 \frac{1}{d_{ex}}   \cdot \frac{1}{k}\cdot \tr \left[(\sigma_{\text{Bath}})^{2}\right]  & \text{ for } k < d_{BH} \\
			 & \qquad \text{ with } \log 2 < S^{(2)}(\sigma_{E}) +\log d_{BH}-S^{(2)}(\sigma_{\text{Bath}})\\
			 \frac{1}{d_{ex}}   \cdot \frac{1}{k^{2}}\cdot \tr \left[(\sigma_{E})^{2}\right]   & \text{ for } d_{BH}  < k \\
			& \qquad \text{ with }S^{(2)}(\sigma_{E}) +\log k-S^{(2)}(\sigma_{\text{Bath}}) < \log d_{in} \\
			\frac{1}{d_{in} \, d_{ex}}   \cdot \frac{1}{k}\cdot \tr \left[(\sigma_{\text{Bath}})^{2}\right]   & \text{ for } d_{BH}  < k \\
			& \qquad \text{ with }S^{(2)}(\sigma_{E}) +\log k-S^{(2)}(\sigma_{\text{Bath}}) > \log d_{in}, \\
		\end{dcases}
	\end{aligned}\label{eq:PurityConnectedReEApprox}
\end{equation}
where we ignored the trivial error case, $S^{(2)}(\sigma_{E}) +\log d_{BH}-S^{(2)}(\sigma_{\text{Bath}}) < \log 2$.

Thus, we obtain the R\'{e}nyi-two entropies for the connected family,
\begin{equation}
	\begin{aligned}
		&\left.  \overline{S^{(2)}\left(\rho_{ref(in),\,ref(ex),\, E}'  \right)} \right|_{\text{Connected}}\\
		&   \approx \begin{dcases}
			 \log d_{ex} + \log k + \log d_{BH}+ S^{(2)}(\sigma_{E}) & \text{ for } k < d_{BH}\\
			& \qquad \text{ with }S^{(2)}(\sigma_{E}) +\log d_{BH}-S^{(2)}(\sigma_{\text{Bath}}) < \log d_{in}\\
			 \log d_{in}+\log d_{ex} + \log k + S^{(2)}(\sigma_{\text{Bath}})     & \text{ for } k < d_{BH} \\
			& \qquad \text{ with }S^{(2)}(\sigma_{E}) +\log d_{BH}-S^{(2)}(\sigma_{\text{Bath}}) > \log d_{in}\\
			 \log d_{ex} + \log k + S^{(2)}(\sigma_{\text{Bath}})  & \text{ for }  d_{BH} < k,
		\end{dcases}	
	\end{aligned}\label{eq:renyiTwoRiReEConnected}
\end{equation}
and 
\begin{equation}
	\begin{aligned}
		&\left.  \overline{S^{(2)}\left(\rho_{ref(ex),\, E}'  \right)} \right|_{\text{Connected}}\\
		& \approx  
		 \begin{dcases}
			\log d_{ex} + \log k + S^{(2)}(\sigma_{\text{Bath}})   & \text{ for } k < d_{BH} \\
			 & \qquad \text{ with } \log 2 < S^{(2)}(\sigma_{E}) +\log d_{BH}-S^{(2)}(\sigma_{\text{Bath}})\\
			\log d_{ex} +2 \log k + S^{(2)}(\sigma_{E})   & \text{ for } d_{BH}  < k \\
			& \qquad \text{ with }S^{(2)}(\sigma_{E}) +\log k-S^{(2)}(\sigma_{\text{Bath}}) < \log d_{in} \\
			\log d_{in} +\log d_{ex} + \log k + S^{(2)}(\sigma_{\text{Bath}})   & \text{ for } d_{BH}  < k \\
			& \qquad \text{ with }S^{(2)}(\sigma_{E}) +\log k-S^{(2)}(\sigma_{\text{Bath}}) > \log d_{in}.		\end{dcases}
	\end{aligned}\label{eq:renyiTwoReEConnected}
\end{equation}
Let us recall that the first, second and third cases in \eqref{eq:renyiTwoRiReEConnected} come from the saddles corresponding to figures \ref{fig:AAWormholeError}, \ref{fig:BBWormholeError} and \ref{fig:FullConneError} respectively, and the first, second, and third cases in \eqref{eq:renyiTwoReEConnected} from the saddles corresponding to figures \ref{fig:BBWormholeError}, \ref{fig:TwoABreplicaWormholesError} and \ref{fig:FullConneError} respectively.

\newpage
\paragraph{Competition between results for disconnected and connected families}
Now that we have evaluated the R\'{e}nyi-two entropies for the disconnected and connected families at early and late times, we evaluate the R\'{e}nyi-two entropies \eqref{eq:renyiTwoThreesystem} and \eqref{eq:renyiTwoTwosystem} by picking up smaller contributions among contributions from disconnected and connected families. Thus, R\'{e}nyi-two entropies are given by
\begin{equation}
	\begin{aligned}
		 &\overline{S^{(2)}\left(\rho_{ref(in),\,ref(ex),\, E}'  \right)}\\
		  & \approx \min\left\{\left. \overline{ S^{(2)}\left(\rho_{ref(in),\, ref(ex),\, E}'  \right) } \right|_{\text{Disconnected}},\left. \overline{ S^{(2)}\left(\rho_{ref(in),\, ref(ex),\, E}'  \right) } \right|_{\text{Connected}} \right\} \\
		  &  \approx \begin{dcases}
			 \min\left\{\log d_{in} +\log d_{ex} +S^{(2)}(\sigma_{E}), \log d_{ex} + \log k + \log d_{BH}+ S^{(2)}(\sigma_{E})\right\} \\
			& \hspace{-7cm} \text{for } k < d_{BH} \text{ with }S^{(2)}(\sigma_{E}) +\log d_{BH}-S^{(2)}(\sigma_{\text{Bath}}) < \log d_{in}\\
			  \min\left\{\log d_{in} +\log d_{ex} +S^{(2)}(\sigma_{E}), \log d_{in}+\log d_{ex} + \log k + S^{(2)}(\sigma_{\text{Bath}})  \right\}    & \\
			& \hspace{-7cm} \text{for } k < d_{BH} \text{ with }S^{(2)}(\sigma_{E}) +\log d_{BH}-S^{(2)}(\sigma_{\text{Bath}}) > \log d_{in}\\
			  \min\left\{\log d_{in} +\log d_{ex} +S^{(2)}(\sigma_{E}), \log d_{ex} + \log k + S^{(2)}(\sigma_{\text{Bath}})\right\}  \\
			  & \hspace{-7cm} \text{for }  d_{BH} < k,
		\end{dcases}\\
		&  = \begin{dcases}
			\log d_{in} +\log d_{ex} +S^{(2)}(\sigma_{E}) \\
			& \hspace{-2cm} \text{for } k < d_{BH} \text{ with } \log d_{BH} -\log d_{in} < I_{c}^{(2)}\left(\frac{1}{d_{BH}}I_{BH},\mathcal{E}\right) \\
			 \log d_{in} +\log d_{ex}+ S^{(2)}(\sigma_{E})& \\
			& \hspace{-2cm} \text{for } k < d_{BH} \text{ with } -\log k <  I_{c}^{(2)}\left(\frac{1}{d_{BH}}I_{BH},\mathcal{E}\right) < \log d_{BH} -\log d_{in} \\
			\log d_{in} +\log d_{ex}+ \log k + S^{(2)}(\sigma_{\text{Bath}})    & \\
			& \hspace{-2cm} \text{for } k < d_{BH} \text{ with } I_{c}^{(2)}\left(\frac{1}{d_{BH}}I_{BH},\mathcal{E}\right) <  -\log k  \\
			\log d_{in} +\log d_{ex} + S^{(2)}(\sigma_{E}) \\
			  & \hspace{-2cm} \text{for }  d_{BH} < k \text{ with } -\log k +\log d_{in} < I_{c}^{(2)}\left(\frac{1}{d_{BH}}I_{BH},\mathcal{E}\right)\\
			 \log d_{ex} +\log k + S^{(2)}(\sigma_{\text{Bath}}) \\
			  & \hspace{-2cm} \text{for }  d_{BH} < k \text{ with } I_{c}^{(2)}\left(\frac{1}{d_{BH}}I_{BH},\mathcal{E}\right) < -\log k +\log d_{in},
		\end{dcases}
	\end{aligned}
\end{equation}
and
\begin{equation}
	\begin{aligned}
		 &\overline{S^{(2)}\left(\rho_{ref(ex),\, E}'  \right)}\\
		  & \approx \min\left\{\left. \overline{ S^{(2)}\left(\rho_{ ref(ex),\, E}'  \right) } \right|_{\text{Disconnected}},\left. \overline{ S^{(2)}\left(\rho_{ref(ex),\, E}'  \right) } \right|_{\text{Connected}} \right\} \\
		  &  \approx \begin{dcases}
		  	\min \left\{\log d_{ex} + S^{(2)}(\sigma_{E}), \, \log d_{ex} + \log k + S^{(2)}(\sigma_{\text{Bath}})  \right\}   &  \\
			 & \qquad \hspace{-6cm} \text{ for } k < d_{BH} \text{ with } \log 2 < S^{(2)}(\sigma_{E}) +\log d_{BH}-S^{(2)}(\sigma_{\text{Bath}})\\
			 \min \left\{\log d_{ex} + S^{(2)}(\sigma_{E}), \, \log d_{ex} +2 \log k + S^{(2)}(\sigma_{E}) \right\} & \\
		  	& \hspace{-6cm} \text{ for } d_{BH}  < k  \text{ with }S^{(2)}(\sigma_{E}) +\log k-S^{(2)}(\sigma_{\text{Bath}}) < \log d_{in}\\
		  	\min \left\{\log d_{ex} + S^{(2)}(\sigma_{E}), \, \log d_{in} +\log d_{ex} + \log k + S^{(2)}(\sigma_{\text{Bath}})  \right\} & \\
		  	& \hspace{-6cm} \text{ for } d_{BH}  < k  \text{ with }S^{(2)}(\sigma_{E}) +\log k-S^{(2)}(\sigma_{\text{Bath}}) > \log d_{in}
		  \end{dcases}\\
		  &  = \begin{dcases}
		  	\log d_{ex} + S^{(2)}(\sigma_{E})  & \\
		  	& \hspace{-3cm} \text{ for } k < d_{BH}\text{ with } -\log k < I_{c}^{(2)}\left(\frac{1}{d_{BH}}I_{BH},\mathcal{E}\right) < \log d_{BH} -\log 2 \\
			 \log d_{ex} + \log k + S^{(2)}(\sigma_{\text{Bath}})  \\
			 & \hspace{-3cm}  \text{ for } k < d_{BH}  \text{ with }  I_{c}^{(2)}\left(\frac{1}{d_{BH}}I_{BH},\mathcal{E}\right) < -\log k\\
			 \log d_{ex} + S^{(2)}(\sigma_{E}) &  \\
		  	& \hspace{-3cm} \text{ for }  d_{BH} < k \text{ with }  \log k-\log d_{in} < I_{c}^{(2)}\left(\frac{1}{d_{BH}}I_{BH},\mathcal{E}\right) \\
		  	\log d_{ex} + S^{(2)}(\sigma_{E}) &  \\
		  	& \hspace{-3cm} \text{ for }  d_{BH} < k \text{ with }  -\log k -\log d_{in} < I_{c}^{(2)}\left(\frac{1}{d_{BH}}I_{BH},\mathcal{E}\right) < \log k-\log d_{in} \\
		  	\log d_{in}  +\log d_{ex} +  \log k + S^{(2)}(\sigma_{\text{Bath}})  & \\
		  	&\hspace{-3cm} \text{ for } d_{BH} < k \text{ with } I_{c}^{(2)}\left(\frac{1}{d_{BH}}I_{BH},\mathcal{E}\right) < -\log k -\log d_{in}, 
		  \end{dcases}
	\end{aligned}
\end{equation}
where $I_{c}^{(2)}\left(\frac{1}{d_{BH}}I_{BH},\mathcal{E}\right)$ is the R\'{e}nyi-two coherent information \eqref{eq:coherentGraviCase}.

By carefully looking at the above expressions under various conditions, we can realize that, when the R\'{e}nyi-two coherent information $I_{c}^{(2)}\left(\frac{1}{d_{BH}}I_{BH},\mathcal{E}\right)$ is bounded by the negative numbers $-\log k$ or $-\log k+\log d_{in}$, the minimum contributions are given by those originated from the disconnected family. Conversely, when the R\'{e}nyi-two coherent information $I_{c}^{(2)}\left(\frac{1}{d_{BH}}I_{BH},\mathcal{E}\right)$ is \textit{not} bounded by the negative numbers $-\log k$ or $-\log k+\log d_{in}$, the minimum contributions are given by those originated from the connected family. Here, we note that by considering the weak subadditivity (see lemma 4.3 of \cite{vanDam:2002ith}), the difference is generally bounded from below\footnote{There is also another lower bound, $I_{c}^{(2)}\left(\frac{1}{d_{BH}}I_{BH},\mathcal{E}\right)=S^{(2)}(\sigma_{\text{Bath}})-S^{(2)}(\sigma_{E}) \geq -S^{(2)}(\sigma_{E})$. Here, since the Hilbert space dimension of the environment system $E$ is $d_{E}$, the R\'{e}nyi-two entropy is bounded as follows $ S^{(2)}(\sigma_{E})\leq \log d_{E}$. Thus, we also have the lower bound by the dimension of the environment system $E$. },
\begin{equation}
	-\log d_{BH} \leq I_{c}^{(2)}\left(\frac{1}{d_{BH}}I_{BH},\mathcal{E}\right).\label{eq:weakSubaddiExpli}
\end{equation}
In appendix \ref{app:weakSubadd}, we give the derivation of this lower bound from the weak subadditivity.

Then, we can simply rewrite the above expressions into the following form,
\begin{equation}
	\begin{aligned}
		 &\overline{S^{(2)}\left(\rho_{ref(in),\,ref(ex),\, E}'  \right)}\\
		  & \approx \min\left\{\left. \overline{ S^{(2)}\left(\rho_{ref(in),\, ref(ex),\, E}'  \right) } \right|_{\text{Disconnected}},\left. \overline{ S^{(2)}\left(\rho_{ref(in),\, ref(ex),\, E}'  \right) } \right|_{\text{Connected}} \right\} \\
		&  = \begin{dcases}
			\log d_{in} +\log d_{ex} +S^{(2)}(\sigma_{E}) \\
			& \hspace{-5cm} \text{for } k < d_{BH} \text{ with } -\log k < I_{c}^{(2)}\left(\frac{1}{d_{BH}}I_{BH},\mathcal{E}\right) \\
			& \hspace{-4.5cm} \text{and for } d_{BH} < k \text{ with } \max\{-\log k + \log d_{in},-\log d_{BH}\} < I_{c}^{(2)}\left(\frac{1}{d_{BH}}I_{BH},\mathcal{E}\right) \\
			\log d_{in} +\log d_{ex}+ \log k + S^{(2)}(\sigma_{\text{Bath}})    & \\
			& \hspace{-5cm} \text{for } k < d_{BH} \text{ with } -\log d_{BH} < I_{c}^{(2)}\left(\frac{1}{d_{BH}}I_{BH},\mathcal{E}\right) <  -\log k  \\
			 \log d_{ex} +\log k + S^{(2)}(\sigma_{\text{Bath}}) \\
			  & \hspace{-5cm} \text{for } d_{BH} < k \text{ with } -\log d_{BH} < I_{c}^{(2)}\left(\frac{1}{d_{BH}}I_{BH},\mathcal{E}\right) < \max\{-\log k + \log d_{in},-\log d_{BH}\} ,
		\end{dcases}
	\end{aligned} \label{eq:renyiTwoRiReEConnectedDisco}
\end{equation}
and 
\begin{equation}
	\begin{aligned}
		 &\overline{S^{(2)}\left(\rho_{ref(ex),\, E}'  \right)}\\
		  & \approx \min\left\{\left. \overline{ S^{(2)}\left(\rho_{ ref(ex),\, E}'  \right) } \right|_{\text{Disconnected}},\left. \overline{ S^{(2)}\left(\rho_{ref(ex),\, E}'  \right) } \right|_{\text{Connected}} \right\} \\
		  &  = \begin{dcases}
		  	\log d_{ex} + S^{(2)}(\sigma_{E})  & \\
		  	& \hspace{-2cm} \text{for } k < d_{BH} \text{ with } -\log k < I_{c}^{(2)}\left(\frac{1}{d_{BH}}I_{BH},\mathcal{E}\right) \\
			& \hspace{-1cm} \text{and for } d_{BH} < k \text{ with } -\log d_{BH} < I_{c}^{(2)}\left(\frac{1}{d_{BH}}I_{BH},\mathcal{E}\right)  \\
			 \log d_{ex} + \log k + S^{(2)}(\sigma_{\text{Bath}})  \\
			 & \hspace{-2cm}  \text{ for } k < d_{BH}  \text{ with } -\log d_{BH} <  I_{c}^{(2)}\left(\frac{1}{d_{BH}}I_{BH},\mathcal{E}\right) < -\log k.
		  \end{dcases}\label{eq:renyiTwoReEConnectedDisco}
	\end{aligned}
\end{equation}
Let us recall which cases in \eqref{eq:renyiTwoRiReEConnectedDisco} and \eqref{eq:renyiTwoReEConnectedDisco} arise from which saddles; the first cases in \eqref{eq:renyiTwoRiReEConnectedDisco} and \eqref{eq:renyiTwoReEConnectedDisco} correspond to the saddles depicted in figure \ref{fig:FullDiscoError} and \ref{fig:TwoABWormholesError} for early and late times respectively, the second case to the one depicted in figure \ref{fig:BBWormholeError}, and the third one of \eqref{eq:renyiTwoRiReEConnectedDisco} to figure \ref{fig:FullConneError} respectively.

Therefore, with these results and \eqref{eq:renyiTwoInFull}, we obtain the R\'{e}nyi-two mutual information
\begin{equation}
	\begin{aligned}
		&\overline{I^{(2)}_{\ket{\Psi'}}(ref(in)\, ;\, ref(ex)\cup E)}\\
		 &\approx
		 \begin{dcases}
		 0	& \hspace{-6cm} \quad \text{for } k < d_{BH}, \\
		 	& \hspace{-6cm} \qquad  \text{ and for } d_{BH} < k \text{ with } \max\{-\log k + \log d_{in},-\log d_{BH}\} < I_{c}^{(2)}\left(\frac{1}{d_{BH}}I_{BH},\mathcal{E}\right)  \\
		 	  \left(-\log k + \log d_{in}\right) - I_{c}^{(2)}\left(\frac{1}{d_{BH}}I_{BH},\mathcal{E}\right) & \\
		 	  & \hspace{-6cm} \text{for } d_{BH} < k \text{ with } -\log d_{BH} \leq  I_{c}^{(2)}\left(\frac{1}{d_{BH}}I_{BH},\mathcal{E}\right) < \max\{-\log k + \log d_{in},-\log d_{BH}\} .
		 \end{dcases}
	\end{aligned} \label{eq:mutualFullEarlyLate}
\end{equation}

This result is analogous to the non-gravitating results (3.25) and (3.31) in \cite{Balasubramanian:2022fiy}.

\section{The R\'{e}nyi-two quantities in the non-gravitating bath}\label{app:NongraRenyi}

In this appendix, we consider the standard West Coast model that is topological, with an error acting on a non-gravitating bath \cite{Balasubramanian:2022fiy}.

\subsection{The R\'{e}nyi-two entropies and mutual information for the decoupling condition}

First, as in  the standard West Coast model, we consider the following physical state, 
\begin{equation}
	\ket{\Psi_{i,i'}}_{\text{NG},phys} =  \frac{1}{\sqrt{N_{\text{NG}}}} \sum_{\alpha=1}^{k} \ket{\psi_{i,i'}^{\alpha}}_{A} \ket{\alpha}_{B}  \qquad (i = 1,\cdots,d_{in}, i'= 1,\cdots , d_{ex}),\label{eq:phsyStateNonGraSym}
\end{equation}
where the state $\ket{\alpha}_{B}$ on the universe $B$, corresponding to the non-gravitating bath, is a non-gravitational state, implying  the state is exactly orthonormal unlike the gravitating state $\ket{\psi^{\alpha}}_{B}$, and $N_{\text{NG}}$ is a normalization factor. The normalization factor $N_{\text{NG}}$ is given by
\begin{equation}
	N_{\text{NG}}=k\,d_{BH},
\end{equation}
where $d_{BH}=e^{S_{0}}$. 

For this physical state, a CPTP error $\mathcal{E}'$ with a Kraus representation $\{K_{m}\}$\footnote{Due to the trace-preserving (TP) property, the Kraus operators satisfy the relation $\sum_{m=1}^{d_{E}}K^{\dagger}_{m}K_{m}$=I.} acts on the universe $B$,
\begin{equation}
	K_{m}\ket{\Psi_{i,i'}}_{\text{NG},phys} =  \frac{1}{\sqrt{N_{\text{NG}}}} \sum_{\alpha=1}^{k} \ket{\psi_{i,i'}^{\alpha}}_{A}\otimes K_{m} \ket{\alpha}_{B}.
\end{equation}
We note that this error acting on the universe $B$ corresponding to the Hawking radiation, whose Hilbert space dimension is equal to $k$. This situation is different from the gravitating bath case, where the error with the Kraus representation $\{E_{m}\}$ acts on the Hilbert space on the universe $B$, but whose dimension is equal to $d_{BH}$.

Again we evaluate the R\'{e}nyi-two mutual information  of the following state,
\begin{equation}
	\begin{aligned}
		\ket{\Psi'}_{\text{NG}} &= \frac{1}{\sqrt{N_{\text{NG}}\, N_{\text{NG},\Psi'} }} \sum_{i=1}^{d_{in}} \sum_{i'=1}^{d_{ex}} \sum_{\alpha=1}^{k} \sum_{m=1}^{d_{E}} \ket{i }_{ref(in)}\otimes \ket{i' }_{ref(ex)}\otimes \ket{\psi_{i,i'}^{\alpha}}_{A}^{*}\otimes (K_{m} \ket{\alpha}_{B})\otimes\ket{e_{m}}_{E}.
	\end{aligned} \label{eq:NonGraState}
\end{equation}
 Here, the normalization factor $ N_{\text{NG},\Psi'}$ is given by $N_{\text{NG},\Psi'} = d_{in}d_{ex}$.

To evaluate the R\'{e}nyi-two mutual information, we need to consider the gravitational path integral of the following purities,
\begin{equation}
	\begin{aligned}
		\tr \left(\rho_{\text{NG};\,ref(in)}'  \right)^{2}&=\frac{1}{(d_{in}d_{ex}\,k\,d_{BH})^{2}} \sum_{\bm{i}=1}^{d_{in}}\sum_{\bm{i}'=1}^{d_{ex}} \sum_{\bm{\alpha}=1}^{k} \braket{\psi_{i_{1},i_{1}'}^{\alpha_{1}}|\psi_{i_{2},i_{1}'}^{\alpha_{1}} } _{A}\braket{\psi_{i_{2},i_{2}'}^{\alpha{2}}|\psi_{i_{1},i_{2}'}^{\alpha_{2}} }_{A},
	\end{aligned}\label{eq:NonGrarenyiTwoOnesystem}
\end{equation}
\begin{equation}
	\begin{aligned}
		\tr \left(\rho_{\text{NG};\,ref(in),\, ref(ex),\, E}'  \right)^{2}&=\frac{1}{(d_{in}d_{ex}\,k\,d_{BH})^{2}} \sum_{\bm{i}=1}^{d_{in}}\sum_{\bm{i}'=1}^{d_{ex}} \sum_{\bm{\alpha},\bm{\beta}=1}^{k} \sum_{\bm{m}=1}^{d_{E}} \braket{\psi_{i_{1},i_{1}'}^{\beta_{1}}|\psi_{i_{2},i_{2}'}^{\alpha_{1}} } _{A}\braket{\psi_{i_{2},i_{2}'}^{\beta_{2}}|\psi_{i_{1},i_{1}'}^{\alpha_{2}} }_{A}\\
		& \hspace{4cm}  \times \braket{\alpha_{1}| K_{m_{2}}^{\dagger} K_{m_{1}} |\beta_{1}}_{B} \braket{\alpha_{2}| K_{m_{1}}^{\dagger} K_{m_{2}} |\beta_{2}}_{B} ,
	\end{aligned}\label{eq:NonGrarenyiTwoThreesystem}
\end{equation}
and
\begin{equation}
	\begin{aligned}
		\tr \left(\rho_{\text{NG};\,ref(ex),\, E}'  \right)^{2}&=\frac{1}{(d_{in}d_{ex}\,k\,d_{BH})^{2}} \sum_{\bm{i}=1}^{d_{in}}\sum_{\bm{i}'=1}^{d_{ex}} \sum_{\bm{\alpha},\bm{\beta}=1}^{k} \sum_{\bm{m}=1}^{d_{E}} \braket{\psi_{i_{1},i_{1}'}^{\beta_{1}}|\psi_{i_{1},i_{2}'}^{\alpha_{1}} } _{A}\braket{\psi_{i_{2},i_{2}'}^{\beta_{2}}|\psi_{i_{2},i_{1}'}^{\alpha_{2}} }_{A}\\
		& \hspace{4cm}  \times \braket{\alpha_{1}| K_{m_{2}}^{\dagger} K_{m_{1}} |\beta_{1}}_{B} \braket{\alpha_{2}| K_{m_{1}}^{\dagger} K_{m_{2}} |\beta_{2}}_{B} ,
	\end{aligned}\label{eq:NonGrarenyiTwoTwosystem}
\end{equation}
where the bold letters again denote the summation with respect to the set of indices.

\begin{figure}[ht]
	\begin{tabular}{cc}
	\begin{minipage}[t]{0.5\hsize}
        \centering
		\includegraphics[width=0.9\linewidth]{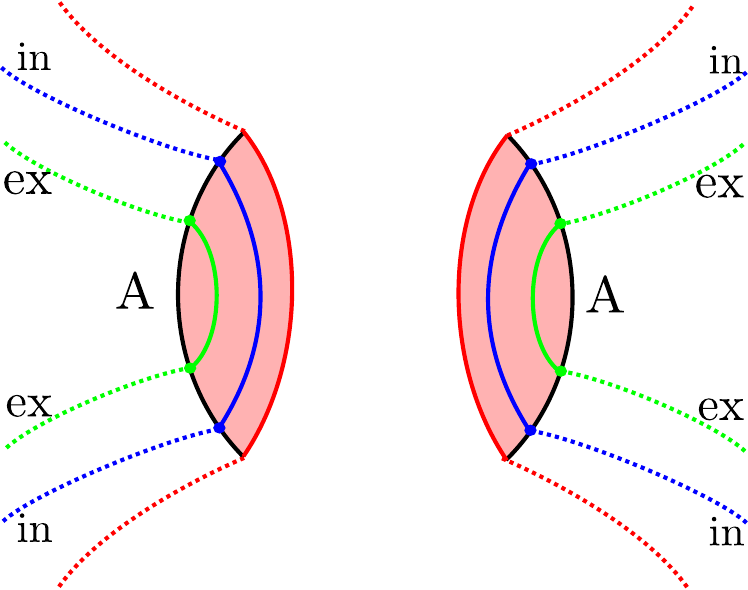}
		\subcaption{Disconnected saddle}
		\label{fig:NonGraFullDiscoError}
		\end{minipage}& 
	\begin{minipage}[t]{0.5\hsize}
        \centering
		\includegraphics[width=0.9\linewidth]{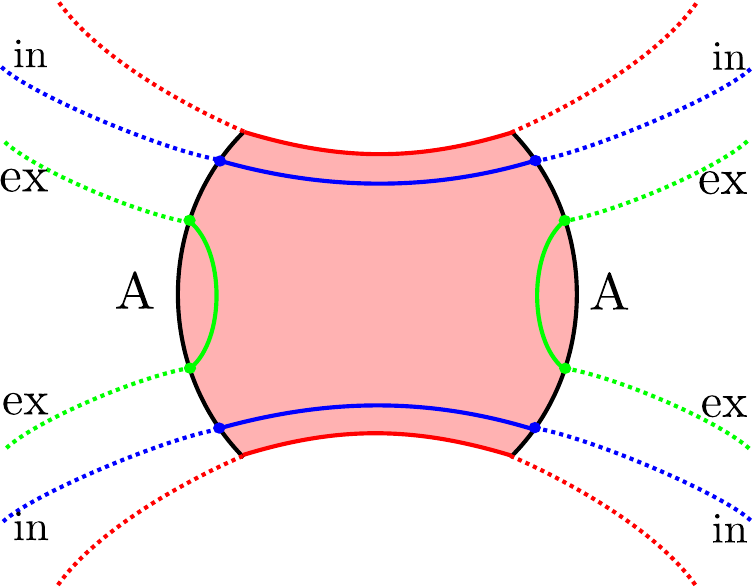}
		\subcaption{Connected saddle}
		\label{fig:NonGraFullConneError}
		\end{minipage}
	\end{tabular}
	\caption{Disconnected and replica wormhole saddle for the gravitational path integral of the R\'{e}nyi-two entropies appearing in the non-gravitating case. In these diagrams, we need to correctly contract code indices so that the resulting contractions are consistent with the  boundary condition given by (\ref{eq:NonGrarenyiTwoOnesystem}), (\ref{eq:NonGrarenyiTwoThreesystem}), and (\ref{eq:NonGrarenyiTwoTwosystem}).  }
	\label{fig:NonGraFullDisconeConneError}
\end{figure}

The gravitational path integral of these purities is done by the standard West Coast model computation \cite{Balasubramanian:2022fiy}, and the results are given by
\begin{equation}
	\begin{aligned}
		\overline{\tr \left(\rho_{\text{NG};\,ref(in)}'  \right)^{2}} = \frac{1}{d_{in}} + \frac{1}{k\, d_{BH}}\approx \frac{1}{d_{in}},
	\end{aligned}\label{eq:NonGraPutityIn}
\end{equation}
\begin{equation}
	\begin{aligned}
		\overline{\tr \left(\rho_{\text{NG};\,ref(in),\,ref(ex),\, E}'  \right)^{2}} = \frac{1}{d_{in}\,d_{ex}} \tr\left[(\tau_{E})^{2}\right]+\frac{1}{d_{ex}}\cdot \frac{1}{d_{BH}}\cdot  \tr\left[(\tau_{\text{Bath}})^{2}\right],
	\end{aligned}\label{eq:NonGraPutityInExtE}
\end{equation}
and
\begin{equation}
	\begin{aligned}
		\overline{\tr \left(\rho_{\text{NG};\,ref(ex),\, E}'  \right)^{2}} = \frac{1}{d_{ex}} \tr\left[(\tau_{E})^{2}\right]+\frac{1}{d_{in}\,d_{ex}}\cdot \frac{1}{d_{BH}}\cdot  \tr\left[(\tau_{\text{Bath}})^{2}\right],
	\end{aligned}\label{eq:NonGraPutityExtE}
\end{equation}
where, in \eqref{eq:NonGraPutityIn} we used $d_{in}\ll k\, d_{BH}$, and $\tau_{E}$ and $\tau_{\text{Bath}}$ are defined by
\begin{equation}
		\tau_{E}= \sum_{m,n=1}^{d_{E}} \dfrac{\tr_{R} \left\{ K_{m}K_{n}^{\dagger} \right\} }{ k } \ket{e_{m}}_{E} \bra{e_{n}},\label{eq:defTauE}
\end{equation}
\begin{equation}
	\tau_{\text{Bath}} = \sum_{m=1}^{d_{E}} K_{m}  \left( \frac{ I_{R} }{k}\right) K_{m}^{\dagger}.\label{eq:defTauBH}
\end{equation}
Here, the first terms in \eqref{eq:NonGraPutityIn}, \eqref{eq:NonGraPutityInExtE} and \eqref{eq:NonGraPutityExtE} come from the disconnected saddles (see figure \ref{fig:NonGraFullDiscoError}), and the second ones from the connected saddles (see figure \ref{fig:NonGraFullConneError}).  

Thus, we have the R\'{e}nyi-two entropies
\begin{equation}
	\begin{aligned}
		\overline{S^{(2)} \left(\rho_{\text{NG};\,ref(in)}'  \right)} \approx  \log d_{in} \qquad (\text{figure } \ref{fig:NonGraFullDiscoError}),
	\end{aligned}\label{eq:NonGraRenyiTwoIn}
\end{equation}
\begin{equation}
	\begin{aligned}
		&\overline{S^{(2)}  \left(\rho_{\text{NG};\,ref(in),\,ref(ex),\, E}'  \right)}\\
		&\approx \begin{dcases}
			\log d_{in} +\log d_{ex} + S^{(2)}\left(\tau_{E}\right) & \text{ for } -\log d_{BH}+\log d_{in} < I_{c}^{(2)}\left(\frac{1}{k}I_{R},\mathcal{E}'\right) \quad   (\text{figure } \ref{fig:NonGraFullDiscoError}) \\
			\log d_{ex}+\log d_{BH}  + S^{(2)}\left(\tau_{\text{Bath}}\right) & \text{ for }  I_{c}^{(2)}\left(\frac{1}{k}I_{R},\mathcal{E}'\right) < -\log d_{BH}+\log d_{in} \quad  (\text{figure } \ref{fig:NonGraFullConneError}),
		\end{dcases}
	\end{aligned}\label{eq:NonGraRenyiTwoInExtEn}
\end{equation}
and
\begin{equation}
	\begin{aligned}
		&\overline{S^{(2)}  \left(\rho_{\text{NG};\,ref(ex),\, E}'  \right)}\\
		&\approx \begin{dcases}
			\log d_{ex} + S^{(2)}\left(\tau_{E}\right) &  \hspace{-3cm} \text{ for } -\log d_{BH}-\log d_{in} < I_{c}^{(2)}\left(\frac{1}{k}I_{R},\mathcal{E}'\right) \quad   (\text{figure } \ref{fig:NonGraFullDiscoError}) \\
			\log d_{in} +\log d_{ex}+\log d_{BH}  + S^{(2)}\left(\tau_{\text{Bath}}\right)& \\
			& \hspace{-3cm} \text{ for }  I_{c}^{(2)}\left(\frac{1}{k}I_{R},\mathcal{E}'\right) < -\log d_{BH}-\log d_{in} \quad  (\text{figure } \ref{fig:NonGraFullConneError}),
		\end{dcases}
	\end{aligned}\label{eq:NonGraRenyiTwoExtEn}
\end{equation}
where $I_{c}^{(2)}\left(\frac{1}{k}I_{R},\mathcal{E}'\right)$ is the R\'{e}nyi-two  coherent information given by
\begin{equation}
	I_{c}^{(2)}\left(\frac{1}{k}I_{R},\mathcal{E}'\right)=S^{(2)}\left(\tau_{\text{Bath}}\right)  - S^{(2)}\left(\tau_{E}\right).
\end{equation}
Here, the figures in the above R\'{e}nyi-two entropies refer to the dominant saddles.
We note that, as in the condition \eqref{eq:coherentInfoBound}, the R\'{e}nyi-two coherent information satisfies the condition,
\begin{equation}
\max\{-\log k,-\log d_{E}\} \leq I_{c}^{(2)}\left(\frac{1}{k}I_{R},\mathcal{E}'\right) \leq \log k. \label{eq:NonGracoherentInfoBound}
\end{equation}

In figure \ref{fig:PhaseDiagramNonGraRenyi}, we give the phase diagrams of the dominant saddle in the R\'{e}nyi-two entropies  on $\log k$-$\left(S^{(2)}\left(\tau_{\text{Bath}}\right)  - S^{(2)}\left(\tau_{E}\right)\right)$ plane. We note that, since we consider the R\'{e}nyi-two entropies for the reference and environment systems, not the Hawking radiation or the black hole, the dominant saddle are different from those for the Hawking radiation or the black hole in some parameter regions. For example, at late times $k> d_{BH}$, there is a possibility that the dominant saddles for  the R\'{e}nyi-two entropies, $\overline{S^{(2)}\left(\rho_{ref(in),\,ref(ex),\, E}'  \right)}$ and $\overline{S^{(2)}\left(\rho_{\,ref(ex),\, E}'  \right)}$ can be the disconnected saddle, while the dominant saddle for the R\'{e}nyi-two entropy of the Hawking radiation, $\overline{S^{(2)}\left(\rho_{B}  \right)}$ is given by a connected saddle.

Thus, the R\'{e}nyi-two mutual information for the non-gravitating is given by
\begin{equation}
	\begin{aligned}
		&\overline{I^{(2)}_{\ket{\Psi'}_{\text{NG}}}(ref(in)\, ;\, ref(ex)\cup E)}\\
		 &\approx
		 \begin{dcases}
		 	0 & \hspace{-5cm} \text{ for } -\log d_{BH}+\log d_{in} < I_{c}^{(2)}\left(\frac{1}{k}I_{R},\mathcal{E}'\right)\\
		 	  \left(-\log d_{BH} + \log d_{in}\right) - I_{c}^{(2)}\left(\frac{1}{k}I_{R},\mathcal{E}'\right) & \\
		 	  & \hspace{-5cm} \text{ for } -\log d_{BH}-\log d_{in} < I_{c}^{(2)}\left(\frac{1}{k}I_{R},\mathcal{E}'\right) < -\log d_{BH}+\log d_{in} \\
		 	2\log d_{in} & \hspace{-5cm} \text{ for }  I_{c}^{(2)}\left(\frac{1}{k}I_{R},\mathcal{E}'\right) < -\log d_{BH}-\log d_{in}.
		 \end{dcases}
	\end{aligned} \label{eq:NonGramutualFull}
\end{equation}
A similar result was obtained in \cite{Balasubramanian:2022fiy} for the $n=1$ case, not the R\'{e}nyi-two case.

In figure \ref{fig:PhaseDiagramNonGraRenyi}, we give the phase diagrams of the value of the R\'{e}nyi-two mutual information  on $\log k$-$\left(S^{(2)}\left(\tau_{\text{Bath}}\right)  - S^{(2)}\left(\tau_{E}\right)\right)$ plane.

\subsection{The R\'{e}nyi-two entropy of the Hawking radiation}
Next, we consider the R\'{e}nyi-two entropy of the Hawking radiation, stored in the universe $B$, in the state \eqref{eq:NonGraState}. In this case, we evaluate the gravitational path integral of the following quantity,
\begin{equation}
	\begin{aligned}
		\tr \left(\rho_{\text{NG};\,B}'  \right)^{2}&=\frac{1}{(d_{in}d_{ex}\,k\,d_{BH})^{2}} \sum_{\bm{i}=1}^{d_{in}}\sum_{\bm{i}'=1}^{d_{ex}} \sum_{\bm{\alpha},\bm{\beta}=1}^{k} \sum_{\bm{m}=1}^{d_{E}} \braket{\psi_{i_{1},i_{1}'}^{\beta_{1}}|\psi_{i_{1},i_{1}'}^{\alpha_{1}} } _{A}\braket{\psi_{i_{2},i_{2}'}^{\beta_{2}}|\psi_{i_{2},i_{2}'}^{\alpha_{2}} }_{A}\\
		& \hspace{4cm}  \times \braket{\beta_{2}| K_{m_{2}}^{\dagger} K_{m_{1}} |\alpha_{1}}_{B} \braket{\beta_{1}| K_{m_{1}}^{\dagger} K_{m_{2}} |\alpha_{2}}_{B},
	\end{aligned}\label{eq:NonGrarenyiTwoHawking}
\end{equation}
and get
\begin{equation}
	\begin{aligned}
		\overline{\tr \left(\rho_{\text{NG};\,B}'  \right)^{2}} = \tr\left[(\tau_{\text{Bath}})^{2}\right]+\frac{1}{d_{in}}\cdot \frac{1}{d_{BH}}\cdot \tr\left[(\tau_{E})^{2}\right],
	\end{aligned}
\end{equation}
where the first term comes from the disconnected saddle (see figure \ref{fig:NonGraFullDiscoError}) and the second term from the connected saddle (see figure \ref{fig:NonGraFullConneError}).
Thus, the R\'{e}nyi-two entropy of the Hawking radiation in the state \eqref{eq:NonGraState} is given by 
\begin{equation}
	\begin{aligned}
		&\overline{S^{(2)}  \left(\rho_{\text{NG};\,B}'  \right)}\\
		&\approx \begin{dcases}
			 S^{(2)}\left(\tau_{\text{Bath}}\right) &  \hspace{-3cm} \text{ for }   I_{c}^{(2)}\left(\frac{1}{k}I_{R},\mathcal{E}'\right) < \log d_{in}+ \log d_{BH} \quad   (\text{figure } \ref{fig:NonGraFullDiscoError}) \\
			\log d_{in} +\log d_{BH}  +  S^{(2)}\left(\tau_{E}\right)& \\
			& \hspace{-3cm} \text{ for } \log d_{in}+ \log d_{BH} <  I_{c}^{(2)}\left(\frac{1}{k}I_{R},\mathcal{E}'\right)  \quad  (\text{figure } \ref{fig:NonGraFullConneError}).
		\end{dcases}
	\end{aligned}
\end{equation}
This is a Page curve of the Hawking radiation stored in the universe $B$, on which the error $\mathcal{E}'$ acts.

\section{Coherent information for Haar random error}
\label{app:coherentHaar}

In this appendix, we show the coherent information \eqref{eq:coherentInf} for (CPTP) Haar random error $\mathcal{E}_{\text{Haar}}$\footnote{See \cite{Balasubramanian:2022fiy} for related discussions.}
  with an input state $\eta$ on a Hilbert space $\mathcal{H}$, whose dimension is $d$. In the main body of this paper, we focus on the maximally mixed state case, $\eta =I/d$.
  Here, the Haar random error $\mathcal{E}_{\text{Haar}}$ means that the Stinespring representation \cite{Nielsen_Chuang_2010} of the error includes a Haar random matrix $U_{\text{Haar}}$,
\begin{equation}
	\mathcal{E}_{\text{Haar}}(\eta)=\tr_{E}\left[U_{\text{Haar}}\left(\eta\otimes\kappa_{E}  \right) U_{\text{Haar}}^{\dagger} \right],\label{eq:StinespringdilationError}
\end{equation}
where $E$ denotes an environment system, whose Hilbert space dimension is $d_{E}$, to implement the error, and $\kappa_{E}$ is a reference state on the environment system.
For simplicity, we assume that $\kappa_{E}$ is given by a pure state on the environment system, and the Hilbert space dimensions are sufficiently large, $d,d_{E}\gg 1$.

For notational convenience, let us define the following state to compute the entropy exchange,
\begin{equation}
	\mathcal{E}_{\text{Haar}}^{c}(\eta)=\tr_{\mathcal{H}}\left[U_{\text{Haar}}\left(\eta\otimes\kappa_{E}  \right) U_{\text{Haar}}^{\dagger} \right]
\end{equation}

For this Haar random error $\mathcal{E}_{\text{Haar}}$ and an input state $\eta$, the R\'{e}nyi-$n$ entropies to the coherent information \eqref{eq:coherentInf} are given by\footnote{One can derive these results by using the Stinespring representation \eqref{eq:StinespringdilationError}. }
\begin{equation}
	\overline{S^{(n)}\left(\mathcal{E}_{\text{Haar}}(\eta)\right)}\approx \frac{1}{1-n}\log \overline{\tr\left[ \left(\mathcal{E}_{\text{Haar}}(\eta)\right)^{n}\right]}\approx \begin{dcases}
		\log d & \text{ for } \log d-\log d_{E} \ll S^{(n)}(\eta)\\
		\log d_{E}+S^{(n)}(\eta) & \text{ for } S^{(n)}(\eta) \ll \log d-\log d_{E} 
	\end{dcases}
\end{equation}
and
\begin{equation}
	\overline{S^{(n)}(\mathcal{E}_{\text{Haar}},\eta)} \approx \frac{1}{1-n}\log \overline{\tr\left[ \left(\mathcal{E}_{\text{Haar}}^{c}(\eta)\right)^{n}\right]} \approx \begin{dcases}
		\log d_{E} & \text{ for }-S^{(n)}(\eta)\ll \log d-\log d_{E} \\
		\log d+S^{(n)}(\eta) & \text{ for }  \log d-\log d_{E}  \ll -S^{(n)}(\eta),
	\end{dcases} 
\end{equation}
where the overlines denote the Haar random average, and $S^{(n)}(\eta)$ is the R\'{e}nyi-$n$ entropy of the state $\eta$, $S^{(n)}(\eta)=\frac{1}{1-n}\log\tr[\eta^{n}]$. 
By taking the von Neumann limit $n\to 1$, the above relation reduces to
\begin{equation}
	\overline{S\left(\mathcal{E}_{\text{Haar}}(\eta)\right)}\approx \begin{dcases}
		\log d & \text{ for } \log d-\log d_{E} \ll S(\eta)\\
		\log d_{E}+S(\eta) & \text{ for } S(\eta) \ll \log d-\log d_{E} 
	\end{dcases}\label{eq:nthErroEntropy}
\end{equation}
and
\begin{equation}
	\overline{S(\mathcal{E}_{\text{Haar}},\eta)}  \approx \begin{dcases}
		\log d_{E} & \text{ for }-S(\eta)\ll \log d-\log d_{E} \\
		\log d+S(\eta) & \text{ for }  \log d-\log d_{E}  \ll -S(\eta),
	\end{dcases} \label{eq:nthEntropyExchange}
\end{equation}
leading to the coherent information,
\begin{equation}
	\overline{I_{c}\left(\eta,\mathcal{E}_{\text{Haar}}\right)} \approx
	 \begin{dcases}
		S(\eta) & \text{ for }   S(\eta) \ll \log d-\log d_{E}\\
		\log d -\log d_{E}  & \text{ for } -S(\eta) \ll \log d-\log d_{E} \ll S(\eta) \\
		-S(\eta) & \text{ for } \log d-\log d_{E} \ll -S(\eta).
	\end{dcases}\label{eq:HaarRandomCoherentInfor}
\end{equation}
As we can see from the R\'{e}nyi-$n$ entropies \eqref{eq:nthErroEntropy} and \eqref{eq:nthEntropyExchange}, the above coherent information can be applied to the $n=2$ case, which we are mainly interested in, by just replacing $S(\eta)$ with  $S^{(2)}(\eta)$.

If the input state $\eta$ is the maximally mixed state, $\eta=\frac{1}{d}I_{d}$, these quantities are reduced to
\begin{equation}
	\overline{S\left(\mathcal{E}_{\text{Haar}}\left(\frac{1}{d}I_{d}\right)\right)}\approx \log d,\label{eq:nthErroEntropyMaximallyMix}
\end{equation}
\begin{equation}
	\overline{S\left(\mathcal{E}_{\text{Haar}},\frac{1}{d}I_{d}\right)}  \approx \begin{dcases}
		\log d_{E} & \text{ for }\log d_{E}\ll 2\log d  \\
		2\log d & \text{ for }  2\log d \ll \log d_{E} ,
	\end{dcases} \label{eq:nthEntropyExchangeMaximallyMix}
\end{equation}
and
\begin{equation}
	\overline{I_{c}\left(\frac{1}{d}I_{d},\mathcal{E}_{\text{Haar}}\right)}\approx \begin{dcases}
		 -\log d & \text{ for } 2\log d \ll \log d_{E}\\
		\log d-\log d_{E} & \text{ for } \log d_{E} \ll 2\log d.
	\end{dcases} 
\end{equation}

\section{Bound from the Weak subadditivity}\label{app:weakSubadd}

In this appendix, we explain that the weak subadditivity  (see lemma 4.3 of \cite{vanDam:2002ith}) gives the bound, 
\begin{equation}
	-\log d_{BH} \leq S^{(2)}(\sigma_{\text{Bath}})-S^{(2)}(\sigma_{E})\leq \log d_{BH},\label{eq:WeakSubAdditivityBound}
\end{equation}
where $\sigma_{\text{Bath}},\sigma_{E}$ are defined by \eqref{eq:defSigmaBH}, \eqref{eq:defSigmaE} respectively.
However, for generality, we show a more general inequality: for a CPTP error $\mathcal{N}$, which has the Kraus representation $\{F_{m}\}$ and the environment system $H_{E}$, acting on the Hilbert space $H_{B}$ with the Hilbert space dimension $d_{B}$ and for an arbitrary state $\rho_{B}$ on the Hilbert space $H_{B}$,
\begin{equation}
	-S^{(0)}(\rho_{B})  \leq S^{(n)}\left(\mathcal{N}(\rho_{B})\right) - S^{(n)}\left(\mathcal{N}^{c}(\rho_{B})\right) \leq S^{(0)}(\rho_{B}) \qquad \text{ for } n \in (0,1)\,  \cup  \, (1,\infty),\label{eq:OurGeneralIneq}
\end{equation}
where $\mathcal{N}^{c}$ is defined by \eqref{eq:complementErrorChannel}. Here, $S^{(n)}(\rho)$ denotes the usual R\'{e}nyi-$n$ entropy,
\begin{equation}
	S^{(n)}(\rho) = \frac{1}{1-n} \log \left( \tr [ \rho^{n}]\right),
\end{equation}
and $S^{(0)}(\rho_{B})$ denotes the max-entropy,
\begin{equation}
	S^{(0)}(\rho)=\log \left(\text{rank}\left[\rho\right]\right).
\end{equation} 
When $\rho_{B}=\frac{1}{d_{B}}I_{B}$ and $n=2$, the above general inequality is reduced to our case.

First, in general, the weak subadditivity (see lemma 4.3 of \cite{vanDam:2002ith}) states that, for an arbitrary bipartite density matrix $\rho_{AB}$ on the Hilbert space $H_{A}\otimes H_{B}$, R\'{e}nyi entropy satisfies the following inequality,
\begin{equation}
	S^{(n)}(\rho_{B})-S^{(0)}(\rho_{A}) \leq S^{(n)}(\rho_{AB}) \leq   S^{(n)}(\rho_{B})+S^{(0)}(\rho_{A}) \qquad \text{ for } n \in (0,1)\,  \cup  \, (1,\infty), \label{eq:weakSubadd}
\end{equation}

To use this weak subadditivity, we introduce an auxiliary system $A$ and consider a state
\begin{equation}
	\ket{\varphi}_{ABE}=\sum_{a=1}^{\text{rank}[\rho]} \sum_{n=1}^{d_{E}} \sqrt{\rho_{A}} \ket{a}_{A}\otimes F_{n}\ket{a}_{B}\otimes \ket{e_{n}}_{E},
\end{equation}
where $\ket{e_{n}}_{E}$ is an orthonormal basis on the environment system, and $\ket{a}_{A},\ket{a}_{B}$ are orthonormal eigenvectors of the density matrix $\rho$ on the Hilbert spaces $H_{A},H_{B}$ respectively, i.e., $\sqrt{\rho_{A}} \ket{a}_{A}=\lambda_{a}^{1/2} \ket{a}_{A} $, $\sqrt{\rho_{B}} \ket{a}_{B}=\lambda_{a}^{1/2} \ket{a}_{B} $ ($\lambda_{a}$: eigenvalue of the density matrix). 
From this state, we can construct density matrices,
\begin{equation}
	\varphi_{A}= \tr_{BE} \left[ \ket{\varphi}_{ABE}\bra{\varphi} \right]=\rho_{A},
\end{equation}
\begin{equation}
	\varphi_{B}= \tr_{AE} \left[ \ket{\varphi}_{ABE}\bra{\varphi} \right]=\sum_{m=1}^{d_{E}}F_{m}\rho_{B}F_{m}^{\dagger}=\mathcal{N}\left(\rho_{B}\right),
\end{equation}
\begin{equation}
	\varphi_{E}= \tr_{AB} \left[ \ket{\varphi}_{ABE}\bra{\varphi} \right]=\sum_{m,n=1}^{d_{E}} \ket{e_{m}}_{E}\bra{e_{n}} \cdot \tr \left[F_{m}\rho_{B} F_{n}^{\dagger} \right] =\mathcal{N}^{c}\left(\rho_{B}\right),
\end{equation}
and
\begin{equation}
	\begin{aligned}
		\varphi_{AB}= \tr_{E} \left[ \ket{\varphi}_{ABE}\bra{\varphi} \right]&= \sum_{a,b=1}^{d_{A}}\sum_{m,n=1}^{d_{E}} \sqrt{\rho_{A}} \ket{a} \bra{a'}_{A} \sqrt{\rho_{A}}\otimes F_{n}\ket{a}_{B} \bra{a'}_{B}F_{n}^{\dagger}
	\end{aligned}
\end{equation}
Here, we note that since the total state $\ket{\varphi}_{ABE}$ is pure, we have $S^{(n)}(\varphi_{E})=S^{(n)}(\varphi_{AB})$.

Then, from the weak subadditivity \eqref{eq:weakSubadd}, we get
\begin{equation}
	S^{(0)}\left( \rho_{A} \right) - S^{(n)}\left( \mathcal{N}\left(\rho_{B}\right) \right) \leq S^{(n)}\left( \mathcal{N}^{c}\left(\rho_{B}\right)\right) \leq S^{(0)}\left( \rho_{A} \right) + S^{(n)}\left( \mathcal{N}\left(\rho_{B}\right) \right).
\end{equation}
Here, since the density matrix $\rho_{A}$ has the same entanglement spectrum as $\rho_{B}$, we have $S^{(0)}\left( \rho_{A} \right) =S^{(0)}\left( \rho_{B} \right) $. Thus, we obtain the inequality,
\begin{equation}
	 S^{(n)}\left(\mathcal{N}(\rho_{B})\right)-S^{(0)}(\rho_{B})\leq S^{(n)}\left(\mathcal{N}^{c}(\rho_{B})\right)  \leq  S^{(n)}\left(\mathcal{N}(\rho_{B})\right)+S^{(0)}(\rho_{B}).
\end{equation}
By rearranging the terms, we get the desired inequality \eqref{eq:OurGeneralIneq}.

\section{Interpretation of coherent information}\label{app:CoherentInformation}

In this appendix, we briefly explain the details of the coherent information $I_{c}(\tau_{B},\mathcal{N})$, ($\tau_{B}$: quantum state, $\mathcal{N}$: quantum channel), \eqref{eq:coherentInf}, in terms of mutual information to get a better understanding of the coherent information.

\begin{figure}[ht]
	\centering
	\includegraphics[scale=0.5]{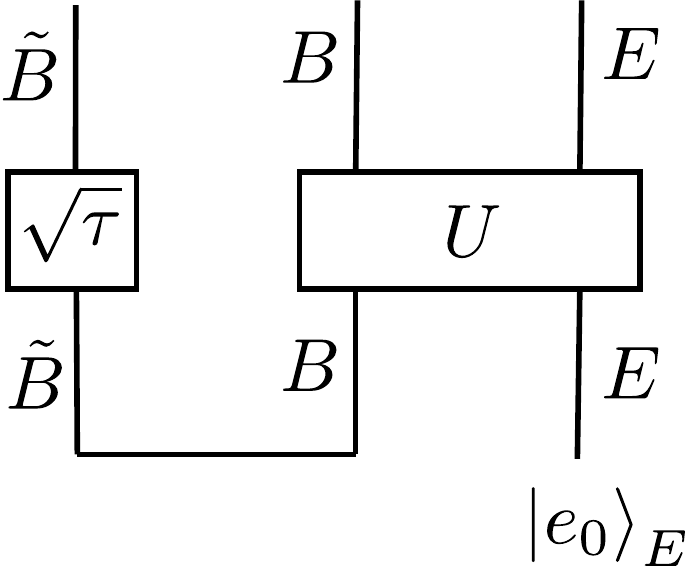}
	\caption{Circuit diagram of the state $\ket{\Phi}_{\widetilde{B}BE}$, (\ref{eq:PhiBBE}).}
	\label{fig:PhiCircuit}
\end{figure}

To this end, it is convenient to use the Stinespring representation of the quantum channel as we briefly explained in the footnote \ref{foot:ComplementChannel}, i.e., for the quantum channel,
\begin{equation}
	\mathcal{N}(\tau_{B}) =\tr_{E} \left[ U_{B,E}\left(\tau_{B} \otimes \ket{e_{0}}\bra{e_{0}}  \right)U^{\dagger}_{B,E}   \right].
\end{equation}
and for the complement channel,
\begin{equation}
	\mathcal{N}^{c}(\tau_{B}) =\tr_{B} \left[ U_{B,E}\left(\tau_{B} \otimes \ket{e_{0}}\bra{e_{0}}  \right)U^{\dagger}_{B,E}   \right]. 
\end{equation}
We can obtain these quantum channels from the following pure state by introducing a purifier system $\widetilde{B}$,
\begin{equation}
	\ket{\Phi}_{\widetilde{B}BE} = \left(I_{\widetilde{B}}\otimes U_{B,E}\right) \ket{\sqrt{\tau}}_{\widetilde{B}B}\otimes \ket{e_{0}}_{E},\label{eq:PhiBBE}
\end{equation}
where $\ket{\sqrt{\tau}}_{\widetilde{B}B}$ is the purified pure state of the input state $\tau$,
\begin{equation}
	\ket{\sqrt{\tau}}_{\widetilde{B}B}=\sum_{b=1}^{d_{B}} \sqrt{\tau}\ket{b}_{\widetilde{B}}  \otimes \ket{b}_{B}.
\end{equation}
Here, $\ket{b}_{\widetilde{B}}, \ket{b}_{B}$ are orthonormal bases for system $\widetilde{B},B$ respectively.
The circuit diagram of this is depicted as in figure \ref{fig:PhiCircuit}.

 This state indeed gives their quantum channels,
\begin{equation}
	\mathcal{N}(\tau_{B}) = \tr_{\widetilde{B}E} \left[ \ket{\Phi}_{\widetilde{B}BE}\bra{\Phi}\right],\quad \mathcal{N}^{c}(\tau_{B}) = \tr_{\widetilde{B}B} \left[ \ket{\Phi}_{\widetilde{B}BE}\bra{\Phi}\right].
\end{equation}

Using the state \eqref{eq:PhiBBE}, we can rewrite the coherent information in terms of the mutual information (see e.g., \cite{Wilde_2013,Wilde:2011npi,Bao:2018msr})
\begin{equation}
	\begin{aligned}
		I_{c}(\tau_{B},\mathcal{N})&=S\left(\mathcal{N}(\tau_{B})\right)-S\left(\mathcal{N}^{c}(\tau_{B})\right)\\
		&=\frac{1}{2}\left\{ I_{\ket{\Phi}}\left(\widetilde{B};B \right) - I_{\ket{\Phi}}\left(\widetilde{B};E \right) \right\},
	\end{aligned}\label{eq:CoherentMutual}
\end{equation}
where $I_{\ket{\Phi}}\left(\widetilde{B};B (E) \right)$ denotes the mutual information between $\widetilde{B}$ and $B (E)$ on the state $\ket{\Phi}$ \eqref{eq:PhiBBE}. One can easily check the above relation.

Thus, based on the above expression, the coherent information measures how effectively the purifier system $\widetilde{B}$ (which contains the purified input information $\tau$) correlates with the original system $B$ compared to the correlation between $B$ and the environment system $E$.

\section{Causality violation and recoverability condition}
\label{app:CausalityViolation}

In this appendix, we relate the semi-classical causality to an error-correcting condition\footnote{For the non-gravitating case, the implications for semi-classical causality are discussed in \cite{Balasubramanian:2022fiy}.}, and show that such a problem does not occur when the (R\'enyi-2) mutual information vanishes.

To make this connection explicit, let us consider the error-correcting condition within the framework of operator algebra quantum error correction (OAQEC) \cite{beny2007generalization,beny2007quantum}, which is equivalent to the decoupling condition\footnote{One can show this equivalence by combining the equivalence between the decoupling condition and the Knill-Laflamme condition \cite{nielsen2007algebraic} and that between the error correcting condition in OAQEC and the Knill-Laflamme condition \cite{beny2007generalization,beny2007quantum}.}, \eqref{eq:decoupling-condition}, quantified by the vanishing of the (R\'enyi-2) mutual information. Concretely, this equivalence can be expressed as
\begin{equation}
	\begin{aligned}
		&\rho_{ref,\, ref(ex),\, E}' = \rho_{ref(in)}' \otimes \rho_{ref(ex),\, E}' \qquad (\text{\eqref{eq:decoupling-condition} : Decoupling condition}) \\
		  &\qquad\Longleftrightarrow \\
		   &[P_{code}E^{\dagger}_{m}E_{n}P_{code},\mathcal{O}_{code(in)}]=0 \qquad \forall m,n, \quad \forall \mathcal{O}_{code(in)} \in \mathcal{B}_{code(in)}\subset \mathcal{B}_{code}.\\
		   &\hspace{6cm} \text{(error correcting condition in OAQEC)}
	\end{aligned}
\end{equation}
Note that, equivalently, the commutator is non-vanishing if and only if the decoupling condition does not hold.
Here, $P_{code}$ is a projector onto the code subspace $H_{code}\subset H_{phys}$, where the code subspace is embedded via the isometric map $V$ in \eqref{eq:phsyStateTwoGraASym1}, and it is given by $P_{code}=VV^{\dagger}$. $\mathcal{B}_{code}$ is an algebra of observables acting on the code subspace, given by $\mathcal{B}_{code}=V\mathcal{B}V^{\dagger}$ where $\mathcal{B}$ is the corresponding algebra in the domain of definition of $V$\footnote{Let $\phi\in \mathcal{B}$ be an operator in the domain of definition of $V$ and explicitly given by 
\begin{equation}
    \phi=\sum_{i,j=1}^{d_{in}}\sum_{i',j'=1}^{d_{ex}}\phi_{i,j;i',j'}\ket{i,i'}\bra{j,j'},
\end{equation}
then the corresponding code operator $\mathcal{O}_{code}$ is given by 
\begin{equation}
	\begin{aligned}
		\mathcal{O}_{code}&=V\phi V^{\dagger}=\sum_{i,j=1}^{d_{in}}\sum_{i',j'=1}^{d_{ex}}\phi_{i,j;i',j'}\ket{\Psi_{i,i'}}_{phys}\bra{\Psi_{j,j'}}.
	\end{aligned}
\end{equation}
}, and $\mathcal{B}_{code(in)}$ is a subalgebra corresponding to the interior part, $\mathcal{B}_{code(in)}=V\mathcal{B}_{(in)}V^{\dagger}$ where $\mathcal{B}_{(in)}\subset \mathcal{B}$\footnote{For an interior operator $\phi_{(in)}$, we have 
\begin{equation}
    \phi_{(in)}=\sum_{i,j=1}^{d_{in}}\sum_{i'=1}^{d_{ex}}\phi_{i,j}\ket{i,i'}\bra{j,i'} \quad \Longrightarrow \quad \mathcal{O}_{code(in)}=V\phi_{(in)}V^{\dagger}=\sum_{i,j=1}^{d_{in}}\sum_{i'=1}^{d_{ex}}\phi_{i,j}\ket{\Psi_{i,i'}}_{phys}\bra{\Psi_{j,i'}}.
\end{equation}
}. Note that for $\mathcal{O}_{code} \in \mathcal{B}_{code}$ (and similarly for $\mathcal{O}_{code(in)} \in \mathcal{B}_{code(in)}$), we have the relation $\mathcal{O}_{code}=P_{code}\mathcal{O}_{code}=\mathcal{O}_{code}P_{code}$\footnote{We can easily show the relation as follows:
\begin{equation}
	\begin{aligned}
		P_{code}\mathcal{O}_{code}=(VV^{\dagger})(V\mathcal{O}V^{\dagger})&=V(V^{\dagger}V)\mathcal{O}V^{\dagger}\\
		&=V\mathcal{O}V^{\dagger}=\mathcal{O}_{code},
	\end{aligned}
\end{equation}
where, in the second equality of the first line, we used the property of the isometric embedding map, $V^{\dagger}V=I$. Similarly, we can show $\mathcal{O}_{code}P_{code}=\mathcal{O}_{code}$.
}.

By using the relation $\mathcal{O}_{code(in)}=P_{code}\mathcal{O}_{code(in)}=\mathcal{O}_{code(in)}P_{code}$, the above error correcting condition in OAQEC is equivalently written as
\begin{equation}
	P_{code}[E^{\dagger}_{m}E_{n},\mathcal{O}_{code(in)}]P_{code}=0 \qquad \forall m,n, \quad \forall \mathcal{O}_{code(in)} \in \mathcal{B}_{code(in)}.
\end{equation}
We again note that this commutator vanishes if and only if the decoupling condition holds, which is equivalent to the vanishing of the mutual information.

This commutation relation implies that when the (R\'enyi-2) mutual information does not vanish, the commutator does not vanish as well, leading to the violation of the semi-classical causality (more precisely, the micro-causality) between the black hole interior and the bath system; the commutator between a semi-classical interior operator on universe $A$ and  a semi-classical bath operator i.e., the Kraus operator, on universe $B$ does not vanish;
\begin{equation}
	\rho_{ref,\, ref(ex),\, E}' \neq  \rho_{ref(in)}' \otimes \rho_{ref(ex),\, E}' \quad  \Longleftrightarrow \quad [\phi_{B},\phi_{in}]\neq 0 \text{ in semi-classical Hilbert space},\label{eq:semi-classical-commu-1}
\end{equation}
where $\phi_{in}$ is a semi-classical interior operator on universe $A$, corresponding to the operator on the code subspace $\mathcal{O}_{code(in)}$, and $\phi_{B}$ is a bath operator on universe $B$, corresponding to the Kraus operators $E^{\dagger}_{m}E_{n}$. 
 On the other hand, when the (R\'enyi-2) mutual information vanishes, the commutator vanishes, and in this case, such a causality violation does not happen;
 \begin{equation}
	\rho_{ref,\, ref(ex),\, E}' =  \rho_{ref(in)}' \otimes \rho_{ref(ex),\, E}' \quad  \Longleftrightarrow \quad [\phi_{B},\phi_{in}]= 0 \text{ in semi-classical Hilbert space}.\label{eq:semi-classical-commu-2}
\end{equation}

As we have seen, if we do not consider the gravitational backreaction from an error and the error has sufficiently negative coherent information, we encounter this violation of causality.
However, once the gravitational backreaction from an error is taken into account, the (R\'enyi-2) mutual information vanishes when the scaling dimension exceeds the critical value, leading to the decoupling condition, and thus the violation of causality does not occur.
This phenomenon occurs only in gravitating bath cases since there are no such gravitational degrees of freedom in non-gravitating bath systems\footnote{Although we expect that this mechanism is unique for gravitating bath cases, it might be interesting to find a similar mechanism from a purely quantum information theoretic perspective.}.
Thus, gravitational degrees of freedom play a crucial role in preserving semi-classical causality by effectively suppressing error effects that would otherwise disturb the interior information.

\section{Details of the backreacted Dilaton profile}
\label{app:detailDilaton}

In this appendix, we give a backreacted dilaton profile from the brane insertion, which is identified with the backreaction of the Kraus operator onto the spacetime.

\subsection{Dilaton profiles}\label{app:backDilaton}
We start with the usual boundary conditions \eqref{eq:bdyMetricDilaton} for the metric and the dilaton profile,
\begin{equation*}
	ds^{2}|_{\partial \mathcal{M}} = \frac{du^{2}}{\epsilon^{2}}, \qquad \phi|_{\partial \mathcal{M}}  = \frac{\phi_{r}}{\epsilon},
\end{equation*}
where $\epsilon$ is an infinitesimal cutoff, $u$ is a physical time with the periodicity $u\sim u+\beta$, and $\phi_{r}$ is a renormalized dilaton value at the boundary. 

In JT gravity, we can treat the dilaton profile by using the embedding coordinates of an (Euclidean) AdS$_{2}$ spacetime,
\begin{equation}
	ds^{2}= -(dX_{-1})^{2}+(dX_{0})^{2}+(dX_{1})^{2}
\end{equation}
where the embedding coordinate $X^{a}$ obeys the condition
\begin{equation}
	X.X\coloneqq-(X_{-1})^{2}+(X_{0})^{2}+(X_{1})^{2} =-1,
\end{equation}
We can choose these coordinates to be e.g., the Euclidean Rindler coordinates
\begin{equation}
	X^{a}=(\cosh \rho,\sinh\rho \sin\tau, \sinh\rho \cos \tau), \qquad \left(\rho\geq 0, \, 0 \leq \tau \leq 2\pi  \right),
\end{equation}
and in this case, the metric is given by
\begin{equation}
	ds^{2}= d\rho^{2}+\sinh^{2}\rho\, d\tau^{2}.\label{eq:rindlerMetric}
\end{equation}
Also, a dilaton profile can be taken to be 
\begin{equation}
	\phi_{C}= Q_{C}.X= \frac{2\pi}{\beta} \phi_{r} \,   \cosh\rho,
\end{equation}
where $Q_{C}$ is a SL$(2,\mathbb{R})$ charge for this dilaton profile
\begin{equation}
	\begin{aligned}
		Q_{C}^{a} &= \frac{2\pi}{\beta} \phi_{r} \,  (1, 0 , 0),
	\end{aligned}\label{eq:dilaProNaive}
\end{equation}
specifying the location of the horizon as
\begin{equation}
		Y_{C}^{a}=(1, 0 , 0).
\end{equation}
In the Euclidean Rindler coordinates, the above vectors correspond to the point $\rho=0$.
 Also, this dilaton profile determines the boundary trajectory $(\rho_{\infty}(\tau),\tau)$ as 
\begin{equation}
	\rho_{\infty}(\tau)\approx \log\left[ \frac{\beta}{\pi \epsilon} \right]=\rho_{\infty},\label{eq:bdytraUndeformed}
\end{equation}
and  the physical boundary time $u$,
\begin{equation}
	u=\frac{\beta}{2\pi} \tau.
\end{equation}

\subsection{Gluing two dilaton profiles along a massive brane}

The above dilaton profile does not include a parameter which can be related to  backreactions from insertions of the brane, related to Kraus operators. To introduce such a parameter, we can utilize the SL$(2,\mathbb{R})$ symmetry of NAdS$_{2}$. First, we consider two embedding coordinates $\tilde{X}^{a}_{(L)}$ and $\tilde{X}^{a}_{(R)}$ which are related to the above one $X^{a}$ by the following SL$(2,\mathbb{R})$ rotations,
\begin{equation}
	\begin{dcases}
		\tilde{X}^{a}_{(L)} \coloneqq (U_{L}^{-1})^{a}_{b}X^{b} & \qquad \left( \frac{\pi}{2} \leq \tau \leq \frac{3\pi}{2} \right) \iff \left( X^{3}\geq 0 \right), \\
		 \tilde{X}^{a}_{(R)} \coloneqq (U_{R}^{-1})^{a}_{b}X^{b} & \qquad \left( 0 \leq \tau \leq \frac{\pi}{2}, \quad \frac{3\pi}{2} \leq \tau \leq2\pi\right) \iff \left( X^{3}\leq  0 \right),
	\end{dcases}
\end{equation}
where $U_{L}$ and $U_{R}$ are given by
\begin{equation}
    U_{L} = \begin{pmatrix}
\cosh \rho_{0} & 0 & -\sinh\rho_{0} \\
0 & 1 & 0 \\
-\sinh\rho_{0} & 0 & \cosh\rho_{0} 
\end{pmatrix}, \qquad U_{R} = \begin{pmatrix}
\cosh \rho_{0} & 0 & \sinh\rho_{0} \\
0 & 1 & 0 \\
\sinh \rho_{0} & 0 & \cosh \rho_{0} 
\end{pmatrix}.
\end{equation}
We note that for these matrices, we have the relation $U_{R}^{-1}=U_{L}$, and these SL$(2,\mathbb{R})$ matrices leave the metric $\eta_{ab}=\text{diag}(-1,1,1)$ invariant,
\begin{equation}
	U_{L(R)}^{T} \, \eta \, U_{L(R)}=\eta,
\end{equation}
which is a consequence of the SL$(2,\mathbb{R})$ invariance. Using these tilde coordinates $\tilde{X}^{a}_{(L)}$ and $\tilde{X}^{a}_{(R)}$, the canonical dilaton profile \eqref{eq:dilaProNaive} corresponds to two dilaton profiles as follows;
\begin{itemize}
	\item for  $\left( \frac{\pi}{2} \leq \tau \leq \frac{3\pi}{2} \right) \iff \left( X^{3}\geq 0 \right)$,
	\begin{equation}
	\phi_{C}= Q_{C}.X= Q_{C}.\left( U_{L}\tilde{X}_{(L)}\right) = \left( U_{L}^{-1}Q_{C} \right).\tilde{X}_{(L)} = -Q_{L}.\tilde{X}_{(L)} \eqqcolon \phi_{L},
\end{equation}
	\item for $\left( 0 \leq \tau \leq \frac{\pi}{2}, \frac{3\pi}{2} \leq \tau \leq2\pi\right) \iff \left( X^{3}\leq  0 \right)$,
	\begin{equation}
	\phi_{C}= Q_{C}.X= Q_{C}.\left( U_{R}\tilde{X}_{(R)}\right) = \left( U_{R}^{-1}Q_{C} \right).\tilde{X}_{(R)} = Q_{R}.\tilde{X}_{(R)} \eqqcolon \phi_{R}.
\end{equation}
\end{itemize}
Here we defined $Q_{L}$ and $Q_{R}$ by
\begin{equation}
	\begin{aligned}
		Q_{L}^{a}&=-\left(U_{L}^{-1}\right)^{a}_{b}Q_{C}^{b}= \frac{2\pi}{\beta} \phi_{r} \,  (\cosh \rho_{0}, 0 , -\sinh\rho_{0}),\\
		Q_{R}^{a}&=\left(U_{R}^{-1}\right)^{a}_{b}Q_{C}^{b}= \frac{2\pi}{\beta} \phi_{r} \,  (\cosh \rho_{0}, 0 , \sinh\rho_{0}).
	\end{aligned} \label{eq:RotatedSL2}
\end{equation}
These rotated SL$(2,\mathbb{R})$ charges specify the horizons in the tilde coordinates as
\begin{equation}
	\begin{cases}
		\tilde{Y}_{L}^{a}= (U_{L}^{-1})^{a}_{b}Y^{b}_{C}=(\cosh \rho_{0}, 0 , -\sinh\rho_{0}) & \text{for } \phi_{L}, \\
		\tilde{Y}_{R}^{a}=(U_{R}^{-1})^{a}_{b}Y^{b}_{C}=(\cosh \rho_{0}, 0 , \sinh\rho_{0}) & \text{for } \phi_{R}.
	\end{cases}
\end{equation}
In the Euclidean Rindler coordinates, these coordinates correspond to points $(\tilde{\rho}_{(L)},\tilde{\tau}_{(L)})=(\rho_{0},\pi)$ and $(\tilde{\rho}_{(R)},\tilde{\tau}_{(R)})=(\rho_{0},0)$ respectively. We note that in these tilde embedding coordinates, the dilaton profiles become
\begin{equation}
	\phi_{L}= \frac{2\pi}{\beta} \phi_{r} \,  (\cosh \rho_{0} \cosh\tilde{\rho}+\sinh\rho_{0}\sinh\tilde{\rho} \cos \tilde{\tau} ) \qquad \left(\frac{\pi}{2} \leq  \tilde{\tau} \leq \frac{3\pi}{2}\right)  \label{eq:dilatonL}
\end{equation}
and 
\begin{equation}
	\phi_{R}=  \frac{2\pi}{\beta} \phi_{r} \,  (\cosh \rho_{0} \cosh\tilde{\rho}-\sinh\rho_{0}\sinh\tilde{\rho} \cos \tilde{\tau}) \qquad \left( 0 \leq  \tilde{\tau} \leq \frac{\pi}{2}, \frac{3\pi}{2} \leq  \tilde{\tau} \leq 2\pi\right). \label{eq:dilatonR}
\end{equation} 
Note that $\phi_{L},\phi_{R}$ are continuous at $\tilde{\tau}=\frac{\pi}{2},\frac{3\pi}{2} $, but their derivatives are not. The discontinuity is related to the brane mass.  This dilaton profile amounts to what is obtained by gluing two spacetimes, described by the two dilaton profiles respectively, along the brane.

In the above discussions, the parameter $\rho_{0}$ is just a free parameter and is not yet related to the backreaction from the Kraus operators, which corresponds to the backreaction from a brane insertion.
 Let us consider their relations.
 Using these SL$(2,\mathbb{R})$ charges and the SL$(2,\mathbb{R})$ charge conservation $Q_{L}^{a}+Q_{R}^{a}+ Q_{B}^{a}=0$ ($ Q_{B}^{a}$:  brane SL$(2,\mathbb{R})$ charge), we can specify the brane SL$(2,\mathbb{R})$ charge as
\begin{equation}
	Q_{B}^{a} = 2\frac{2\pi}{\beta} \phi_{r} \,  (0, 0 , \sinh\rho_{0}). \label{eq:branecharge}
\end{equation}
The norm of this brane charge is related to the scaling dimension or the brane mass $\Delta$, \cite{Kourkoulou:2017zaj,Goel:2018ubv,Bulycheva:2019naf},
\begin{equation}
	2\frac{2\pi}{\beta} \phi_{r} \sinh\rho_{0} =\Delta. \label{eq:branePosiMass}
\end{equation}
This equation relates the parameter $\rho_{0}$ to the brane mass $\Delta$. 
Once given the brane charge, we can also specify the brane trajectory $\tilde{Y}_{B}$ by 
\begin{equation}
	Q_{B}.\tilde{Y}_{B}=0 \qquad \rightarrow \tilde{Y}_{B}^{a} = (\cosh \tilde{\rho},\pm  \sinh\tilde{\rho} ,0) \quad 0\leq \tilde{\rho}.\label{eq:branePoint}
\end{equation}

By contracting this vector $\tilde{Y}$ with the charges \eqref{eq:RotatedSL2} defining the dilaton profiles, we obtain the dilaton profile along the brane,
\begin{equation}
	\begin{aligned}
		\left.\phi_{C}\right|_{\text{Brane}}=\left.\phi_{L}\right|_{\text{Brane}}=\left.\phi_{R}\right|_{\text{Brane}}&= \frac{2\pi}{\beta} \phi_{r} \cosh \rho_{0} \cosh\tilde{\rho}\\
		&=\sqrt{\left(\frac{2\pi}{\beta} \phi_{r}\right)^{2}+\left(\frac{\Delta}{2}\right)^{2}}\cosh\tilde{\rho},
	\end{aligned}
\end{equation}
where, in the second line, we used the relation \eqref{eq:branePosiMass}.
From this expression, we can find that the minimum value of the dilaton along the brane,
\begin{equation}
	\begin{aligned}
		\left.\phi_{C}\right|_{\text{Brane,Min}}= \min_{x\in \text{Brane trajectory}} \{\left.\phi_{C}\right|_{\text{Brane}}\} &= \sqrt{\left(\frac{2\pi}{\beta} \phi_{r}\right)^{2}+\left(\frac{\Delta}{2}\right)^{2}}	.\label{eq:dilatonMinAlongBrane}
		\end{aligned}
\end{equation}

\section{Upper bound of the scaling dimension of the error}\label{app:UpperBoundScaling}

From the relation \eqref{eq:angleDeltaPhib}, $\Delta$ can not be greater than $\Delta_{\max}=2\phi_{b}$ since $\sin \theta \leq 1$. 
In this appendix, we discuss the meaning of the upper bound. Note that this discussion is not directly related to the structure of the phase diagram, but rather to the validity of the semi-classical approximation used in computing the phase diagram.

\begin{figure}[ht]
	\centering
	\includegraphics[scale=1]{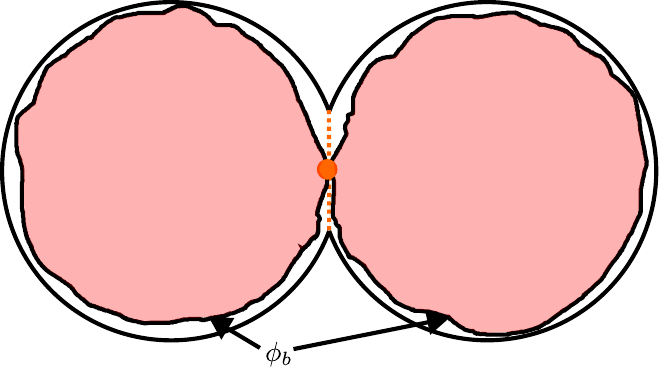}
	\caption{Schematic picture of a brane trajectory at the threshold brane mass. In this case, the brane length approaches zero. This picture is simplified to explain the limiting brane configuration.}\label{fig:saddle_Limiting_cusp}
\end{figure}

First, let us briefly explain the situation where $\sin \theta \lesssim 1$. In this case, we have $\theta \lesssim \pi/2$. This means that the cusp angle approaches its maximal value, and the two cusps come very close to each other, leading to a configuration where the two black hole spacetimes are almost pinched out\footnote{See \S 6.1 of \cite{Kourkoulou:2017zaj} for the similar discussion.}. More precisely, the corresponding brane trajectory (almost) goes out outside the cutoff surface. See figure \ref{fig:saddle_Limiting_cusp}.
Thus, the contribution from the brane action \eqref{eq:bulkMassive} becomes very small or vanishes due to the cutoff. Consequently, when $\Delta$ exceeds the specific value, the gravitational backreaction from the error brane loses its meaning, since the brane action does not contribute regardless of how large the scaling dimension $\Delta$ (which appears as its coefficient in the brane action) becomes.

Next, let us check that the above consideration indeed gives the upper bound. 
For the brane trajectory to make sense, we need to require that the brane trajectory is located inside the cutoff surface. Otherwise, we can not see the trajectory where two spacetimes are glued, implying they are effectively disconnected. This condition requires that the minimum value of the dilaton along the brane is smaller than the boundary dilaton value.
As derived in appendix \ref{app:detailDilaton}, the minimum value of the dilaton along the brane is given by \eqref{eq:dilatonMinAlongBrane},
\begin{equation}
	\begin{aligned}
		\left.\phi_{C}\right|_{\text{Brane,Min}} &= \sqrt{\left(\frac{2\pi}{\beta} \phi_{r}\right)^{2}+\left(\frac{\Delta}{2}\right)^{2}}.
		\end{aligned}
\end{equation}
This minimum value must be smaller than the boundary dilaton value;
\begin{equation}
	\begin{aligned}
		\phi_{b} \geq \left.\phi_{C}\right|_{\text{Brane,Min}} &= \sqrt{\left(\frac{2\pi}{\beta} \phi_{r}\right)^{2}+\left(\frac{\Delta}{2}\right)^{2}}.
	\end{aligned}
\end{equation}
We can also rewrite the relation as
\begin{equation}
	0\leq \left(\frac{\phi_{\text{Horizon}}}{\phi_{b}} \right)^{2}+\left(\frac{\Delta}{2\phi_{b}}\right)^{2} \leq 1,\label{eq:conditonBdyHorizonDelta}
\end{equation}
where $\phi_{\text{Horizon}}$ is the dilaton value at the horizon, 
\begin{equation}
	\phi_{\text{Horizon}}=\frac{2\pi}{\beta} \phi_{r}.
\end{equation}
Since, in physically reasonable situations, the dilaton value at the horizon, $\phi_{\text{Horizon}}$, is smaller than the boundary dilaton value, $\phi_{b}$,
\begin{equation}
	0\leq \frac{\phi_{\text{Horizon}}}{\phi_{b}}\leq 1,
\end{equation}
the relation \eqref{eq:conditonBdyHorizonDelta} implies the inequality 
\begin{equation}
	0\leq \frac{\Delta}{2\phi_{b}} \leq 1,\label{eq:GeometricCondition}
\end{equation}
consistent with the relation \eqref{eq:angleDeltaPhib}. This indeed gives the upper bound, $\Delta\leq \Delta_{\max}=2\phi_{b}$.

From the above analysis, we can also say that, with a scaling dimension $\Delta$ greater than $2\phi_{b}$, we would lose the validity of the semi-classical gravitational description, since the brane is located outside the cutoff surface, $\phi_{b} \leq \left.\phi_{C}\right|_{\text{Brane,Min}}$.

\subsection{Hierarchy among parameters}
\label{app:Hierarchy}

In addition to \eqref{eq:GeometricCondition}, there are other conditions to hold,  in order to ensure the validity of the semi-classical approximation. Since JT gravity is obtained by the  dimensional reduction of the effective action  describing slight excitations of  4D  extremal  black hole,  the dominant contribution comes from  the topological part rather than the dynamical part,
\begin{equation}
	S_{0} \gg \phi_{b} \gg \phi_{r}.
\end{equation}
Here, we note that the boundary dilaton value $\phi_{b}$ plays a role of the cutoff for the dynamical dilaton part. We note that dominant contributions of the black hole entropy come from the topological part of the JT action, and the dynamical part gives a correction to the topological contribution,
\begin{equation}
	S_{BH}=S_{0}+\frac{2\pi}{\beta}\phi_{r}\approx S_{0}.
\end{equation}

As in the beginning of section \ref{sec:QECNobackreaction}, the entropy of the bulk semi-classical excitations, $\log d_{in}, \log d_{ex}$, are assumed to be smaller than that of the black hole entropy, $\log d_{BH}=S_{BH}$;
\begin{equation}
	\log d_{in}, \log d_{ex} \ll \log d_{BH}=S_{BH}.
\end{equation}
Furthermore, since the semi-classical excitations should not give a large backreaction, their entropies should also be smaller than the dynamical contribution of the JT gravity, 
\begin{equation}
\log d_{in}, \log d_{ex} \lesssim \phi_{b}.
\end{equation}

Next, the semi-classical entropy of the Hawking radiation, $\log k$, can be comparable or exceed the black hole entropy;
\begin{equation}
	\log k \lessgtr \log d_{BH}=S_{BH}.
\end{equation}

The two (R\'enyi-two) entropies related to the error, $S^{(2)}(\sigma_{\text{Bath}}),S^{(2)}(\sigma_{E})$ are at most of order of the black hole entropy,
\begin{equation}
	S^{(2)}(\sigma_{\text{Bath}}) \lesssim  \log d_{BH}=S_{BH},
\end{equation} 
\begin{equation}
	S^{(2)}(\sigma_{E}) \lesssim  2\log d_{BH}=2S_{BH},
\end{equation}
and 
\begin{equation}
	\left| S^{(2)}(\sigma_{\text{Bath}})-S^{(2)}(\sigma_{E}) \right| \leq \log d_{BH}=S_{BH}.
\end{equation}

Finally, the contribution of the gravitational backreaction from the error is much smaller than the above contribution as discussed in the previous subsection,
\begin{equation}
	\Delta \lesssim 2 \phi_{b} \ll S_{0} \approx S_{BH}.
\end{equation}

Under the above hierarchy of the parameters, we can consider the parameter region, where the (R\'enyi-two) mutual information for the decoupling condition vanishes.

\bibliographystyle{JHEP}
\bibliography{island_new}

\end{document}